\documentclass[12pt,letterpaper]{article}
\usepackage{jheppub}

\usepackage{graphicx}
\usepackage{bbm}
\usepackage{amsmath}
\usepackage{amssymb}
\usepackage{mathtools}
\usepackage{amsthm}
\usepackage{physics}
\usepackage{float}
\usepackage{upgreek}

\usepackage[labelformat=simple]{subcaption}

\newcommand{\calF}{\mathcal{F}}

\newcommand{\ztwo}{{\mathbb{Z}_2}}
\newcommand{\thetaztwo}{\theta_{\mathbb{Z}_2}}
\newcommand{\thetaFL}{\theta_{F_L}}
\newcommand{\df}{\mathrel{\equiv}}

\newcommand{\RR}{\mathbb{R}}
\newcommand{\ZZ}{\mathbb{Z}}

\newcommand{\AM}{{A_\textsc{M}}}

\newcommand{\rmd}{\mathrm{d}}
\newcommand{\bfq}{\vec{q}}
\newcommand{\tildeq}{Q}
\newcommand{\bfQ}{\vec{Q}}

\newcommand{\U}[1]{\mathrm{U}\!\left(#1\right)}

\DeclareMathOperator{\Out}{Out}

\newcommand{\msplit}{m_{\mathrm{split}}}
\newcommand{\Tsplit}{T_{\mathrm{split}}}
\newcommand{\Tsplitp}[1]{T_{\mathrm{split},{#1}}}
\newcommand{\Tsplitstr}{T_{\mathrm{split},1}}
\newcommand{\Tsplitfive}{T_{\mathrm{split},5}}
\newcommand{\Tflux}{T_{\mathrm{flux}}}
\newcommand{\Tfluxp}[1]{T_{\mathrm{flux},{#1}}}
\newcommand{\Tfluxstr}{T_{\mathrm{flux},1}}
\newcommand{\Tfluxfive}{T_{\mathrm{flux},5}}
\newcommand{\tstr}{t_1}
\newcommand{\tfive}{t_5}
\newcommand{\tildetstr}{{\tilde t}_1}
\newcommand{\tildetfive}{{\tilde t}_5}
\newcommand{\tildetstrfive}{{\tilde t}_{1,5}}
\newcommand{\Tstrfive}{T_{1,5}}
\newcommand{\vecalphasplitstr}{\vec \alpha_{\mathrm{split},1}}
\newcommand{\vecalphafluxstr}{\vec \alpha_{\mathrm{flux},1}}
\newcommand{\vecalphasplitfive}{\vec \alpha_{\mathrm{split},5}}
\newcommand{\vecalphafluxfive}{\vec \alpha_{\mathrm{flux},5}}

\def\be{\begin{equation}}
	\def\ee{\end{equation}}

\newtheoremstyle{named}{0.75\baselineskip}{0.75\baselineskip}{\itshape}{}{\bfseries}{.}{.5em}{#3}
\theoremstyle{named}

\begin{document}
	\begin{titlepage}

		\setcounter{page}{1} \baselineskip=15.5pt \thispagestyle{empty}
		{\flushright {ACFI-T25-01} \\ }
		
		\bigskip\
		
		\vspace{0.5cm}
		\begin{center}
				{\LARGE \bfseries  Confined monopoles and failure of the \vspace{.24cm}\\ Lattice Weak Gravity Conjecture}
				
			\end{center}
			\vspace{0.5cm}
			
			\begin{center}
				{\fontsize{14}{30}\selectfont Muldrow Etheredge$^{a}$, Ben Heidenreich$^a$, Nicholas Pittman$^a$,\\ Sebastian Rauch$^a$, Matthew Reece$^b$, and Tom Rudelius$^c$}
			\end{center}

			\begin{center}
				\vspace{0.25 cm}
				\textsl{$^a$Department of Physics, University of Massachusetts, Amherst, MA 01003 USA}\\
				\textsl{$^b$Department of Physics, Harvard University, Cambridge, MA, 02138, USA}\\
				\textsl{$^c$Department of Mathematical Sciences, Durham University, Durham, DH1 3LE, UK}\\

				\vspace{0.25cm}
				
			\end{center}
			\vspace{1cm}
			\noindent

			Almost all known theories of quantum gravity satisfy the Lattice Weak Gravity Conjecture (LWGC), which posits that a consistent theory of quantum gravity must have a superextremal particle at every site in the charge lattice. However, a number of theories have been observed to violate the LWGC; such theories exhibit only a (finite index) sublattice of superextremal particles. This paper aims to identify universal features and patterns associated with LWGC violation across numerous examples in effective field theory, string theory, and M-theory. Some of these examples have appeared previously in the literature, while others are novel. In all such examples, we observe that LWGC failure is accompanied by the existence of fractionally charged monopoles confined by flux tubes, where superextremal particles exist everywhere in the sublattice dual to the superlattice of fractional confined monopole charges. The confining flux tubes become light when the failure of the LWGC becomes more extreme, so monopoles deconfine in the limit where LWGC-violating particles become infinitely massive. We also identify similarities between these confined monopoles, non-invertible symmetries, and the Hanany-Witten effect.
			
			\vspace{.9cm}
			
			\bigskip
			\noindent\today

			\end{titlepage}
			
			\setcounter{tocdepth}{2}
			
			\hrule
			\tableofcontents
			
			\bigskip\medskip
			\hrule
			\bigskip\bigskip

			\section{Introduction}	The Weak Gravity Conjecture (WGC) \cite{Arkanihamed:2006dz} is a simple yet profound statement about effective field theories coupled to quantum gravity. Roughly speaking, the WGC states that gravity should not be strong enough to stop black holes from decaying. More precisely, in any consistent theory of quantum gravity with a massless abelian gauge boson, the WGC requires a ``superextremal'' particle whose charge-to-mass ratio is greater than or equal to that of a large extremal black hole:
			\begin{equation}
			\text{WGC: }~~~ \frac{|q|}{m} \geq \frac{|Q|}{M}\biggr|_{\text{large ext BH}} \equiv Z_{\rm ext}.
			\label{WGCbound}
			\end{equation}
			
			In the nearly twenty years since its proposal, the original idea of the WGC stated above has undergone a series of refinements and generalizations.\footnote{ See \cite{Harlow:2022ich} for a review and \cite{Rudelius:2024mhq} for a pedagogical introduction.} These statements have important applications to particle physics, black holes, cosmology, mathematics, CFTs, and more. Thus, it is crucial to understand which versions of the WGC, if any, are correct.
			
			One promising version of the WGC, called the Lattice WGC (LWGC), was put forth in \cite{Heidenreich:2015nta}. This statement holds that \emph{every} site in the charge lattice of a U$(1)$ gauge theory in quantum gravity must have a particle that satisfies the Weak Gravity bound \eqref{WGCbound}. In theories with multiple U$(1)$s, the LWGC compares the charge-to-mass ratio of each particle to that of a large extremal black hole in the same direction in the charge lattice. 
			
			The LWGC is satisfied in many theories within the string landscape. However, counterexamples to the LWGC do exist. The first explicit examples were found in \cite{Heidenreich:2016aqi}, based on earlier ideas in \cite{Arkanihamed:2006dz}. In all of these examples, although the LWGC is violated, there exists a finite-index sublattice of the electric charge lattice with particles that satisfy the WGC bound. This led to the formulation of a slightly weaker conjecture called the sublattice WGC (sLWGC), which holds that there exists an integer $k$ such that for any $\bfq$ in the electric charge lattice $\Gamma$, there exists a superextremal particle of charge $k \bfq$. We refer to this integer $k$ as the \emph{coarseness} of the sublattice. In \cite{Heidenreich:2016aqi}, all examples studied had coarseness $k=1$, 2, or 3.
			
			The sLWGC can be proven to hold within the context of perturbative string theory using modular invariance \cite{Heidenreich:2016aqi, Montero:2016tif, Heidenreich:2024dmr}. Away from the perturbative limit, the conjecture has been tested using BPS states~\cite{Grimm:2018ohb, Gendler:2020dfp, Bastian:2020egp,  Alim:2021vhs, Gendler:2022ztv} and emergent string limits of F-theory~\cite{Lee:2018urn, Lee:2018spm, Klaewer:2020lfg}. There are also examples in which the sLWGC has been neither verified nor falsified. For example, in certain 4d F-theory compactifications, an elliptic genus computation reveals superextremal towers that do not lie in a sublattice~\cite{Lee:2019tst}, but this does not preclude the possibility of other superextremal states that are not detected by the elliptic genus. Examples along these lines (see also~\cite{Cota:2022yjw, Cota:2022maf, Casas:2024ttx}) led to the proposal of the Minimal Weak Gravity Conjecture \cite{FierroCota:2023bsp}, which postulates that the sLWGC must hold only in cases that allow for circle compactifications at arbitrarily small radius. Our assessment is that, at this time, there are strong lines of evidence in favor of the sLWGC, and no strong evidence against it. Even for advocates of the Minimal WGC, it is interesting to understand the sLWGC better in the large class of examples in which it holds. Thus, for the remainder of this paper, we will focus on theories that satisfy the sLWGC but not the LWGC. This raises a number of questions. Why is the LWGC not true in full generality?  Is there an upper bound on the index and/or coarseness of the sLWGC-satisfying sublattice?  What is the more fundamental feature that underlies LWGC violation?
			
			In this paper, we propose that the answer to this last question is \emph{confined, fractionally charged monopoles}. We will show that in a wide variety of theories with LWGC violation---including several novel examples---LWGC violation is accompanied by monopoles of fractional magnetic charge confined by flux tubes. These confined monopole charges live in a superlattice of the magnetic charge lattice, and the dual of this superlattice is a sublattice of the electric charge lattice. In the examples studied below, the charge sites in this sublattice contain superextremal particles, ensuring that the sLWGC is satisfied. The coarseness of the sLWGC-satisfying sublattice $\Gamma_{\rm ext}$ is thus no larger than the coarseness of the magnetic charge lattice, viewed as a sublattice of the confined monopole superlattice.
			
			This relationship between superextremal particles and fractional monopoles offers a satisfying explanation as to why LWGC is violated in certain theories. At the same time, it also explains why the sLWGC remains satisfied in these theories, distinguishing a particular sublattice $\Gamma_{\rm ext}$ where superextremal particles are required. It is still possible, however, for superextremal particles to exist outside $\Gamma_{\rm ext}$, as in the heterotic orbifold examples of \cite{Heidenreich:2016aqi} and described briefly for $\ZZ_p$ discrete gauge theory at the end of \S$\ref{subsec:lwgcZp}$. In other words, all sites in $\Gamma_{\rm ext}$ should contain superextremal particles, but not all superextremal particles necessarily lie in $\Gamma_{\rm ext}$.

            To make a precise, quantitative connection between LWGC violation and fractional monopole confinement, we observe that in the simplest examples, the mass-squared $m^2(q)$ of the lightest particle of a given charge $q \notin \Gamma_{\rm ext}$ is heavier than the WGC bound $m_{\text{ext}}^2(q) = q^2 / Z_{\text{ext}}^2(\hat{q})$ by a $q$-independent (but moduli-dependent) amount, even for very large charges. With this in mind, we define the ``sublattice splitting'' of the extremal sublattice $\Gamma_{\rm ext}$ as:
            \begin{align}
			\msplit^2 \equiv \text{max} \biggl( 0, \inf_{q \in \Gamma_{\text{big}} \setminus \Gamma_{\text{ext}}} [m^2(q) - m_{\rm ext}^2(q)] \biggr)\,,
			\label{Deltadef}
			\end{align}
            where $\Gamma_{\text{big}}$ is the ``large charge'' part of the charge lattice, excluding the region of order-one charges where smaller splittings (or even superextremal particles) might coincidentally appear off $\Gamma_{\rm ext}$.\footnote{A more precise definition of $\Gamma_{\text{big}}$ would be to consider only charges $q\in\Gamma$ satisfying $q^2 \ge \Lambda^2$ then afterwards take the limit $\Lambda \to \infty$.} Note that $\msplit$ vanishes if there are superextremal particles of arbitrarily large charge outside of $\Gamma_{\rm ext}$.\footnote{For example, the heterotic orbifold models studied in 
            \S3.2 of \cite{Heidenreich:2016aqi} feature LWGC violation at a nonzero, finite number of sites in the charge lattice. The definition \eqref{Deltadef} thus gives $\msplit = 0$ for these models, and \eqref{geninv} is satisfied trivially.}

            In what follows, we argue that if $\msplit$ is large (i.e., $\msplit \gg 1$ in Planck units), then the tensions of the flux tubes confining the fractional monopoles must be small (i.e., $\Tflux \ll 1$ in Planck units). In the limit where the masses of subextremal particles diverge and are effectively eliminated from the spectrum of the theory, the tensions of the flux tubes vanish, and the associated monopoles become unconfined. In a sense, this may be viewed as a dynamical change of the global structure of the gauge group, $G \rightarrow \tilde G = G/K$, as the electric charge lattice shrinks to a sublattice and the magnetic charge lattice expands to include the charges of the confined monopoles. The Completeness Hypothesis \cite{polchinski:2003bq} then requires that these monopoles deconfine in the limit where the masses of subextremal particles diverge. Conversely, Dirac quantization requires that these particle masses diverge whenever the fractionally charged monopoles deconfine.
            
            Alternatively, we may think of the $\msplit \rightarrow \infty$, $\Tflux \rightarrow 0$ limit as a limit where a $K$ 1-form symmetry emerges as LWGC-violating charged particles become infinitely heavy. This seems to imply the existence of an approximate 1-form symmetry when $\msplit/M_d \gg 1$, which violates the standard lore that approximately global symmetries must be badly broken at or below the Planck scale \cite{Fichet:2019ugl, Nomura:2019qps, Cordova:2022rer}. This apparent discrepancy is evidently resolved by the fact that the emergent symmetry $K$ is a gauge symmetry rather than a global symmetry. In this limit, the magnetic charge lattice is extended, and Dirac quantization implies that either (a) {\em all} objects with electric charge in $\Gamma \setminus \Gamma_\mathrm{ext}$ must become infinitely heavy 
            or else (b) the charge lattice must increase in dimension such that the particles with charge in $\Gamma \setminus \Gamma_\mathrm{ext}$ develop an integral Dirac pairing with the asymptotically deconfined monopoles. In the former case, even black hole states with charges in $\Gamma \setminus \Gamma_\mathrm{ext}$ must become arbitrarily heavy, rather than lying along the classical extremality bound.

			Magnetic monopole confinement is closely linked with the spontaneous breaking of a gauge symmetry, as encapsulated by the adage that ``Higgsing is dual to confinement.'' Indeed, as shown previously in \cite{Saraswat:2016eaz}, a $U(1)^n$ gauge theory that satisfies the LWGC can, under certain conditions, be Higgsed to a $U(1)^{n-1}$ gauge theory that violates the LWGC. Likewise, all of the examples of LWGC violation that we encounter feature massive gauge bosons---a hallmark of the Higgs mechanism. These examples satisfy a further quantitative relation between the gauge boson mass $m_{\gamma}$, the flux tube tension $\Tflux$, and the sublattice splitting $\msplit$: 
			\begin{equation}
			\msplit \cdot \Tflux \leq c  \, m_{\gamma} M_d^{d-2}\,,
			\label{geninv}
			\end{equation}
			where $c$ is an order-one number, and $M_d$ is the reduced Planck mass in $d$ dimensions. In other words, for fixed photon mass $m_\gamma$, \eqref{geninv} implies that the degree of LWGC violation is inversely proportional to flux tube tension. 

            In the above discussion, we assumed that the degree of splitting off of $\Gamma_{\rm ext}$ could characterized by a single number. In more complicated examples there may be different characteristic splittings in different parts of the charge lattice, corresponding to more than one fractional monopole confinement scale. Such cases can be accounted for by defining the sublattice splitting with respect to an arbitrary proper sublattice $\Gamma_1 \subset \Gamma$:
            \begin{align}
			\msplit^2(\Gamma_1) \equiv \text{max} \biggl( 0, \inf_{q \in \Gamma_{\text{big}} \setminus \Gamma_1} [m^2(q) - m_{\rm ext}^2(q)] \biggr)\,.
			\label{Deltadef2}
			\end{align}
            Per the sublattice WGC, this vanishes unless $\Gamma_1 \supseteq \Gamma_{\rm ext}$ and matches our previous definition when $\Gamma_1 = \Gamma_{\rm ext}$. However, for a \emph{proper} superset $\Gamma_1 \supset \Gamma_{\rm ext}$ the sublattice splitting may be larger, in which case all the preceding statements can be applied to place more stringent bounds on the tension $\Tflux$ of flux tubes confining those fractional monopoles that lie within $\Gamma_1^\ast$.
            
			Our work does not directly address the question of how large the coarseness of the sLWGC sublattice may be in a consistent theory of quantum gravity. However, in an example in \S\ref{sec:Zp}, we show that LWGC violation with coarseness $k$ in $d$ dimensions can be achieved by compactifying a $\mathbb{Z}_k$ theory in $D=d+1$ dimensions on a circle with a $\mathbb{Z}_k$ Wilson line turned on, provided that the parent $D$-dimensional theory has sufficiently many charge sites without massless matter.\footnote{More precisely, the coarseness of the sLWGC sublattice will be $n$, where $n$ is a divisor of $k$ that is determined by the spectrum of massless matter in the $D$-dimensional $\mathbb{Z}_k$ gauge theory.} This means that the question of the maximum allowed sLWGC coarseness is intimately related to the longstanding question of the maximum allowed order of a cyclic gauge group in string theory \cite{Aspinwall:1998xj, Cianci:2018vwv, Collinucci:2019fnh, Taylor:2018khc, Anderson:2019kmx, Hajouji:2019vxs}. Accordingly, the example of a $U(1)$ gauge theory in 6d with massless matter of charge 21 discovered in \cite{Raghuram:2018hjn} seems to give rise (after Higgsing to $\mathbb{Z}_{21}$ and compactification to 5d with a discrete Wilson line) to an LWGC-violating theory with coarseness 21! More work is needed to understand this example in greater detail.
			
			The remainder of our paper is structured as follows. In \S\ref{sec:Zp}, we explore the aforementioned example of LWGC violation arising through circle compactification of a $\mathbb{Z}_p$ gauge theory. In \S\ref{sec:exchangeEFTexample}, we explore a similar example of LWGC violation involving $U(1) \times U(1) \rtimes \mathbb{Z}_2$ gauge theory compactified on a circle with a $\mathbb{Z}_2$ Wilson line, and generalizations thereof. In \S\ref{s.9dexample}, we consider a UV-complete example of LWGC violation for charged strings in 9d, discovered previously in \cite{Montero:2022vva}. In \S\ref{s.8d}, we consider a UV-complete example for charged particles and strings in 8d IIA string theory, and also interpret the result geometrically in M-theory. In \S\ref{s.7dexample}, we consider a UV-complete example for charged particles in 7d. In \S\ref{s.Higgsing}, we explore LWGC violation and monopole confinement in the context of Higgsing models like the one studied previously in \cite{Saraswat:2016eaz}. In \S \ref{s.noninvertibles}, we discuss similarities between the phenomena of confined monopoles, non-invertible symmetries, and the Hanany-Witten effect. In \S\ref{s.conclusions}, we conclude with a summary of our work and possible next steps. Some useful dimensional reduction formulas and conventions for the string theory examples can be found in appendices.
            
			\section{Discrete gauge theory example}
			\label{sec:Zp}
			
			In this section, we give our first example of how the LWGC can fail in a simple EFT, namely a compactification of $\ZZ_p$ gauge theory on a circle with a nontrivial Wilson line. We will see that this LWGC violation automatically comes with confined, fractionally-charged magnetic monopoles. If we invoke further assumptions (which can only be justified in a UV completion, not within the EFT itself), we see that the confinement scale for monopoles becomes lower as the LWGC violation becomes more severe. 
			
			\subsection{LWGC violation in compactified $\ZZ_p$ gauge theory}
			\label{subsec:lwgcZp}
			
			Consider a $D$-dimensional $\mathbb Z_p$ discrete gauge theory, dimensionally reduced to $d = D-1$ dimensions on $S^1$ with coordinate $y \cong y + 2\pi R$, with a discrete Wilson line $2\pi / p$ for $\mathbb Z_p$ along this circle. Unless the original theory contains massless particles of every possible $\mathbb{Z}_p$ charge, the reduced theory violates the LWGC in $d$ dimensions, as we now show.
			
			Consider a field $\phi_q$ that has charge $q$ under $\mathbb Z_p$. The Wilson line imposes a $\ZZ_p$ holonomy for $\phi_q$ around the circle:
			\begin{align}
			\phi_q(y+2\pi R)= e^{2\pi i q/p}\phi_q(y).
			\end{align}
			Because the field is non-periodic, the translation operator around the circle, when acting on the field of charge $q$, must satisfy
			\begin{align}
			e^{i k_q (2\pi  R)}=e^{2\pi i q/p},
			\end{align}
			which means the compact momentum around the circle $k_q$ is quantized according to
			\begin{align}
			2\pi k_q R=\frac{2\pi q}{p}+2\pi n,\qquad \text{or}\qquad k_q=\frac{1}R\left(n+\frac{q}{p}\right),
			\end{align}
		where $n\in \ZZ$. Therefore the charge $Q_\text{KK}=k_q R$ of the KK modes of  $\phi_q$ under the graviphoton,
			\begin{align} \label{eq:fracKKcharges}
			Q_\text{KK}=n+\frac{q}{p}\in \frac{1}{p} \ZZ,
			\end{align}
		is fractional in the conventional normalization where graviton KK modes carry integer charges! As a result, the electric charge lattice of the graviphoton is finer than in the theory without a Wilson line, with the new sites populated by the dimensional reduction of fields of nontrivial $\ZZ_p$ charge.

        We can now compute the charge to mass ratios of these fields and compare to the extremal values. The mass formula in the reduced theory is given by $m_d^2=m_D^2+k_q^2$, or
			\begin{align} \label{eq:ZpKKmass}
			m_d^2=m_D^2+\frac{1}{R^2}\left(n+\frac{q}{p}\right)^2.
			\end{align}
			Meanwhile, the extremality bound for a black hole of mass $M$ and charge $Q_\text{KK}$ in Kaluza-Klein theory (taking into account the radion-mediated force; see, e.g.,~\cite{Heidenreich:2015nta}) is
			\begin{equation}
			M \geq \frac{Q_\text{KK}}{R}, 
			\end{equation}
			so it is from~\eqref{eq:ZpKKmass} that the only superextremal $d$-dimensional particles are those with $m_D = 0$, which are precisely extremal. KK modes of the graviton thus supply extremal particles for any integer charge $Q_\text{KK} \in \ZZ$. However, for fractional $Q_\text{KK}$, we will find extremal $d$-dimensional particles only if there were massless $\ZZ_p$-charged particles of charge $p Q_\text{KK} \bmod p$ in the $D$-dimensional theory. There is no reason to expect such massless particles to exist, in general. Thus, the LWGC generically does not hold in the $d$-dimensional theory. If all $\ZZ_p$-charged particles in $D$ dimensions were massive, then the sLWGC holds with coarseness $p$ (in the KK $\mathrm{U}(1)$ direction) relative to the full LWGC.\footnote{Even if there are massless $\ZZ_p$-charged particles, the extremal sublattice may still have coarseness $p$, depending on which $\ZZ_p$ charges appear in the massless spectrum. If so, there will be superextremal particles outside of $\Gamma_{\text{ext}}$.}
			
			\subsection{A KK monopole puzzle}  \label{subsec:KKpuzzle}
			
			We have found that, in the presence of a Wilson line for $\ZZ_p$ on the circle, there are now states of fractional electric Kaluza-Klein charge $1/p$, and thus the electric charge lattice is finer than in the standard Kaluza-Klein theory without the Wilson line. Dirac quantization implies that the magnetic charge lattice should become correspondingly coarser relative to the theory without the Wilson line in response; it suggests magnetic monopoles should only exist with charges that are multiples of $p$. However, in Kaluza-Klein theory we are accustomed to the existence of a KK monopole of magnetic charge 1, so upon introducing the Wilson line the theory naively appears to violate Dirac quantization. The resolution to this conundrum is that the KK monopole of charge 1 is confined. In the presence of the Wilson line it must now be attached to a flux tube, as we explain below. 
            However, first we must resolve an even more significant puzzle, which is that in the presence of the Wilson line, it is not clear that KK monopoles of \emph{any} charge can exist.

\begin{figure}[h]
				\centering
				\includegraphics[width = .75\linewidth]{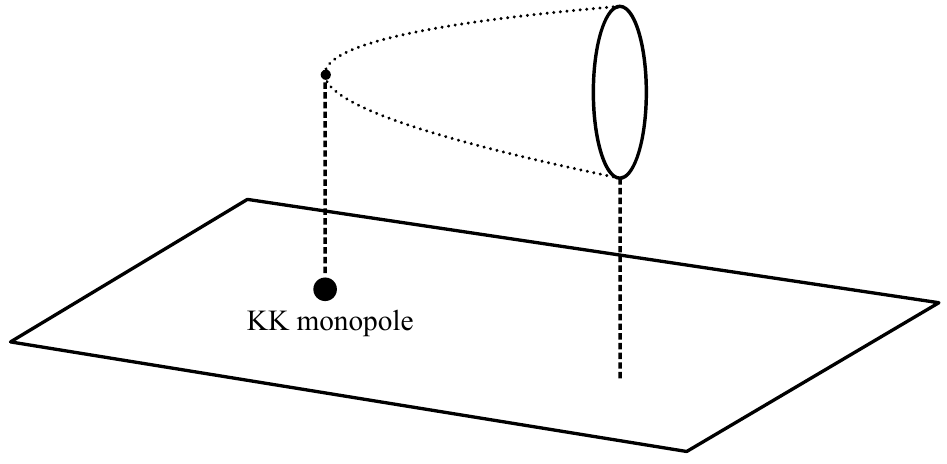}
			\caption{The KK monopole geometry is the Taub-NUT space, which is an $S^1$ fibration over a base which degenerates at the location of the monopole. However, the total space is simply connected, and therefore this background does not have a topologically nontrivial $S^1$ needed to support a discrete Wilson line.}
			\label{f.S1}
			\end{figure}

			In standard Kaluza-Klein theory, an object carrying magnetic charge 1 under the KK $\mathrm{U}(1)$ gauge symmetry can be explicitly constructed as a smooth solution in $D$-dimensional general relativity~\cite{Gross:1983hb, Sorkin:1983ns}. The spatial geometry of the KK monopole is a Taub-NUT solution times $\mathbb{R}^{d-4}$, centered at $r = 0$. Away from $r = 0$, the Taub-NUT geometry can be viewed as an $S^1$ fiber bundle over a $3$-dimensional space. However, near $r = 0$, the fiber degenerates and the Taub-NUT geometry is locally $\RR^4$ (see Figure~\ref{f.S1}). In fact, the Taub-NUT space is simply connected:
			\begin{align}
			\pi_1(\text{Taub-NUT})=0.
			\end{align}
			This means that, unlike the Kaluza-Klein theory on $\RR^{d-1,1} \times S^1$ without the monopole, in the KK monopole configuration the space is distorted so that there is \textit{no} nontrivial cycle around which the Wilson line can be wrapped at all! The circle on which we would attempt to establish the Wilson line has effectively dissolved. In the presence of a KK-monopole, we must therefore revisit the meaning of the Wilson line.
			
			This leads to a puzzle: it appears that the Wilson line is an obstruction to finding magnetically charged objects, and vice versa. But the compactification with a Wilson line turned on defines a consistent vacuum,  so the Completeness Hypothesis for quantum gravity tells us that magnetically charged objects should exist~\cite{polchinski:2003bq}. How do we reconcile these points of view? To lay the groundwork for the answer, we first review the spectrum of electrically and magnetically charged objects in $\ZZ_p$ gauge theory.
			
			\subsection{Charged objects and compactification in $\ZZ_p$ gauge theory}			
			Here we summarize some basic facts about $\ZZ_p$ gauge theory; for more detailed reviews, see~\cite{Banks:2010zn, Gaiotto:2014kfa, Heidenreich:2020pkc}. A $\ZZ_p$ gauge theory can admit electrically charged particles, labeled by a charge $q \in \ZZ_p$, as well as magnetically charged dynamical objects known as twist vortices. These are codimension-two objects, i.e., $(D-3)$-branes in $D$-dimensional spacetime. The twist vortex is labeled by a charge $m \in \ZZ_p$, and has the defining property that a charge $q$ particle circling it acquires a phase $e^{2\pi i q m/p}$. 
			
			One can view $\ZZ_p$ gauge theory as a BF-type theory, with a 1-form $\mathrm{U}(1)$ gauge field $A$ and a $(D-2)$-form $\mathrm{U}(1)$ gauge field $B$ coupled through a Chern-Simons term $\frac{p}{2\pi} B \wedge \rmd A$. From this perspective, electrically charged particles are charged under $A$ and break a $\ZZ_p$ 1-form electric global symmetry, and twist vortices are charged under $B$ and break a $\ZZ_p$ $(D-2)$-form magnetic global symmetry. A magnetic monopole of $A$ is a codimension 3 object or $(D-4)$-brane, and the Chern-Simons term indicates that $p$ twist vortices can end on such a monopole. The object magnetically charged under $B$ is a 0-dimensional object or instanton, and the Chern-Simons term indicates that the worldlines of $p$ electrically charged particles can end on such an instanton. The existence of such objects reflects the fact that charge is conserved only modulo $p$.

			Now, we consider the compactification of $\ZZ_p$ gauge theory on a circle. Each $D$-dimensional $p$-form gauge field gives rise to both a $p$- and $(p-1)$-form gauge field in $d$ dimensions, depending on whether one leg is taken along the $S^1$ or not. Similarly, a $p$-brane in $D$ dimensions descends to both a $p$- and a $(p-1)$-brane in $d$ dimensions, depending on whether it wraps the circle or not.
			
			A quantity that will be of interest below is the dependence of the tension of a $d$-dimensional object on the radion field $\rho$ for the compactification. We take $\rho$ to be canonically normalized; to dimensionally reduce $D$-dimensional Einstein gravity to $d$ dimensions, we use the ansatz
			\begin{equation}
			ds_D^2 = \exp\left[-\frac{2\kappa_d}{\sqrt{(d-1)(d-2)}} \rho\right] ds_d^2 + \exp\left[2\kappa_d\sqrt{\frac{d-2}{d-1}} \rho\right] (dy - R\ C)^2, 
			\end{equation}
			where $C$ is the Kaluza-Klein gauge field normalized so that standard KK modes have integer charges. The mass formula~\eqref{eq:ZpKKmass} should be interpreted as the mass at $\rho = 0$. 
			
			A $p$-brane of tension $T$ in the $D$-dimensional theory descends in the $d$-dimensional theory, depending on whether it does or does not wrap the compact dimension, to branes of respective tensions (see e.g. \cite{Heidenreich:2019zkl})
            \begin{equation} \label{eq:wrapped}
			T_{(p-1);d} = \exp\left[\frac{d-p-2}{\sqrt{(d-1)(d-2)}} \kappa_d \rho\right] 2\pi R T_{p;D},
			\end{equation}
			\begin{equation} \label{eq:unwrapped}
			T_{p;d} = \exp\left[-\frac{p+1}{\sqrt{(d-1)(d-2)}} \kappa_d \rho\right] T_{p;D}.
			\end{equation}

			The compactified gauge theory includes Chern-Simons terms involving the KK gauge field. Let us denote the $D$-dimensional gauge fields $A$ and $B$ with a superscript $(D)$ for clarity. Then we can decompose them into $d$-dimensional gauge fields as follows:
			\begin{align} \label{eq:gaugeansatz}
			A^{(D)} &= A + \frac{a}{2\pi} \left(\frac{dy}{R} - C\right), \nonumber \\
			B^{(D)} &= B + \frac{b}{2\pi} \wedge \left(\frac{dy}{R} - C\right).
			\end{align}
			Thus, in the $d$-dimensional theory, $A$ is a 1-form gauge field, $a$ is a 0-form gauge field, $B$ is a $(d-1)$-form gauge field, and $b$ is a $(d-2)$-form gauge field. The BF term $\frac{p}{2\pi} B^{(D)} \wedge \rmd A^{(D)}$ in the $D$-dimensional theory matches to two different terms, $\frac{p}{2\pi} b \wedge \rmd A$ and $\frac{p}{2\pi} B \wedge \rmd a$, in the $d$-dimensional theory. However, the $C$-dependence in the ansatz~\eqref{eq:gaugeansatz} leads to additional structure: $A$ transforms under an $a$ gauge transformation, and similarly $B$ transforms under a $b$ gauge transformation. The gauge-invariant field strengths are
			\begin{equation} \label{eq:modifiedbianchi}
			\widetilde{F} = \rmd A - \frac{a}{2\pi} \rmd C \quad \text{and} \quad \widetilde{H} = \rmd B - \frac{b}{2\pi} \wedge \rmd C.
			\end{equation}
			A careful dimensional reduction reveals that these combinations show up in the effective $d$-dimensional Chern-Simons couplings. In particular, the $b \wedge \rmd A$ term is actually
			\begin{equation} \label{eq:fullbdA}
			\frac{p}{2\pi} \int b \wedge \left(\rmd A - \frac{a}{2\pi} \rmd C\right) = \frac{p}{2\pi} \int b \wedge \widetilde{F}.
			\end{equation}
			The $b \wedge a\, \rmd C$ portion of this term is crucial for our resolution of the puzzle posed in \S\ref{subsec:KKpuzzle}.
			
			\subsection{Domain walls separating regions of distinct Wilson loops} \label{subsec:domainwalls}

            \begin{figure}
			\begin{center}
				\includegraphics[width = .7\linewidth]{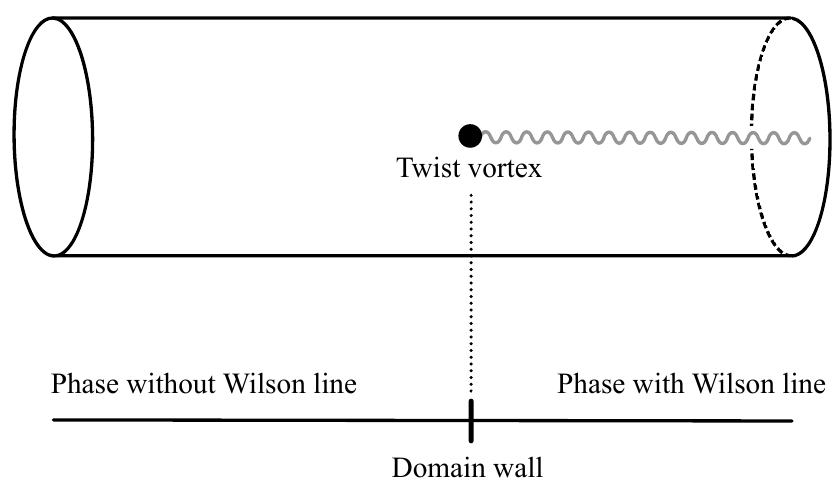}
			\end{center}
			\caption{The twist vortex is a codimension 2 object in the parent theory. Circling the vortex implements a monodromy. The wavy gray line indicates the ``branch cut" of the vortex, which is an unphysical locus (similar to a Dirac string) at which we impose the monodromy as a boundary condition. Here we show a twist vortex localized on the $S^1$. This unwrapped $D$-dimensional twist vortex behaves as a domain wall in the $d$-dimensional theory.}
			\label{f.twistS1}
			\end{figure}   
			
			A twist vortex in the $D$-dimensional theory gives rise to two different kinds of objects in the $d$-dimensional theory: a (codimension two) twist vortex charged under $b$, and a (codimension one) domain wall charged under $B$. The domain wall arises when a twist vortex of the $D$ dimensional theory is localized at a point on the $S^1$, rather than wrapping it, yielding a codimension 1 object in the daughter theory, as illustrated in Figure~\ref{f.twistS1}. The effect of a domain wall is to separate the $d$-dimensional space into regions with different values of the field strength that is magnetically dual to $\widetilde{H}$. Thanks to the existence of a $\frac{p}{2\pi} a \,\rmd B$ Chern-Simons term, this field strength can be identified as $h = \frac{p}{2\pi} a$, which takes integer values. Thus, the value of $a$ jumps by $2\pi/p$ across a $B$ domain wall. Another way to understand this is that the Wilson loop wrapping $S^1$ becomes the operator $\exp(i a)$ in the $d$-dimensional theory. The linking relationship between the Wilson loop and the twist vortex in $D$ dimensions descends to the statement in $d$ dimensions that $\exp(i a)$ is multiplied by $\exp(2\pi i/p)$ when crossing a domain wall carrying $B$ charge. 
			
			In a region with $a = 0$, the $d$-dimensional theory has a Chern-Simons term $\frac{p}{2\pi} b \wedge \rmd A$ inherited from the $D$-dimensional theory. In this case, the gauge group is a product $\U1 \times \ZZ_p$ of the Kaluza-Klein $\U1$ gauge group and a $\ZZ_p$ gauge group. However, in a region where a minimal charge Wilson loop is wrapped on $S^1$, we have $a = 2\pi/p$ (or $h = 1$). The Chern-Simons coupling~\eqref{eq:fullbdA} becomes
			\begin{equation} \label{eq:massivephoton}
			\frac{1}{2\pi} \int b \wedge \left(p\, \rmd A - \rmd C\right).
			\end{equation}
			This effectively Higgses the would-be $\U1$ generated by $A$ and the $\U1$ Kaluza-Klein gauge group down to a single $\U1$, with no surviving discrete factor.\footnote{We emphasize that, in a UV completion, it may not be the case that the $\ZZ_p$ gauge theory in $D$ dimensions can be interpreted as a Higgsed $\U1$ gauge theory. We have only used the topological BF term, not kinetic terms for $A$ and $B$, to derive the structure of the low-energy EFT. Nonetheless, the language of Higgsing is useful for gaining intuition.} This $\U1$ couples to both the Kaluza-Klein charge and to the $\ZZ_p$ charge in the UV theory. Roughly speaking, we have set $A = C/p$, so the low-energy $\U1$ has $p$ times the periodicity of the usual Kaluza-Klein $\U1$. This is closely analogous to familiar examples such as axion alignment~\cite{kim:2004rp, Choi:2014rja}, past discussions of Higgsing and the WGC~\cite{Saraswat:2016eaz}, or the fractional quantum Hall effect~\cite{Wen:2007joe}. More generally, a region with a given value of $h$ has effective gauge group $\U1 \times \ZZ_{\gcd(h, p)}$. 
			
			To relate this to our earlier discussion in \S\ref{subsec:lwgcZp}, observe from~\eqref{eq:massivephoton} that a particle with charge $q$ under $ A$ and charge $n$ under $C$ will couple only to the massive combination when $n + q/p = 0$, i.e., when the charge $Q_\text{KK}$ defined in~\eqref{eq:fracKKcharges} vanishes. This shows that viewing the EFT through its Chern-Simons terms reproduces the result of a simple mode analysis.
			
			\subsection{The fate of magnetic monopoles}
			
\begin{figure}[h]
				\centering
				\includegraphics[width = .7\linewidth]{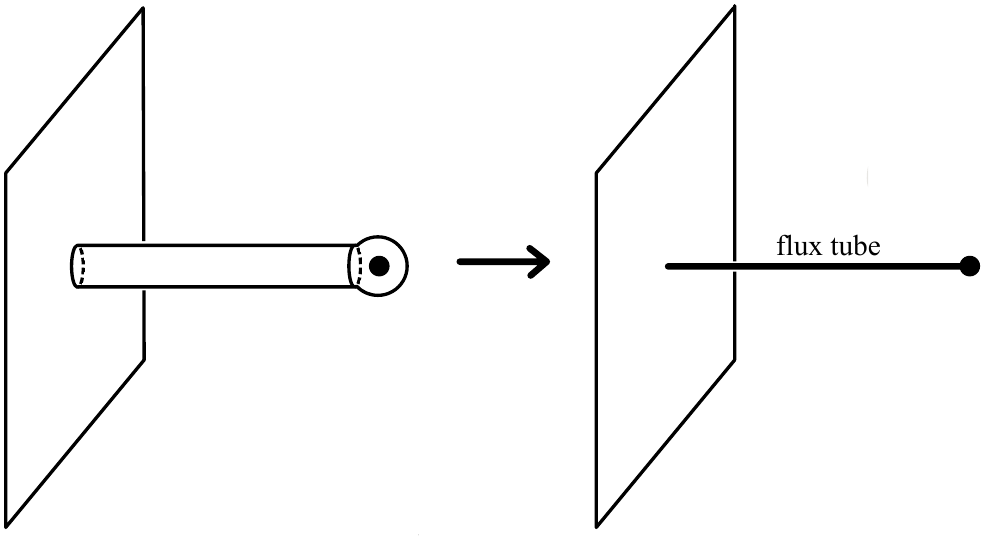}
				\caption{Upon pushing a KK-monopole across a domain wall to a phase with a $\mathbb{Z}_p$ Wilson line turned on, it becomes attached to a flux tube.
				}\label{f.flux tubes.push}
            \end{figure}

			Consider a configuration of the compactified theory in which a domain wall separates a region with trivial Wilson loop and a region with a Wilson loop of charge 1 on the circle. In the region with trivial Wilson loop, we can have a KK monopole. In the region with a nontrivial Wilson loop, as argued in~\S\ref{subsec:KKpuzzle}, we cannot. What happens, then, if we begin with a KK monopole in the trivial region and try to push it into the nontrivial region? The KK monopole cannot actually cross into the domain where the Wilson line is turned on, but the domain wall can wrap around it,  forming a narrow tube connecting the KK-monopole to the rest of the domain wall, as depicted in Figure~\ref{f.flux tubes.push}.

\begin{figure}[h]
            \begin{subfigure}{.55\textwidth}
                \centering
				\includegraphics[width = \linewidth]{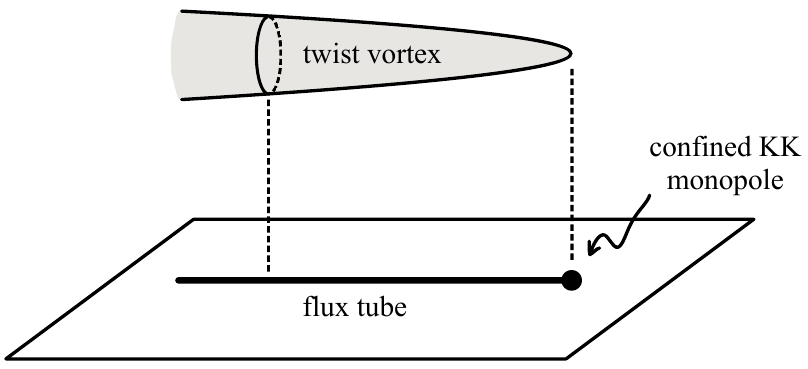}
                \caption{KK monopole, confined}
                \label{f.kkmonopolewrapped}
            \end{subfigure}
			\begin{subfigure}{.35 \textwidth}
				\centering
				\includegraphics[width = \linewidth]{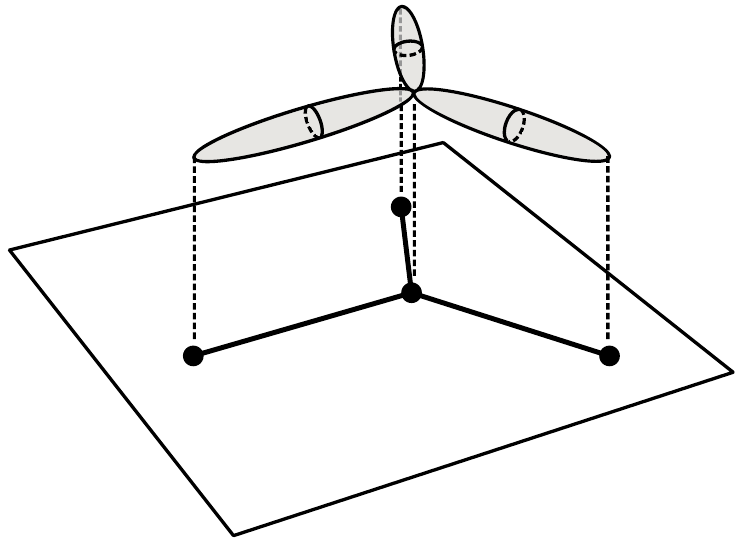}
                \caption{$p$ KK monopoles, unconfined}
                \label{f.monopolescombined}
			\end{subfigure}
			\caption{(a) The charge 1 KK monopole is confined by a flux tube formed from a twist vortex wrapped on the $S^1$. (b) $p$ charge 1 monopoles confined together by flux tubes combine to make a charge $p$ monopole, which is unconfined. ($p=4$ in this illustration).}
			\end{figure}

			This resolves the puzzle about Dirac quantization. One can have a fractionally charged monopole in the $d$-dimensional theory with the Wilson line, but such a fractionally charged monopole must be confined by a ``flux tube'' that can end on a domain wall. In fact, such a flux tube can be formed from a twist vortex of the $D$-dimensional theory wrapped on $S^1$ to produce a vortex in the $d$ dimensional theory, charged under $b$, as shown in Figure~\ref{f.kkmonopolewrapped}. The Stueckelberg structure in~\eqref{eq:modifiedbianchi} ensures that the $b$ charge can be dissolved into a domain wall charged under $B$, and hence that the picture of pushing out a tube of $B$ domain wall is physically equivalent to ending a monopole on a $b$-charged vortex. From a different perspective, the Chern-Simons coupling in~\eqref{eq:massivephoton} ensures that vortices charged under $b$ can end on any monopole carrying a magnetic charge under the combination $p A - C$ of ordinary gauge fields. 

			What about a KK monopole with charge a multiple of $p$, which {\em is} allowed by Dirac quantization? In this case, we have $p$ flux tubes attached to the charge-$p$ KK monopole, which have zero net $\ZZ_p$ charge. Thus, there is a junction on which they can end, which is the magnetic monopole for $A$ in the BF theory viewpoint. One can think of the unconfined object allowed by Dirac quantization, then, as made up of $p$ charge 1 KK monopoles, $p$ flux tubes, and one $A$ monopole (descending from the $D$-dimensional theory). This composite object carries no net magnetic charge under $p A - C$, and hence is unconfined. Another closely related configuration can be obtained by moving the $A$ monopole onto one of the KK monopoles; since it wraps the shrinking $S^1$ fiber, this results in an unconfined configuration involving only twist vortices wrapping cycles in the Taub-NUT geometry, as illustrated in Figure~\ref{f.monopolescombined}.
            
            In a given theory, these complicated combinations of monopoles and flux tubes may recombine into a simpler bound state. We will not attempt to describe it in more detail here, since our primary concern is the relationship between the flux tube tension and LWGC violation, to which we will now turn.
            
			\subsection{Relationship between LWGC failure and flux tube tensions}
            \label{subsec:splitvstension}
			
			Restoring the dependence on $\rho$ to~\eqref{eq:ZpKKmass}, we have
			\begin{align} \label{eq:ZpKKmassRho}
			m_d^2 &=\exp\left[-\frac{2\kappa_d  \rho}{\sqrt{(d-1)(d-2)}}\right]m_D^2+\exp\left[-2\kappa_d \rho \sqrt\frac{d-1}{d-2}\right]\frac{1}{R^2}\left(n+\frac{q}{p}\right)^2,
			\end{align}
			where the sublattice splitting \eqref{Deltadef} is measured by the first term, 
			\begin{align} \label{eq:LWGCviolZp}
			\msplit^2 = \exp\left[-\frac{2\kappa_d\rho}{\sqrt{(d-1)(d-2)}}\right]m_D^2.
			\end{align}
			The dependence on $\rho$ follows from squaring~\eqref{eq:unwrapped} in the case of $p = 0$. The confining flux tube is a $(d-3)$-brane arising from a wrapped $(D-3)$-brane, so~\eqref{eq:wrapped} in the case $p = d - 2$ tells us that in fact its tension is independent of $\rho$:
			\begin{equation} \label{eq:TfluxZp}
			\Tflux = 2\pi R T_{(D-3);D}.
			\end{equation}
			
			Purely working in effective field theory with no further assumptions, this is as far as we are able to go. We have no information that relates the size of the LWGC violation as measured by~\eqref{eq:LWGCviolZp} to the flux tube tension~\eqref{eq:TfluxZp}. However, by making additional assumptions about the $D$-dimensional theory, we can obtain an interesting relationship that we can then check in UV-complete examples. Suppose that the $\ZZ_p$ gauge theory in $D$ dimensions is described by a $BF$ theory with kinetic terms for the gauge fields $A$ and $B$ (with field strengths $F$ and $H$ respectively):
			\begin{equation} \label{eq:ZpAction}
			S = \int \left(-\frac{1}{2e^2} F \wedge \star F - \frac{1}{2g^2} H \wedge \star H + \frac{p}{2\pi} B \wedge F\right).
			\end{equation}
			This is now the theory of a massive photon, with mass
			\begin{equation}
			m_{\gamma;D} = \frac{p}{2\pi} e g.
			\end{equation}
			In this theory, we see that, {\em holding the photon mass fixed}, there is an inverse relationship between the coefficients $e$ and $g$. Let us now further suppose that, viewing $A$ and $B$ as $\mathrm{U}(1)$ gauge fields, there are superextremal (WGC-satisfying) minimally charged objects. (Some remarks on why we might expect the WGC to apply to theories with a massive photon appear in~\cite{Reece:2018zvv}.) This means that we have a particle of charge $1$ under $\ZZ_p$ with 
			\begin{equation} \label{eq:mDfromWGC}
			m_D^2 \leq \gamma_A e^2 M_D^{D-2}
			\end{equation}
			and a twist vortex of magnetic $\ZZ_p$ charge 1 with 
			\begin{equation} \label{eq:TvortexfromWGC}
			T_{(D-3);D}^2 \leq \gamma_B g^2 M_D^{D-2}. 
			\end{equation}
			The coefficients $\gamma_A$ and $\gamma_B$ depend on scalar forces in the theory (and in particular, $\gamma_B \to \infty$ in the absence of such forces~\cite{Heidenreich:2015nta}). 
			
			Under dimensional reduction, the photon mass in the lower-dimensional theory becomes
			\begin{equation} \label{eq:photonmassZp}
			\left(m_{\gamma;d}\right)^2 = \left(\frac{p}{2\pi} e g \exp\left[-\frac{\kappa_d\rho}{ \sqrt{(d-1)(d-2)}}\right]\right)^2 + \left(\frac{1}{2\pi} e_\mathrm{KK} g \exp\left[-\kappa_d \rho \sqrt{\frac{d-1}{d-2}}\right]\right)^2.
			\end{equation}
			The first term is the direct dimensional reduction of~\eqref{eq:mDfromWGC} converted to the lower-dimensional Einstein frame, while the second is an additional contribution from the kinetic term of the Kaluza-Klein gauge field (thanks to the structure of~\eqref{eq:massivephoton}), with $e_\mathrm{KK}^2 = 2\kappa_d^2/R$.
			Furthermore, the Planck scales in the theories are related by $M_d^{d-2} = 2\pi R M_D^{D-2}$. Putting together~\eqref{eq:LWGCviolZp},~\eqref{eq:TfluxZp}, and~\eqref{eq:photonmassZp}, we find that
			\begin{equation} \label{inverserelationZp}
			\msplit \Tflux \leq \frac{2\pi}{p} \sqrt{\gamma_A \gamma_B} \, m_{\gamma; d} M_d^{d-2}\,,
			\end{equation}
            in accordance with \eqref{geninv}.
			Said differently, up to constant factors, the maximum possible tension of the flux tubes confining magnetic monopoles and the mass parameter characterizing violation of the LWGC are inversely related, with proportionality constant (in Planck units) bounded above by the mass of the photon. In other words, the more severe the violation of the LWGC, the closer the theory must come to having free, fractionally-charged monopoles.
			
			We have made several assumptions to derive this result, but there are reasons to believe that it may be pointing us toward a general fact. First, we will see below that a similar quantitative relationship holds in actual examples of LWGC-violating string vacua. Second, the absence of global symmetries in quantum gravity requires that the $\ZZ_p$ gauge theory have charged particles and twist vortices. We expect that they cannot be too heavy, because even approximate global symmetries should be forbidden at the quantum gravity scale~\cite{Cordova:2022rer}. This gives some reason to hope that relationships along the lines of~\eqref{eq:mDfromWGC} and~\eqref{eq:TvortexfromWGC} may hold more generally. There is no guarantee that a $\ZZ_p$ gauge theory should be described by an action of the form~\eqref{eq:ZpAction} at all, but we might hope that the scaling relations persist with $m_{\gamma;D}$ replaced by a more general UV scale at which the BF theory description breaks down. 

            In the Introduction, we suggested, in addition to an inverse relation of the form~\eqref{inverserelationZp} at fixed photon mass, that the flux tube tension should become small in Planck units whenever the sublattice splitting $\msplit$ becomes large in Planck units. From~\eqref{eq:LWGCviolZp}, we see that one way to take $\msplit$ large is through the $\rho \to -\infty$ limit. The flux tube tension~\eqref{eq:TfluxZp} is independent of $\rho$, in apparent contradiction to our claim. However, the EFT analysis above is unreliable in the small radius limit $\rho \ll 0$. In subsequent sections, when we discuss explicit examples in string theory, we will avoid such difficulties by working in different duality frames in different regions of the moduli space. A different way to take $\msplit$ large is to take $m_D$ large; again, this has no obvious implications for $\Tflux$ without a further assumption such as fixed $m_{\gamma;D}$. Again, working in explicit UV completions will allow us to make more definite statements about the relationships among different mass and tension scales.
            
            Finally, we note that in the limit of extreme violation of the LWGC, the decoupling of electric states leads to an approximate 1-form electric symmetry that should be forbidden in quantum gravity. The obvious route to removing such a symmetry is to have electrically charged states, but the alternative is to {\em gauge} the 1-form symmetry, which corresponds to changing the global form of the gauge group (in this case, the quantum of $\mathrm{U}(1)$ charge). In that case, the magnetic $(D-3)$-form symmetry is enlarged, and one expects to find additional monopoles. Thus, we could view a limit in moduli space where the LWGC is badly broken as a limit in which the global form of the gauge group is shifting from one with electrically charged particles with charge a multiple of $1/p$ and monopoles of magnetic charge a multiple of $p$, to one with electrically charged particles with charge a multiple of $1$ and monopoles of magnetic charge a multiple of $1$ (i.e., the electric particles at fractional charges are becoming heavier until they drop out of the theory, and the confined monopoles of charges 1 through $p-1$ are becoming unconfined, genuine objects). The only way for this to happen continuously is that the monopoles of charge $1$ were present all along, but confined, with the confinement scale gradually dropping to zero in the asymptotic part of moduli space. We will comment below on how this qualitative picture holds up in further examples.
			
			\section{$\boldsymbol{[\mathrm{U}(1)\times \mathrm{U}(1)]\rtimes \mathbb Z_2}$ gauge theory example}\label{sec:exchangeEFTexample}
			
			In this section, we consider a second EFT example of LWGC failure, in a theory with two $\U1$ gauge groups exchanged by a $\ZZ_2$ gauge group, compactified with a nontrivial $\ZZ_2$ Wilson line. Again, we find that there are fractionally charged magnetic monopoles.

            The findings of this section will prove to be quite general, and are particularly applicable to the string theory examples discussed in later sections.
			
			\subsection{LWGC violation in compactified $\boldsymbol{[\mathrm{U}(1)\times \mathrm{U}(1)]\rtimes \mathbb Z_2}$ gauge theory}

			We consider a theory with two gauge fields, $A_1$ and $A_2$, in $D$ dimensions, together with a $\ZZ_2$ gauge symmetry that exchanges them. In particular, we will work in a $\ZZ_2$-symmetric phase where the gauge coupling $e_D$, spectra of charged particles, and all other features of $A_1$ and $A_2$ are identical. We assume that the LWGC is satisfied in this theory, i.e., for any $(q_1, q_2) \in \ZZ^2$, there is a field $\phi_{(q_1,q_2)}$ with charges $q_1$ under $A_1$ and $q_2$ under $A_2$ and mass
			\begin{equation} \label{eq:extbdorig}
			m_{(q_1,q_2)}^2 \leq \left(m^{\mathrm{ext};D}_{(q_1,q_2)}\right)^2 =  \gamma (q_1^2 + q_2^2) e_D^2 M_D^{D-2},
			\end{equation}
			where $\gamma$ is a constant that depends on the moduli couplings in the $D$-dimensional theory, with $\gamma=\frac{D-2}{D-3}$ in the absence of moduli. For $q_1 \neq q_2$, the gauged $\ZZ_2$ exchange symmetry requires that there is an additional field $\phi_{(q_2,q_1)}$ of equal mass.
			
			We consider a compactification of the theory on a circle $y \cong y + 2\pi R$ with a $\ZZ_2$ Wilson line. In particular, this means that the fields are exchanged after transport around the $y$ circle:
			\begin{equation}
			A_1(x, y + 2\pi R) = A_2(x, y), \quad A_2(x, y + 2\pi R) = A_1(x, y).
			\end{equation}
			Carrying out a mode expansion of the fields, $A_i(x,y) = \sum A_{i,n}(x) \exp(i k_n y)$, we find solutions when $k_n R = \frac{1}{2} n \in \frac{1}{2} \ZZ$, which contains two classes of modes: $k_n R$ integer and half-integer ($1/2,3/2,...$). Modes with integer $k_n R$ must have $A_{1,n}(x) = A_{2,n}(x)$ to satisfy the above pseudo-periodicity constraint, whereas modes with half-integer $k_n R$ must have $A_{1,n}(x) = -A_{2,n}(x)$. Thus, of the original $\U1 \times \U1$, the only mode of the parent gauge fields that survives as a massless $d$-dimensional $\U1$ gauge field is the zero mode along $y$, which has $A_{1,0}(x) = A_{2,0}(x)$ since $k=0$ is an integer. We will denote this gauge field as $A(x) = A_{1,0}(x) = A_{2,0}(x)$ (which is independent of the $y$ direction by virtue of having been a zero mode along that direction). The $d$-dimensional kinetic term for $A$ receives equal contributions from the $D$-dimensional kinetic terms of both $A_1$ and $A_2$, determining the $d$-dimensional coupling as
			\begin{equation} \label{eq:couplingmatchtwist}
			\frac{1}{e_d^2} = \frac{4\pi R}{e_D^2}.
			\end{equation}  
			There is also a Kaluza-Klein $\U1$ gauge field $C$ which, as in~\S\ref{sec:Zp}, admits fields of half-integer KK charge due to the $\ZZ_2$ Wilson line. 
			
			Consider a pair of fields $\phi_{(q_1,q_2)}$ and $\phi_{(q_2,q_1)}$ in the $D$-dimensional theory. Just as for $A_{1}$ and $A_2$, the $\ZZ_2$ twist on the circle relates their KK modes in $d$ dimensions. There is a mode with zero KK momentum and $\phi_{(q_1,q_2)} = \phi_{(q_2,q_1)}$, which from the $d$-dimensional viewpoint has charge
			\begin{equation}
			q = q_1 + q_2
			\end{equation}
			under $A$, and mass $m_{(q_1,q_2)}$ inherited from the $D$-dimensional theory. This mode obeys the WGC if
			\begin{equation} \label{eq:extbdcomp}
			m_{(q_1,q_2)}^2 \leq \left(m^{\mathrm{ext};d}_q\right)^2 = \gamma q^2 e_d^2 M_d^{d-2} = \gamma \frac{(q_1 + q_2)^2}{2} e_D^2 M_D^{D-2},
			\end{equation}
			where we used~\eqref{eq:couplingmatchtwist} and the usual relationship between $D$- and $d$-dimensional Planck scales. The coefficient $\gamma$ in the extremality bound is invariant under dimensional reduction, once we take the radion into account~\cite{Heidenreich:2015nta}.
			
			Now, we can ask: does a theory that satisfies the LWGC~\eqref{eq:extbdorig} in $D$ dimensions  necessarily satisfy the LWGC~\eqref{eq:extbdcomp} in $d$ dimensions? In other words, is $\bigl(m^{\mathrm{ext};D}_{(q_1,q_2)}\bigr)^2 < \bigl(m^{\mathrm{ext};d}_q\bigr)^2$? To answer this, we compute:
			\begin{align} \label{eq:LWGCfailure}
			\bigl(m^{\mathrm{ext};d}_q\bigr)^2  - \bigl(m^{\mathrm{ext};D}_{(q_1,q_2)}\bigr)^2 &= \gamma \biggl[\frac{1}{2}(q_1 + q_2)^2 - (q_1^2 + q_2^2) \biggr] e_D^2 M_D^{D-2}  \nonumber \\
			&= - \gamma (q_1 - q_2)^2 e_d^2 M_d^{d-2}  \leq 0.
			\end{align}
			This shows that the zero mode of a WGC-saturating state in $D$ dimensions does {\em not} satisfy the WGC in $d$ dimensions except in the special case $q_1 = q_2$. However, in that case, $q = q_1 + q_2 = 2 q_1$ is even. Thus, a spectrum that saturates the LWGC in $D$ dimensions will (focusing only on the $A$ direction in charge space) obey the sLWGC with coarseness 2 in $d$ dimensions.

\begin{figure}[h]
            \begin{subfigure}{.45\textwidth}
                \centering
				\includegraphics[width = \linewidth]{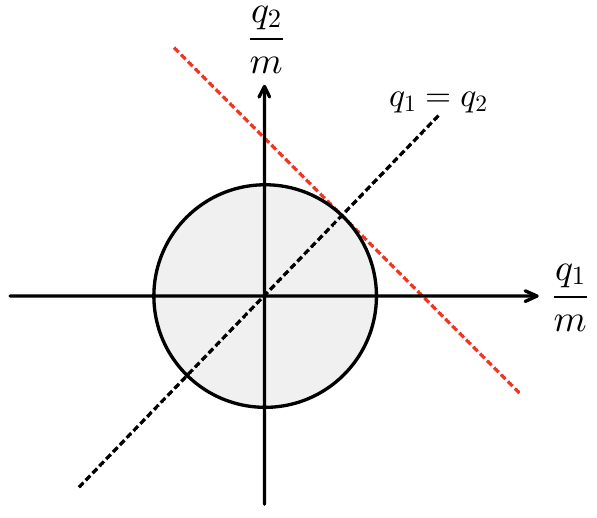}
                \caption{BH region that is strictly convex}
                
            \end{subfigure}
            \hspace{0.05\textwidth}
			\begin{subfigure}{.45 \textwidth}
				\centering
				\includegraphics[width = \linewidth]{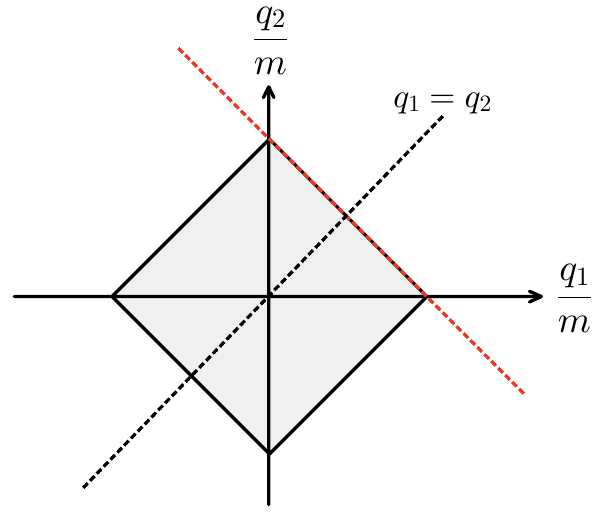}
                \caption{BH region that is not strictly convex}
                \label{f.nonconvexBHregion}
			\end{subfigure}
			\caption{(a) If the black hole region is strictly convex then a state saturating the $D$-dimensional WGC does not satisfy the $d$-dimensional WGC unless $q_1 = q_2$. (b) Otherwise, WGC-saturating states with $q_1 \ne q_2$ may satisfy the $d$-dimensional WGC. In the pictured example, this is true for any state with $q_1, q_2$ both the same sign, hence the LWGC is not violated.\label{f.BHregion}}
			\end{figure}

            Above we assumed the constant $\gamma$ was independent of $\vec{q}$. This is true if the gauge kinetic terms are $\mathbb{Z}_2$ invariant and diagonal, $\mathcal{L}_{\text{kin}} = - \frac{1}{2e_D^2} (F_1^2 + F_2^2)$, across the moduli space. More generally, there could be $\mathbb{Z}_2$-odd moduli that couple oppositely to the two gauge fields and are activated in black hole spacetimes with $q_1 \ne q_2$. There can also be kinetic mixing $\mathcal{L}_{\text{kin}} = - \frac{1}{2e_D^2} (F_1^2 + F_2^2+2 \beta F_1 \cdot F_2)$ even without such moduli. With these more general couplings, $\gamma = \gamma(\hat{q})$ depends on the charge direction $\hat{q}=\vec{q}/|\vec{q}|$, i.e., the region covered by large black holes in the $\vec{q}/m$ plane is no longer a disk. Depending on the shape of this region, there are two possibilities, see Figure~\ref{f.BHregion}.

If the tangent to the boundary point where $q_1 = q_2$ intersects the black hole region at only one point, i.e., if the region is \emph{strictly convex} at this point, then the above conclusions are unchanged, and the LWGC is violated with coarseness 2 in the $d$-dimensional theory.\footnote{This is because extremal $q_1 = q_2$ black holes remain extremal in the presence of the $\mathbb{Z}_2$ Wilson line. (Here we assume there is a single, $\mathbb{Z}_2$-symmetric extremal $q_1 = q_2$ solution, rather than a pair of solutions exchanged by $\mathbb{Z}_2$; see Figure 19 in \cite{Harlow:2022ich} for a sketch of how the latter situation could arise.)} However, if this tangent line intersects the black hole region anywhere else then some WGC-saturating states with $q_1 \ne q_2$ will still satisfy the WGC in the $d$-dimensional theory, and the LWGC may not be violated, as in the example shown in Figure~\ref{f.nonconvexBHregion}.

Henceforward, we focus on examples of the former kind, where the LWGC is indeed violated with coarseness two in the compactified theory with $\mathbb{Z}_2$ Wilson line turned on.

			\subsection{Confined monopoles and flux tubes}
			
			The intuition that we developed in \S\ref{sec:Zp} leads us to expect that, since the LWGC fails for odd charges in the $d$-dimensional theory, we will find magnetic monopoles of ``half-integer" charge that are confined by flux tubes.\footnote{The normalization here is that the unconfined monopoles are charge 1, and confined monopoles are charge $1/2$, rather than $p$ and 1 respectively as we defined in the previous section.} In the $D$-dimensional theory, there are magnetic monopoles with magnetic charges labeled by pairs of integers $(p_1, p_2)$. Because magnetic monopoles are objects of codimension 3, to obtain a magnetic monopole for $A$ in the $d$-dimensional theory, we must wrap a magnetic monopole of the $D$-dimensional theory on the circle. However, because of the $\ZZ_2$ Wilson line, this is not always possible to do straightforwardly. Before investigating this further, let us make some remarks on the matching of charges between $D$- and $d$-dimensional monopoles.
			
			In the $D$-dimensional theory, we can define magnetic gauge fields $\AM_i$ (with $i \in \{1,2\}$) that satisfy $\frac{1}{2\pi} \rmd \AM_i = \frac{1}{e_D^2} \star \rmd A_i$. In the $d$-dimensional theory, we similarly define a magnetic gauge field $\AM$ that satisfies $\frac{1}{2\pi} \rmd \AM = \frac{1}{e_d^2} \star \rmd A$. The fact that $A = A_{1,0} = A_{2,0}$ and the coupling matching condition~\eqref{eq:couplingmatchtwist} then imply that a magnetic monopole in the $D$-dimensional theory coupled to $p_1 \AM_1 + p_2 \AM_2$, wrapped $w$ times on the compact circle, descends to a magnetic monopole in the $d$-dimensional theory coupled to $\frac{w}{2}(p_1 + p_2) \AM$. Thus, after dimensional reduction, a magnetic monopole with odd $p_1 + p_2$ in $D$ dimensions, wrapped once on the circle, has half-integral charge. Therefore these monopoles have non-integral Dirac pairing with electrically charged matter of integral charge, and are inconsistent with $d$-dimensional Dirac quantization.
			
			One way to understand this phenomenon is to view the $\ZZ_2$ Wilson line as introducing a coordinate defect localized at a point on the $S^1$ (and extended in all $d$ dimensions), at which fields and objects obey a boundary condition with a $\ZZ_2$ twist, as depicted in Figure~\ref{f.twistS1}. This defect is not a physical object, and does not have a well-defined location, but is an artifact of attempting to represent a nontrivial bundle in a single coordinate chart (much like a Dirac string). A $(p_1,p_2)$-monopole becomes a $(p_2,p_1)$-monopole on the other side of the defect, so it cannot wrap back to itself unless $p_1 = p_2$. It could, however, be wrapped twice, converting to a $(p_2, p_1)$ monopole on the first crossing of the defect and back to a $(p_1,p_2)$ monopole on the second crossing, as illustrated in Figure~\ref{f.(1,1)}. This shows that wrapping a magnetic monopole around the $S^1$, in the presence of the Wilson line, will only produce configurations with integer values of $\frac{w}{2}(p_1 + p_2)$, i.e., properly quantized magnetic charges for the $d$-dimensional U(1) gauge theory. 

			\begin{figure}[h]
				\begin{center}
					\includegraphics[width = 15cm]{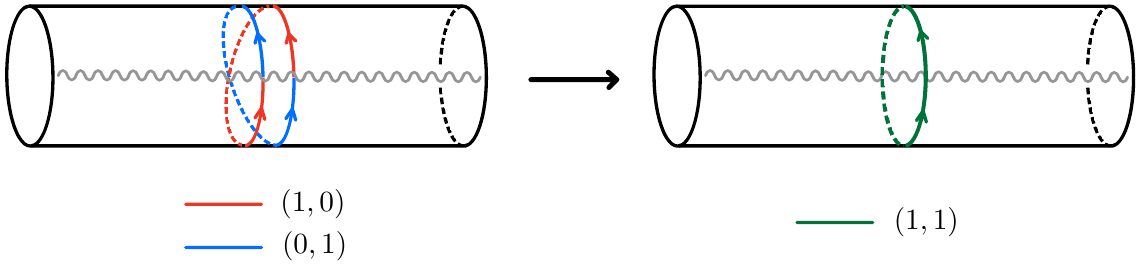}
				\end{center}
				\caption{A $(1,0)$ magnetic brane cannot singly-wrap the compact circle because the monodromy (which acts at the gray branch cut) changes it into a $(0,1)$ brane, preventing the two ends of the brane from joining together. Wrapping it twice avoids this problem, which is equivalent after brane recombination to a singly-wrapped $(1,1)$ brane, with integral magnetic charge in the reduced theory.}    
				\label{f.(1,1)}
			\end{figure}            
			
\begin{figure}[h]
				\begin{center}
					\includegraphics[width = 15cm]{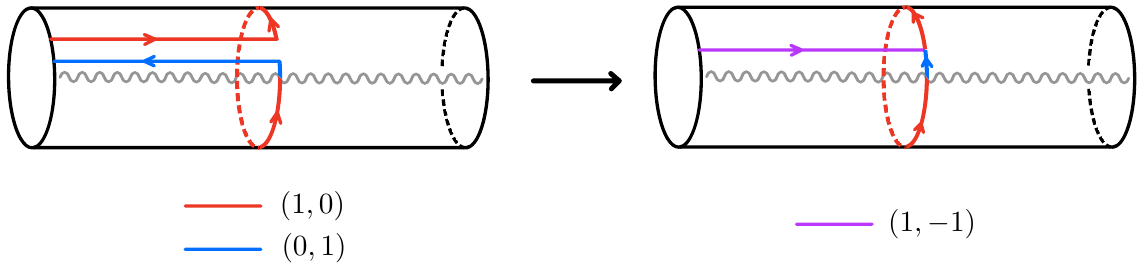}
				\end{center}
				\caption{A $(d-3)$-dimensional charge $(1,0)$ monopole is attached to a $(d-2)$-dimensional flux tube of charge $(1,-1)$. The flux tube is formed from the combination of the ingoing $(1,0)$ brane, plus its outgoing image which has been pushed across the coordinate defect, which is a $(0,1)$ string oriented oppositely to the ingoing string. The two branes then combine into a $(1,-1)$ flux tube, shown in purple. The wavy gray line is the branch cut corresponding to the monodromy.}
				\label{f.T3fractional}
			\end{figure}

			In principle, a configuration with an odd value of $w(p_1 + p_2)$, and hence a half-integer magnetic charge in $d$ dimensions, could be allowed if it is attached to a flux tube. Given the ingredients we have introduced, the only candidate for such a flux tube is an {\em unwrapped} $D$-dimensional magnetic monopole. Indeed, such configurations are possible, and moreover arise naturally when one attempts to introduce a singly-wrapped half-integer magnetic charge brane. For example, consider an extended $(1,0)$-monopole that is oriented transverse to the $S^1$ and parallel to the coordinate defect. Then, at some point, this monopole turns and almost wraps the circle. However, before fully wrapping the circle and crossing the defect, the monopole is then sent off back in a direction transverse to the $S^1$ (and parallel to the defect). If we pass the outgoing part of this monopole across the defect and join it with the incoming part, then we have a $(d-3)$-dimensional charge-$(1,0)$ monopole in the $d$-dimensional theory, but this monopole is attached to a $(d-2)$-dimensional flux tube of charge $(1,-1)$; see Figure~\ref{f.T3fractional}. 
			
\begin{figure}[h]
				\begin{center}
					\includegraphics[width = 15cm]{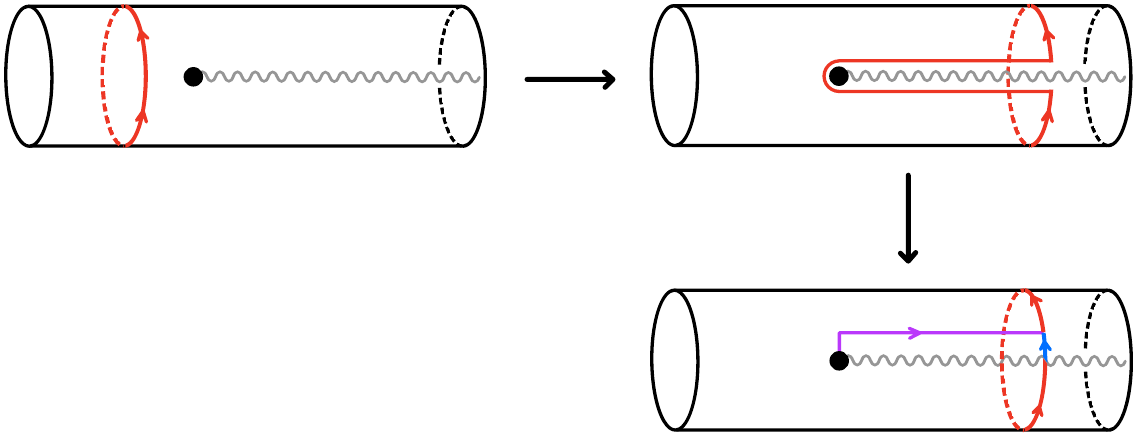}
				\end{center}
				\caption{A complementary perspective on how a $(1,0)$ brane can be introduced into the confining phase of the theory. Here, we consider the brane as beginning in the unconfined phase and being pushed into the confining phase to the right of the twist vortex, where a monodromy exists upon moving around the circle. We deform the brane to wrap it around the twist vortex, leaving two long legs. As in the previous figure, one of these is brought across the monodromy and combines with the other leg, forming a flux tube (shown in purple) attached to the $(1,0)$ brane.}
				\label{f.T3fractionalViaPushingThroughDomainWall}
			\end{figure}

            Along the same lines as in Section~\ref{subsec:domainwalls}, another way to understand the confined monopoles is to consider a domain wall separating a region in which the $\ZZ_2$ Wilson line is turned off and a region in which it is turned on. Such a domain wall corresponds to an unwrapped $\ZZ_2$ twist vortex of the $D$-dimensional theory. We can then try to push a $(1,0)$ monopole across the domain wall into the region with the $\ZZ_2$ Wilson line on, which then gets ``stuck" on the twist vortex, developing two long tails which then recombine into a flux tube connecting the would-be monopole back to the wall, as shown in Figure~\ref{f.T3fractionalViaPushingThroughDomainWall}.

\begin{figure}[h]
				\begin{center}
					\includegraphics[width = 10cm]{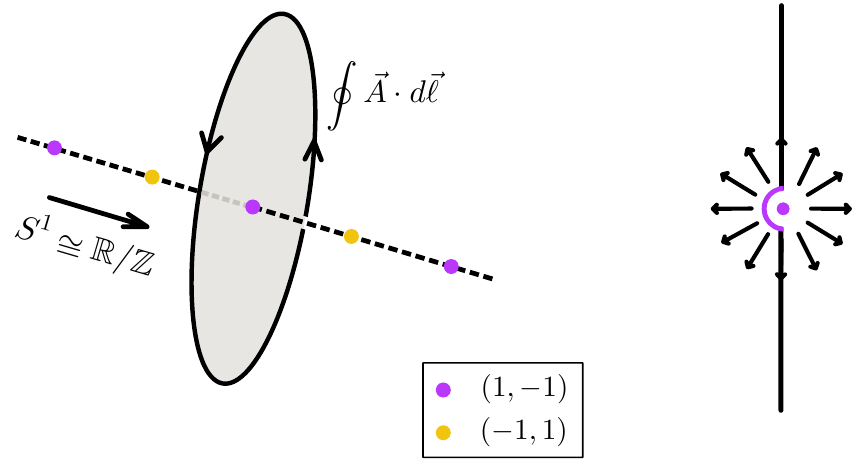}
				\end{center}
				\caption{The Aharonov-Bohm (AB) phase carried by the flux tube can be computed by lifting the $S^1$ compactification to its universal cover, where the unwrapped $(1,-1)$ brane becomes a string of alternating charges. Stokes' theorem relates the AB phase to the magnetic flux through a large-diameter disk intersecting the string of charges. Placing this disk in the same plane as one of the charges, symmetry dictates that the only flux contribution comes from a small hemispherical bump around the charge itself, picking up half the fields lines emanating from it. Thus, the AB phase is $\pi$.}
				\label{f.FluxTubeABPhase}
			\end{figure}

While the fractional monopole is a somewhat complicated object due to the $(p,q)$ monopole junction joining it to the flux tube, etc., the flux tube itself is easier to study as it is merely an unwrapped brane. The flux tube is necessarily uncharged under the remaining massless gauge fields, since it can terminate on a confined monopole, but it does carry an Aharonov-Bohm phase that can be detected by electric charges in theory. To determine this phase, note that the higher-dimensional monopole that makes up the flux tube is \textit{not} invariant under the $\mathbb{Z}_2$ Wilson line, which instead maps it to its antiparticle. Formally lifting the $S^1$ direction to its universal cover $\mathbb{R}$, the flux tube becomes an infinite chain of monopoles and antimonopoles placed alternately every $2 \pi R$ along the unrolled $S^1$ direction, see Figure~\ref{f.FluxTubeABPhase}.

Using this unrolled picture, it is straightforward to compute the Aharonov-Bohm phase using magnetostatics and appropriate symmetries, with the result that $\oint \vec{A} \cdot d\vec{\ell} = \pi$ for the flux tube attached to a charge $1/2$ fractional monopole, i.e., odd charges pick up a sign upon circling the flux tube, whereas even charges do not.

By the usual argument leading to Dirac quantization, the $\pi$ Aharonov-Bohm phase carried by the flux tube implies that the monopole confined by it must carry half-integer magnetic charge.\footnote{We take the ``charge'' of a confined particle to be defined by the electromagnetic flux it generates \emph{outside} the flux tube that confines it. The Aharonov-Bohm phase carried by the flux tube, on the other hand, can be thought of as representing the (fractional part of) the electromagnetic flux ``inside'' the flux tube, where the latter is linked to the fractional part of the former by Dirac quantization.} We have, of course, already deduced this, but the Aharonov-Bohm calculation provides an important cross check.

            \subsection{Large splitting $\Rightarrow$ low tension flux tubes}

Now that we have confirmed the existence of fractional monopoles, we study how their confinement scale (i.e., the tension scale of the flux tubes confining them) is related to the size of the LWGC-violating splitting.

A $D$-dimensional monopole of charge $(p,q)$ respects the magnetic WGC if it has tension 
\begin{equation} \label{eq:magWGCDdim}
	T_D^2 \leq \gamma_\textsc{M} \frac{4\pi^2}{e_D^2} (p^2+q^2) M_D^{D-2},
\end{equation}
for some constant $\gamma_M$ that depends on the moduli couplings in the $D$-dimensional theory. The flux tube confining a charge $1/2$ monopole in the $d$-dimensional theory arises from an unwrapped monopole of charge $(1,-1)$, with tension
\begin{equation}
\Tflux = T_D \exp\left[-\kappa_d \rho \sqrt{\frac{d-2}{d-1}}\right],
\end{equation}
where we now track the radion dependence as in~\eqref{eq:unwrapped}. Restoring the radion dependence to~\eqref{eq:LWGCfailure}, we find that the sublattice splitting \eqref{Deltadef} for electrically charged states of odd charge is given by 
\begin{align} \label{eq:LWGCviolZ2exchange}
	\msplit^2 \le \gamma  e_d^2 M_d^{d-2} \exp\left[-\frac{2\kappa_d\rho}{\sqrt{(d-1)(d-2)}}\right]\,,
\end{align}
where the inequality is saturated when the LWGC is saturated in the $D$-dimensional theory, in which case
the infimum in \eqref{Deltadef} is realized by states with $|q_1-q_2|=1$.    
			
Using~\eqref{eq:couplingmatchtwist} and defining the modulus-dependent Kaluza-Klein scale
			\begin{equation}
				m_\text{KK} \equiv \frac{1}{R} \exp\left[-\kappa_d \rho \sqrt{\frac{d-1}{d-2}}\right],
			\end{equation}
			we find that
			\begin{equation}
				\msplit \Tflux \leq \sqrt{\gamma \gamma_\textsc{M}} m_\text{KK} M_d^{d-2}.
                \label{eq:finsw}
			\end{equation}
			This is very similar  to~\eqref{geninv} and~\eqref{inverserelationZp}, with the Kaluza-Klein scale $m_\text{KK}$ in place of $m_{\gamma}$. Indeed, there is a massive photon at the KK scale, arising from the linear combination of the two $D$-dimensional U(1) gauge fields with no zero mode.

The preceding discussion assumes that the gauge kinetic terms $\mathcal{L}_{\text{kin}} = - \frac{1}{2e_D^2} (F_1^2 + F_2^2)$ are diagonal. More generally, $\mathcal{L}_{\text{kin}} = - \frac{1}{2e_D^2} (F_1^2 + F_2^2+2 \beta F_1 \cdot F_2)$, where $\beta$ controls the kinetic mixing between the two $U(1)$s. Accounting for such mixing, a particle of charge $(q_1, q_2)$ in $D$ dimensions obeys the WGC in $D$ dimensions if
	\begin{equation} \label{eq:mixextbdorig}
			m^2 \leq \Bigl(m^{\mathrm{ext};D}_{(q_1,q_2)}\Bigr)^2 = \gamma \frac{q_1^2 + q_2^2 - 2 \beta q_1 q_2}{1-\beta^2} e_D^2 M_D^{D-2} = 2\gamma \frac{q_1^2 + q_2^2 - 2 \beta q_1 q_2}{1-\beta} e_d^2 M_d^2,
			\end{equation}
where the relation between the gauge couplings is now $\frac{1}{e_d^2} = \frac{4 \pi R}{e_D^2} (1+\beta)$, generalizing \eqref{eq:couplingmatchtwist}.
Likewise, it obeys the WGC in $d$ dimensions if
\begin{equation} \label{eq:mixextbdcomp}
			m_{(q_1,q_2)}^2 \leq \left(m^{\mathrm{ext};d}_q\right)^2 = \gamma q^2 e_d^2 M_d^{d-2} = \gamma (q_1+q_2)^2 e_d^2 M_d^{d-2}\,.
\end{equation}
Thus, the $d$-dimensional WGC bound is strictly stronger when $q_1 \neq q_2$:
\begin{align} \label{eq:mixLWGCfailure}
			\left(m^{\mathrm{ext};d}_q\right)^2  - \left(m^{\mathrm{ext};D}_{(q_1,q_2)}\right)^2 
			&= - \gamma (q_1 - q_2)^2 \frac{1+\beta}{1-\beta} e_d^2 M_d^{d-2}  \leq 0\,,
			\end{align}
generalizing \eqref{eq:LWGCfailure}. Likewise, a monopole of charge $(p,q)$ satisfying the WGC in $D$ dimensions has tension:
\begin{equation}
T_D^2 \leq \gamma_\textsc{M} \frac{4\pi^2}{e_D^2}(p^2+q^2+ 2 \beta p q) M_D^{D-2} = \gamma_\textsc{M} \frac{p^2+q^2+ 2 \beta p q}{1+\beta} \frac{1}{2 e_d^2 R^2} M_d^{d-2}\,. \label{eq:mixMonTension}
\end{equation}
Combining \eqref{eq:mixLWGCfailure} and \eqref{eq:mixMonTension} and accounting for the fact that the lightest flux tubes arise from $(1,-1)$ monopoles, all the $\beta$-dependent factors cancel and one recovers
\begin{equation}
				\msplit \Tflux \leq \sqrt{\gamma \gamma_\textsc{M}} m_\text{KK} M_d^{d-2}.
                \label{eq:finsw1}
\end{equation}
as in \eqref{eq:finsw}. This demonstrates that the inverse relationship between the flux tube tension and the sublattice splitting (at fixed photon mass) persists even when the sublattice splitting is parametrically different from the WGC scale $e_d^2 M_d^{d-2}$, as happens when the kinetic mixing parameter $\beta \in [-1,1]$ approaches $\pm 1$.

Since the lightest KK mode of the projected-out gauge field has mass  $m_\gamma = \frac{1}{2R} = \frac{m_{\text KK}}{2}$, we conclude that \eqref{geninv} is satisfied with $c = 2 \sqrt{\gamma \gamma_M}$. Moreover, $\gamma, \gamma_\textsc{M} \le \frac{D-2}{D-3}$ because couplings to moduli can only \emph{decrease} the masses of extremal black holes (see, e.g., appendix A in \cite{Harlow:2022ich}). Thus, we conclude that $c \le 2\frac{D-2}{D-3}$.\footnote{We have shown this under the assumption that $\gamma$ does not depend on the charge direction $\hat{q}$, since the computation in \eqref{eq:mixLWGCfailure} relied upon this supposition. More generally, the sublattice splitting is related to the curvature of the boundary of the black hole region at the point where $q_1 = q_2$, see figure~\ref{f.BHregion} and the associated discussion. Whether this is inversely related to the flux tube tension for arbitrary moduli couplings is an interesting question for future research.} In simple string theory examples, such the one discussed in~\S\ref{s.9dexample}, one finds $\gamma = \gamma_\textsc{M} = 1/2$ and so $c=1$ in such cases. It would be interesting to formulate sharper bounds on $c$ through a more careful study of the string landscape.

Note that the effective field theory description used to obtain the above results is only valid at large $S^1$ radius, where $m_\text{KK} \ll M_d$. In this controlled regime, \eqref{eq:finsw} implies that whenever the splitting is large ($\msplit \gg M_d$) there are low-tension flux tubes ($\Tflux \ll M_d^{d-2}$). However, this leaves open the question of what happens when the $S^1$ radius is small. Even if \eqref{eq:finsw} remains true in this regime, it would no longer imply that large sublattice splitting must come with low-tension flux tubes. To find out what actually happens at small radii we need a full string theory embedding, see~\S\ref{s.9dexample}.

\subsection{Flux tube stability and decay via pair creation}\label{ss.fluxtubestabilitygeneral}

We have argued that whenever the sublattice splitting is large, there are low tension flux tubes confining fractionally charged monopoles. These flux tubes are necessarily unstable, since they can break apart by pair-creating monopoles. If this process is rapid then the existence of the flux tubes might have little discernible effect on the low-energy EFT, regardless of their tension.

To address this, we now estimate the rate at which flux tubes break apart via nucleation of monopole anti-monopole pairs. We will find that this rate strongly depends on whether the flux tube tension scale is above or below the monopole mass/tension scale.

\begin{figure}[h]
				\begin{center}
					\includegraphics[width = 10cm]{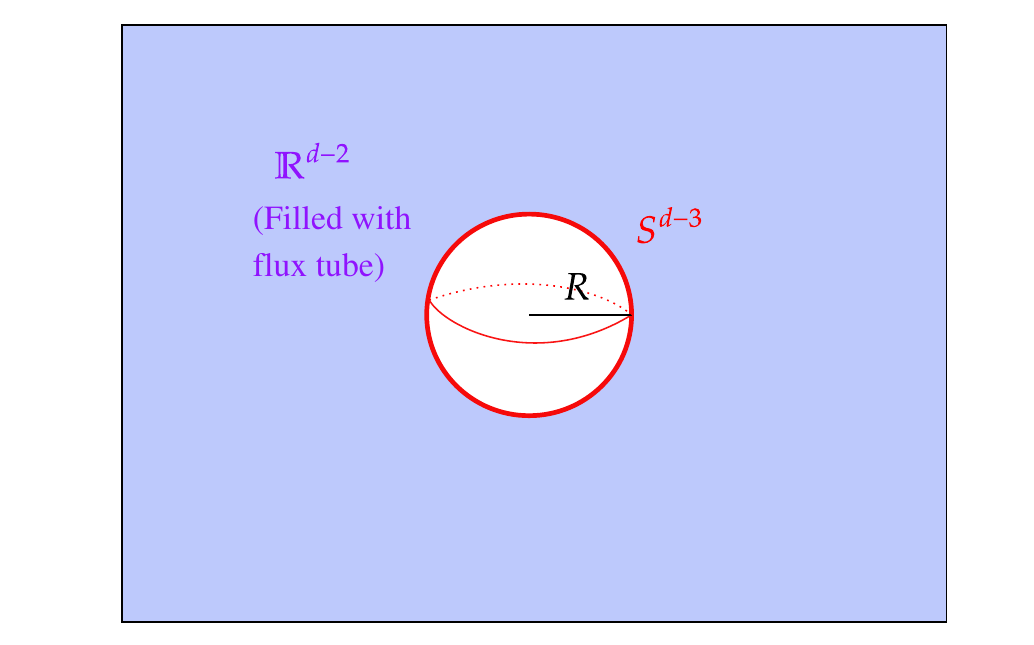}
				\end{center}
				\caption{Bounce solution describing the opening of a monopole-lined hole of radius $R$ (shown in red) inside the flux tube worldvolume (shaded blue). The actions of such a configuration controls the rate at which flux tubes fragment via monopole pair creation.}
				\label{f.bounce}
			\end{figure}

To compute this rate, we consider the Euclidean bounce solution shown in Figure \ref{f.bounce}. The transition rate per volume is controlled by the Euclidean action of this configuration relative to that of the unbroken flux tube, $\Gamma\sim\alpha e^{-\Delta S}$ where $\Delta S = S_{\text{bounce}} - S_{\text{tube}}$. 
            The monopole bubble makes a positive contribution to $\Delta S$, whereas the removal of the flux tube inside the bubble makes a negative contribution. For a $(p+1)$-dimensional flux tube, we find:
            \begin{align}   
                \Delta S = T_{\mathrm{mon}} A(R) - T_{\mathrm{flux}} V(R)\sim T_{\mathrm{mon}} R^p - T_{\mathrm{flux}} R^{p+1}.
            \end{align}
            Extremizing $\Delta S$, we find a saddle point at $R \sim T_{\mathrm{mon}}/T_{\mathrm{flux}}$, with
            \begin{align}
                (\Delta S)_{\text{saddle}} \sim \dfrac{T_{\mathrm{mon}}^{p+1}}{T_{\mathrm{flux}}^p}\sim\left(\frac{m_{\mathrm{mon}}}{m_{\mathrm{flux}}}\right)^{p(p+1)}
            \end{align}
            where in the last step we express the tensions in terms of the associated mass scale for each brane, $m_{\mathrm{mon}}^p\equiv T_{\mathrm{mon}}$, $m_{\mathrm{flux}}^{p+1}\equiv T_{\mathrm{flux}}$.\footnote{Note that we suppressed various factors in our calculation, such as the area of a unit $p$ sphere and the volume of a unit $p+1$ ball. A more careful calculation yields $(\Delta S)_{\text{saddle}} = \bigl(k \frac{m_{\mathrm{mon}}}{m_{\mathrm{flux}}}\bigr)^{p(p+1)}$ where $1 < k \le \sqrt{\pi}$ is an order-one numerical factor that we suppress for simplicity.} Thus, so long as $m_{\mathrm{mon}}\gg m_{\mathrm{flux}}$, the flux tubes are exponentially long-lived.

In the example studied above, the monopole and flux tube tensions roughly satisfy $T_{\text{mon}} \sim (2\pi R) T_{\text{tube}}$ since they both original from $D$-dimensional monopoles (albeit with slightly different charges), one being wrapped and the other not. Thus, the criterion $m_{\mathrm{mon}}\gg m_{\mathrm{flux}}$ is essentially equivalent to the requirement that the KK scale is far below the tension scale of the higher-dimensional monopoles. Since the latter are typically fundamental branes in the theory, their tension scale has to be above the quantum gravity scale and thus the flux tubes are long-lived whenever our EFT description is valid. We study this point more carefully in a specific landscape example in~\S\ref{subsec:9dfluxtubestability}.

            \subsection{Relation to LWGC violation in $T^3/\mathbb{Z}_2$ compactifications}
			In \cite{Heidenreich:2016aqi}, an example of LWGC failure was found in pure gravity compactified on an orbifolded three-torus. That example is closely related to our $[\mathrm{U}(1)\times \mathrm{U}(1)]\rtimes \mathbb Z_2$ example, as we will now briefly explain. That example was presented in terms of a $T^3$ parametrized by three $2\pi$-periodic angles $w, y, z$, quotiented by a freely acting group $\ZZ_2 \times \ZZ'_2$ where
			\begin{alignat}{3}
				& \ZZ_2: \quad && w \mapsto w + \pi, && \quad y \mapsto y + \pi,  \\
				& \ZZ'_2: \quad && w \mapsto -w, && \quad z \mapsto z + \pi.
			\end{alignat}
			This presentation was convenient for analyzing the modes of the orbifold theory, but as that reference remarked in a footnote, $T^3/\ZZ_2$ is again a 3-torus, so we can alternatively present the space as follows:
			\begin{equation}
				a = y + w, \quad b = y - w, \quad c = z.
                \label{twoz2involution}
			\end{equation}
			One can check that, taking these to be individually invariant under $2\pi$ shifts (which follows from the action of the first $\ZZ_2$ above), the torus parameterized by $(a,b,c)$ is equivalent to the original $T^3/\ZZ_2$. Then the action of $\ZZ'_2$, translated to the new variables, is
			\begin{equation}
				\ZZ'_2: \quad  a \leftrightarrow b,  \quad c \mapsto c + \pi.\label{eq.singleZ2involution}
			\end{equation}
			If we first dimensionally reduce pure gravity on the two-torus parametrized by $a$ and $b$, we obtain a $[\mathrm{U}(1)\times \mathrm{U}(1)]\rtimes \mathbb Z_2$ gauge theory, where the $\ZZ_2$ acts as $a \leftrightarrow b,c \mapsto c + \pi$. Further compactifying the $c$ direction with the $\ZZ'_2$ orbifold then manifestly involves a Wilson line for the $\ZZ_2$ exchange symmetry of the $a$- and $b$-cycles (and their corresponding $\mathrm{U}(1)$ gauge fields).

We will see this connection in a landscape example in~\S\ref{subsec:Mlift}.

\subsection{Generalization to other gauge groups}\label{ss.generalization}

The mechanism discussed above is easily generalized by considering discrete Wilson lines in more general gauge groups. Let $G$ denote the connected portion of the $0$-form gauge group, and suppose there is a discrete $\mathbb{Z}_k$ gauge symmetry that acts on $G$ via some element $w \in \Out(G)$ of order $k$. We now compactify on $S^1$ with a $w$ Wilson line turned on. Only the photons that are $w$-invariant
will be massless in the reduced theory. For an electric charge $\vec{q} \in
\Gamma$ in the original electric charge lattice $\Gamma$, the charge under
these photons is:
\begin{equation}
  \vec{q}_{\text{inv}} = \Pi (\vec{q}) \df \frac{1}{k} \sum_{n = 0}^k w^n
  (\vec{q}),
\end{equation}
which is the projection onto the $w$-invariant plane $\Sigma_{\text{inv}}$.
The reduced electric charge lattice is therefore $\Pi (\vec{\Gamma})$. By
contrast, since magnetic monopoles come from higher-dimensional monopoles
wrapping the compact circle, the higher-dimensional monopoles must have a
$w$-invariant charge $\vec{Q} \in \tilde{\Gamma}$ to be unconfined, $\vec{Q} =
w (\vec{Q})$, so the unconfined magnetic charge lattice is $\tilde{\Gamma}
\cap \Sigma_{\text{inv}}$.

First, we show that this results in a complete spectrum of unconfined objects,
i.e., we show that
\begin{equation}
  \Pi (\Gamma)^{\vee} = \tilde{\Gamma} \cap \Sigma_{\text{inv}} \qquad
  \text{where} \qquad \tilde{\Gamma} = \Gamma^{\ast},
\end{equation}
where $(\cdots)^{\vee}$ denotes the dual within the invariant plane and
$(\cdots)^{\ast}$ denotes the dual in the ambient space. To do so, we note
that:
\begin{equation}
  \vec{Q} \cdot \Pi (\vec{q}) = \frac{1}{k} \sum_{n = 0}^k \vec{Q} \cdot w^n
  (\vec{q}) = \frac{1}{k} \sum_{n = 0}^k w^{- n} (\vec{Q}) \cdot \vec{q} = \Pi
  (\vec{Q}) \cdot \vec{q},
\end{equation}
where we use the fact that $w$ must be an orthogonal transformation since it
is a symmetry of the charge lattice. Now we can reason as follows: suppose
that $\vec{Q}$ is chosen such that
\begin{equation}
  \vec{Q} \cdot \Pi (\vec{q}) \in \mathbb{Z} \qquad \text{for any} \quad
  \vec{q} \in \Gamma .
\end{equation}
This implies that
\begin{equation}
  \Pi (\vec{Q}) \cdot \vec{q} \in \mathbb{Z} \qquad \text{for any} \quad
  \vec{q} \in \Gamma \qquad \Rightarrow \qquad \Pi (\vec{Q}) \in
  \tilde{\Gamma} .
\end{equation}
Thus, requiring both $\vec{Q} = \Pi (\vec{Q})$ as well as $\vec{Q} \cdot \Pi
(\Gamma) \in \mathbb{Z}$, we arrive at $\vec{Q} \in \tilde{\Gamma} \cap
\Sigma_{\text{inv}}$, which is exactly what we wanted to show. Note that the
same argument also implies that
\begin{equation}
  \Pi (\tilde{\Gamma})^{\vee} = \Gamma \cap \Sigma_{\text{inv}},
\end{equation}
which we will use later.

The reduced theory theory features both splitting and confinement. Under the assumption of a strictly convex black hole region (see \ref{f.BHregion} and the accompanying text), an
extremal electric charge in the parent theory will remain extremal if $w
(\vec{q}) = \vec{q}$, i.e., if $\vec{q} \in \Gamma \cap \Sigma_{\text{inv}}$.
Otherwise it will become subextremal. Thus, the extremal sublattice is
$\Gamma \cap \Sigma_{\text{inv}}$; off this sublattice, there is a
splitting in the reduced theory.

Meanwhile, confining flux tubes come from
unwrapped branes with charges of the form
\begin{equation}
  \vec{Q} - w (\vec{Q}),
\end{equation}
for some $Q \in \Gamma$. Similar to figure~\ref{f.FluxTubeABPhase}, we can compute the Aharonov-Bohm phase associated to these flux tubes by unwrapping the $S^1$ to its $\mathbb{R}$ universal cover. This yields a sequence of $k$ charges
\[ \vec{Q} - w (\vec{Q}), w (\vec{Q}) - w^2 (\vec{Q}), \cdots, w^{k - 1}
   (\vec{Q}) - \vec{Q} \]
spaced out by distances $2 \pi R / k$, repeating infinitely. Each time we
cross one of these charges, the flux through a disk perpendicular to it jumps
by the charge, hence this is consistent with flux
\[ \vec{\Phi}_n = \vec{\phi}_{\vec{Q}} - w^n (\vec{Q}) \qquad \text{in
   between} \qquad w^{n - 1} (\vec{Q}) - w^n (\vec{Q}) \qquad \text{and}
   \qquad w^n (\vec{Q}) - w^{n + 1} (\vec{Q}), \]
for some constant $\vec{\phi}_{\vec{Q}}$ to be determined. In the linear
approximation, which is valid at large radius $R$, the average flux should be zero,
i.e.,
\begin{equation}
  0 = \sum_{n = 0}^{k - 1} \vec{\Phi}_n = k \vec{\phi}_{\vec{Q}} - \sum_{n =
  0}^{k - 1} w^n (\vec{Q})  \qquad \Rightarrow \qquad \vec{\phi}_{\vec{Q}} =
  \frac{1}{k} \sum_{n = 0}^{k - 1} w^n (\vec{Q}) \,.
\end{equation}
This is because averaging the flux is the same as averaging over constant
shifts in the positions of the charges, which in the linear approximation is the
same as smearing out the charge distribution. Since the net charge vanishes,
the smeared charge density also vanishes, so
\begin{equation}
  \bar{\Phi} \df \frac{1}{2 \pi R} \int_0^{2 \pi R} \Phi (y) \rmd y = 0\,.
\end{equation}
Thus we have,
\begin{equation}
  \vec{\Phi}_n = - w^n (\vec{Q}) + \frac{1}{k} \sum_{\ell = 0}^{k - 1}
  w^{\ell} (\vec{Q}) = - w^n (\vec{Q}) + \Pi (\vec{Q})\,.
\end{equation}
The Aharonov-Bohm phase for a particle of electric
charge $\vec{q}$ circling the flux tube is then given by:\footnote{As noted, this result is valid when the compactification radius $R$ is sufficiently large to justify the linear approximation above. It would be interesting to understand better whether/when this result extends to smaller radii. For $k=2$, however, the argument given in figure~\ref{f.FluxTubeABPhase} applies, irrespective of the radius.}
\begin{equation}
  \exp [2 \pi i \vec{q} \cdot \vec{\Phi}_n]  = \exp [2 \pi i \vec{q} \cdot \Pi (\vec{Q})] \,,
\end{equation}
where we have used the fact that $\vec q \cdot w^n(\vec{Q}) \in \mathbb{Z}$, since $\vec q \in \Gamma$, $w^n(\vec{Q}) \in \tilde \Gamma$.

Let us now perform some important checks on this result. First, note that if
we shift $\vec{Q} \rightarrow \vec{Q} + \delta \vec{Q}$ with $\delta \vec{Q}
\in \Gamma_{\text{mag}}^{\text{inv}}$ then $w^n (\vec{Q}) - w^{n + 1}
(\vec{Q})$ does not change, so we are talking about the same flux tube.
This modifies the Aharonov-Bohm phase by a multiplicative factor of
\begin{equation}
  \exp [2 \pi i \vec{q} \cdot \Pi (\delta \vec{Q})] = \exp [2 \pi i \vec{q}
  \cdot \delta \vec{Q}] = 1,
\end{equation}
so the phase is unaffected by the shift. Moreover, we can rewrite the
Aharonov-Bohm phase as:
\begin{equation}
  \exp [2 \pi i \vec{Q} \cdot \Pi (\vec{q})]
\end{equation}
so it manifestly only depends on the unbroken gauge charge of the particle
$\vec{q}_{\text{inv}} = \Pi (\vec{q})$. This can be further rewritten as:
\begin{equation}
  \exp [2 \pi i \Pi (\vec{Q}) \cdot \Pi (\vec{q})] .
\end{equation}
From this we see that the lattice of confined monopoles is just the projection
of $\Pi (\tilde{\Gamma})$ of $\tilde{\Gamma}$ onto the invariant plane.
\begin{table}\renewcommand{\arraystretch}{1.5}
\begin{center}
\begin{tabular}{|c|c|c|} \hline
 \textbf{Symbol} & \textbf{Description} & \textbf{Relationships}  \\ \hline
     $\Pi (\Gamma)$ &  Electric charge lattice & $= (\tilde{\Gamma} \cap \Sigma_{\text{inv}})^\vee$ \\\hline
     $\Gamma \cap \Sigma_{\text{inv}}$ & Extremal sublattice & $= \Pi (\tilde{\Gamma})^\vee$ \\\hline
   $\tilde{\Gamma} \cap \Sigma_{\text{inv}}$ &   Magnetic charge lattice & $=\Pi (\Gamma)^\vee$  \\\hline
    $\Pi (\tilde{\Gamma})$ &  Confined monopole superlattice &  $=( \Gamma \cap \Sigma_{\text{inv}})^\vee$ \\\hline
   \end{tabular}
   \end{center}
   \caption{List of lattices. The superextremal sublattice is dual to the superlattice of confined monopoles.}
   \label{glossarydisc}
   \end{table}
From this we conclude that, indeed, the superextremal sublattice is the dual of the
superlattice of confined monopoles. See Table \ref{glossarydisc} for the full list of lattices and the relationships between them.

We can generalize this yet further by taking $G$ to be a $p$-form gauge symmetry for $p > 0$ acted on by a discrete $0$-form gauge symmetry $\mathbb{Z}_k \subseteq \Out(G)$. All the arguments and computations are analogous, except that upon reduction we get both a $p$-form gauge symmetry and a $(p - 1)$-form gauge
symmetry, and under the latter there are confined \emph{electric} charges,
as well as monopole splitting. We will see a landscape example of this in the next section.

\section{9d string theory example\label{s.9dexample}}
We now investigate a 9d string theory example of LWGC failure discovered by Montero and Parra de Freitas in \cite{Montero:2022vva}. In this theory, the LWGC for strings fails, and we find associated fractionally-charged magnetic fourbranes confined by fivebrane flux tubes. We also discover that the LWGC for fivebranes fails, and we similarly find fractionally-charged magnetic particles confined by string flux tubes. For both sets of objects, in regions of the moduli space where the LWGC failure is more extreme (i.e., the lightest string or fivebrane of a given charge becomes very subextremal), the tensions of the associated flux tubes become light in 9d Planck units.

This section will share many general features with \S\ref{sec:exchangeEFTexample}, but we will also find several novel features not present in the EFT examples. Crucially, string dualities will allow us to extend our analysis into strongly-coupled and small-radius regions of moduli space inaccessible to EFT descriptions. Indeed, we will find light, stable flux tubes accompanying large sublattice splitting across the entirety of the moduli space.

\subsection{DP$_-$ phase review}
The 9d theory discovered by Montero and Parra de Freitas in \cite{Montero:2022vva} is obtained by compactifying 10d Type IIB string theory on a circle with an $\Omega T$ Wilson line, where $\Omega$ is worldsheet parity and $T$ is the $T$-generator of $\mathrm{SL( 2,\mathbb{Z}})$. This forces a nonzero axion background $C_0=-\frac{1}{2}$ to be consistent with the Wilson line.\footnote{In \S\ref{sec:Zp} and \S\ref{sec:exchangeEFTexample}, we followed a convention for the normalization of gauge fields where fluxes took values in $2\pi\mathbb{Z}$, so that axions are $2\pi$-periodic. In this section, we have shifted conventions so that fluxes are valued in $\mathbb{Z}$ and axions have period $1$, for consistency with earlier literature~\cite{Montero:2022vva}.} From the 9d perspective the backgrounds with Wilson line $\Omega$, $C_0=0$ and Wilson line $\Omega T$, $C_0=-\frac1 2$ correspond to a choice of discrete $\theta$ angle.\footnote{Note that our conventions differ slightly from those in \cite{Montero:2022vva}, where $C_{0}^{\text{there}}=+1/2$. Following the conventions  explained in detail in Appendix \ref{s.conventions}, $\Omega T$ acts on the axio-dilaton as $\tau \to -1 - \tau^\star$, thus, we are forced to set $C_0^\text{here}=-1/2$ to avoid a moduli gradient around the circle, in effect introducing a discrete $\theta$ angle.}

The component of moduli space of this 9d theory with $\theta \neq 0$ is two-dimensional, which we parametrize using the canonically normalized dilaton and radion $\phi$ and $\rho$, the normalization conventions of which are given in \S\ref{ss.coordinatesof9dphases}. This moduli space contains three distinct phases,
 \begin{align}
\text{DP$_-$:}~ \phi < 0\,,~ \rho > -\frac{1}{\sqrt{7}}\phi\,;~~~\text{AOB$_-$:}~ \phi >0 \,,~ \rho > \frac{1}{\sqrt{7}}\phi\,;~~~\text{O8$^\pm_-$:} ~  \frac{1}{\sqrt{7}}\phi <  \rho < - \frac{1}{\sqrt{7}}\phi\,.
 \label{e.regions}
\end{align}
The DP$_-$ phase can be viewed as the previously-known DP (Dabholkar Park) \cite{Dabholkar:1996pc} background, but with a discrete $\theta$ angle turned on. The AOB$_-$ phase is the S-dual of the DP$_-$ phase and can be viewed as the previously-known AOB (Asymmetric Orbifold of Type IIB) \cite{Hellerman:2005ja} background with a discrete $\theta$-angle turned on, also with $C_0 = - \frac{1}{2}$. The AOB$_-$ phase is self T-dual at $\rho=\phi/\sqrt 7$ and as in \cite{Montero:2022vva} this line actually extends across the moduli space. The specific element of SL$(2,\ZZ)$ that exchanges DP$_-$ and AOB$_-$ is given in Appendix \ref{s.conventions}. Finally, the O8$^\pm_-$ phase is the T-dual of the DP$_-$ phase and can be viewed as the previously-known O8$^\pm$ \cite{Aharony:2007du} background with a discrete $\theta$ angle in the form of the holonomy $\int C_1 \neq 0$. The O8$^\pm_-$ phase is self-dual at $\rho=\phi/\sqrt 7$. Following \cite{Aharony:2013hda}, the subscript $-$ on the labels DP$_-$, AOB$_-$, and O$8^\pm_-$ of the phases indicate that the discrete $\theta$ angle is turned on, and the traditional phases with the discrete theta-angles set to 0 are labeled AOB$_+$, DP$_+$, and O$8^\pm_+$.

In this section, we will use the DP$_-$ phase to describe our analysis. We now review the features of the $\text{DP}_-$ background. The action of $\Omega T$ on the two-form gauge fields $C_2$ and $B_2$, as well as on $(p_1,q_1)$-strings and $(p_5,q_5)$-fivebranes\footnote{The convention used here is that $p_1$ and $q_1$ respectively refer to the number of F1 and D1 strings, while $p_5$ and $q_5$ respectively refer to the number of D5- and NS5-branes.}, is
\begin{align}
\Omega T :& \ \begin{matrix}
					C_2&\mapsto & B_2 + C_2\\
					B_2&\mapsto &- B_2\\
					C_0&\mapsto &-C_0-1\\
					(p_1, q_1)&\mapsto &(q_1-p_1, q_1)\\
					(p_5,q_5)  &\mapsto &(p_5 - q_5, - q_5). 
\end{matrix} \label{e.OmegaTaction}
\end{align}

The 1-form gauge group\footnote{The 1-form gauge group is part of a higher-group structure due to the presence of Chern-Simons terms. Here, we focus our attention on the 1-form symmetries of this larger structure.} in AOB$_-$ and DP$_-$ is $U(1)$, not $U(1)\times\mathbb{Z}_2$ as one might naively expect. Indeed, in DP$_+$, the 1-form gauge group is $U(1)\times\mathbb{Z}_2$, with $C_2$ invariant under the Wilson line and $B_2$ projected to a $\mathbb{Z}_2$ discrete field. Once we turn on the discrete theta angle, however, the Wilson line now mixes $B_2$ with $C_2$ and the gauge group is the extension of $U(1)$ by $\mathbb{Z}_2$. Hence the discrete remnant of $B_2$ now implements $(-1)\in U(1)$. Additionally, one can read off the (abelian) gauge group as the Pontryagin dual of the conserved charge lattice. In the present example we see the basis of $(p,q)$-string states given by $\{(1,1),(0,1)\}$ integrally generates the charge lattice and under the Wilson line transforms as $(1,1)\leftrightarrow(0,1)$. Thus the conserved charges are generated by two elements $a,b$ with relation $a\sim b$, and so the charge lattice is $\mathbb{Z}$ with dual gauge group $U(1)$.

The action of $\Omega T$ leaves invariant the linear combination
\begin{align}
    A_2 \equiv B_2 + 2 C_2,
\end{align}
whereas $B_2$ acquires a minus sign under the action of $\Omega T$ and is therefore massive in the 9d theory. Thus, the charge under $A_2$ is \emph{conserved}, whereas the charge under $B_2$ is \emph{broken}. The conserved and broken charges of a $(p_1,q_1)$-string can be obtained by rewriting the string's worldvolume coupling to the gauge potentials as
\begin{equation}
			 \int(p_1C_2+q_1B_2)=\int \underbrace{\frac{q_1}{2}}_{\mathrm{\begin{array}{c}  \text{conserved charge}\end{array}}} \times \underbrace{A_2}_{\mathrm{\begin{array}{c}  \mathrm{massless} \end{array}}} + \underbrace{\left( p_1 - \frac{q_1}{2} \right)}_{\begin{array}{c}\text{broken charge}\end{array}}  \times \underbrace{B_2}_{\begin{array}{c}\text{massive}\end{array}.}
\end{equation}
			Which indicates the charges of a string under the invariant gauge field $A_2$ and the odd field $B_2$ are
			\begin{align}
				Q_\text{1}= \frac{q_1}2,\qquad \mathcal{Q}_\text{1}= p_1-\frac{q_1}{2}.\label{e.1charge}
			\end{align}

            We see that for states of $q_1$ odd (i.e., the half-integer invariant charge $Q_1$) the charge under the odd field $\mathcal{Q}_1$ is necessarily non-zero. $\mathcal{Q}_1$ contributes to the tension of the state; however, it is no longer a gauge charge and hence does not contribute to the extremality bound. As such, these states will necessarily be subextremal, and we violate the LWGC with coarseness 2.

			Meanwhile, the charge of a $(p_5,q_5)$-fivebrane is defined by the flux integrals:\footnote{The reason for the minus sign, as explained in the appendix, is a conventional choice that results in simplifications in other $SL(2,\mathbb{Z})$ expressions.}
			\begin{align}
				\begin{pmatrix}
					p_5\\q_5 
				\end{pmatrix}\propto 
				\begin{pmatrix}
					\oint_{S^3}G_3\\-\oint_{S^3} H_3
				\end{pmatrix},
			\end{align}
			where $G_3=\rmd B_2$ and $H_3 = \rmd C_2$. 
            
            The charges of the $(p_5,q_5)$-fivebrane under the invariant gauge field $A_2$ and odd field $B_2$ are given respectively by
			\begin{align}
				Q_\text{5}=\oint F_3=\oint H_3+2G_3=2p_5-q_5,\qquad 
				\mathcal{Q}_\text{5}=\oint H_3=-q_5\,, \label{e.5charge}
			\end{align}
			where locally $F_3 =\rmd A_2$ gives the field strength of the remaining gauge field. Similarly to the case of strings, fivebranes are extremal if and only if they are uncharged under the odd field, $\mathcal{Q}_5=-q_5=0$; however, since the invariant charge is given by $Q_\text{5}=2p_5-q_5$, the odd-$Q_5$ sites in the charge lattice only contain fivebranes with $q_5$ odd and therefore nonzero. These sites possess only sub-extremal states, and thus the LWGC is violated with coarseness 2.

			We lastly remark that the broken 2-form gauge field $B_2$ leaves behind a massive 2-form ``gauge field" of mass
			\begin{align}
				m_{\gamma}=\frac 12 \exp\left (- \sqrt{\frac{8}{7}}\rho\right)M_9
                \label{mgamma9d}
			\end{align}
			coming from the lowest KK mode that survives the projection, which in this case is the $1/2$ mode. $M_9$ denotes the 9-dimensional Planck scale. Most of the quantities in this section will be stated in terms of $M_9$, in order to make comparisons in terms of the daughter theory's Planck scale without referencing the higher dimensional theory.\footnote{However, we will suppress factors of $\kappa_9$ (or $\kappa_d$, when we are working in terms of a $\rho$ canonically normalized to a dimension other than 9) in exponents for ease of reading.}
			
\subsection{Quantifying LWGC violation \label{s.9dexample.quantifying}}

In this section we calculate sublattice splitting for the subextremal objects in the AOB$_-$ and DP$_-$ phases to quantify the degree of LWGC violation. To begin, we may obtain the extremality bound by noting that branes with zero $\mathcal Q_{1,5}$ are BPS. The tension of these BPS branes does not vanish anywhere in moduli space, so they are exactly extremal \cite{Harlow:2022ich}, and their charge-to-tension ratios determine the extremality bound.

        The tension of a $(p,q)$ string or fivebrane is given by
			\begin{align}
				\Tstrfive =\tildetstrfive \frac{|p+\tau q|}{\sqrt{\tau_2}}\,, \qquad \text{where} \qquad \tilde{t}_1 = \frac{(2\pi)^{3/4}}{(2\kappa_{10}^2)^{1/4}} ,\qquad \tilde{t}_5 = \frac{(2\pi)^{1/4}}{(2\kappa_{10}^2)^{3/4}}\,,
			\end{align}
            where
            $
            \tau = \tau_1 + i \tau_2 = C_0+ie^{-\Phi} = C_0 + \frac{i}{g_s} 
            $.
            In particular, measured in 9d Planck units the dimensionful prefactors depend on the radion:
			\begin{align}
				\tildetstr \propto \exp\left(-\frac{1}{\sqrt{14}}\rho\right)\tstr,\qquad \tildetfive \propto \exp\left(-\frac{3}{\sqrt{14}}\rho \right) \tfive\,,
                \label{tildedef} 
			\end{align}
			with $\rho$ the canonically normalized radion, and $t_{1,5}$ are $M_9^{2,6}$ times dimensionless numbers.
        
        The presence of the Wilson line in the AOB$_-$ or DP$_-$ phase fixes $C_0 = -1/2$, so in this case the tension formula gives
        \begin{align}
		   \Tstrfive=\tildetstrfive\frac{\sqrt{\left(p-\frac{q}{2}\right)^2+q^2 \tau_2^2}}{\sqrt{\tau_2}}.
		\end{align}
        This formula can of course be rewritten in terms of the conserved and broken charges $Q_{1,5}$ and $\mathcal{Q}_{1,5}$, which yields
        \begin{align}
				T_{1}=\tildetstr \frac{\sqrt{\mathcal{Q}_1^2+4Q_1^2 \tau_2^2}}{\sqrt{\tau_2}}
			\end{align}
            for onebranes and
            \begin{align}
				T_{5}=\tildetfive \frac{\sqrt{4Q_5^2+\mathcal{Q}_5^2 \tau_2^2}}{\sqrt{\tau_2}}
			\end{align}
            for fivebranes.
            
        We obtain the extremality bounds for 1- and 5-branes by setting $\mathcal{Q}_{1,5}=0$ respectively:
        \begin{subequations}
				\begin{align}
					T_\text{ext}(Q_\text{1})&=2\tildetstr |Q_\text{1}|\sqrt{\tau_2},\\
					T_\text{ext}(Q_\text{5})&=\tildetfive \frac{|Q_\text{5}|}{2\sqrt{\tau_2}}\,,
				\end{align}
                \label{exteq}
			\end{subequations}
        where we now describe states by their invariant charges $Q_{1}=q_1/2$ and $Q_5 = 2p_5 - q_5$ rather than their $(p,q)$.
            
        However, for half-integer string charge $Q_\text{1}$ and for odd-integer fivebrane charge $Q_\text{5}$, the lightest strings and fivebranes cannot have $\mathcal{Q}=0$, and instead have tensions
			\begin{subequations}
				\begin{align}
					T_\text{min}(Q_\text{1})&=2 \tildetstr \sqrt{\tau_2 Q_\text{1}^2+\frac{1}{16 \tau_2}},\\
					T_\text{min}(Q_\text{5})&=\tildetfive \frac{\sqrt{(Q_\text{5})^2+4\tau_2^2}}{2 \sqrt{\tau_2}}
				\end{align}
                \label{mineq}
			\end{subequations}
            As expected, for half-integer string charge, or odd-integer fivebrane charge, the lightest strings and fivebranes are subextremal, and the LWGC is violated.

			To quantify the failure of the LWGC, we generalize the splitting \eqref{Deltadef} from masses $m$ to tensions $T$, giving
            \begin{subequations}
            \begin{align}
				(\Tsplitstr)^2&=T_\text{min}(Q_\text{1})^2-T_\text{ext}(Q_\text{1})^2 = 
				\frac{\tildetstr^2}{4\tau_2}, \\ 
				(\Tsplitfive)^2 &=T_\text{min}(Q_\text{5})^2-T_\text{ext}(Q_\text{5})^2	 = \tildetfive^2 \tau_2.
			\end{align}
            \label{DeltaT9d}
            \end{subequations}
			Written in terms of the canonically normalized radion $\rho$ and the canonically normalized dilaton $\phi= (\Phi-\log 2)/\sqrt{2}$ (of the DP$_-$ phase),\footnote{Here, the $\log 2$ offset is chosen so that $\phi = 0$ marks the boundary between the AOB$_-$ and DP$_-$ phases, where the AOB string tension is equal to the DP string tension. See Appendix \ref{ss.coordinatesof9dphases} for further details.}
           these become
			\begin{align}
				\Tsplitstr\propto 
				\frac{\tstr}{2}\exp\left(-\frac{1}{\sqrt{14}}\rho+\frac 1{\sqrt 2}\phi\right),\qquad 
				\Tsplitfive\propto 
				\tfive\exp\left(-\frac{3}{\sqrt{14}}\rho-\frac 1{\sqrt 2}\phi\right).\label{e.Tsplit}
			\end{align}
			Thus, the sublattice splitting for strings and fivebranes respectively is large when
			\begin{align}
				-\frac{1}{\sqrt{14}}\rho+\frac 1{\sqrt 2}\phi \gg1,\qquad 
				-\frac{3}{\sqrt{14}}\rho-\frac 1{\sqrt 2}\phi \gg 1.
			\end{align}

            \subsection{Flux tubes in 9d string theory}\label{ss.moncon}
            
			We now show that regions of moduli space where the LWGC failure is large also exhibit low-tension flux tubes that confine fractionally charged monopoles. We first construct the flux tubes in the manner of \S\ref{sec:exchangeEFTexample}. The Wilson line can be viewed as introducing a coordinate defect localized at a point on the $S^1$ and extended in all $9$ remaining dimensions, at which fields and objects obey a boundary condition with an $\Omega T$ twist, as in Figure \ref{f.T3fractional}. A $(p,q)$ string or fivebrane becomes an $\Omega T(p,q)$ string or fivebrane on the other side of the defect, so it cannot connect back to itself in isolation unless $(p,q)=\Omega T(p,q)$. The minimal charge branes do not satisfy this, namely the confined monopoles, so instead of reconnecting the brane to itself, we direct the ingoing and outgoing tails which recombine into a flux tube:
			\begin{align}
				(p,q)-\Omega T(p,q)=\begin{cases}
					(2p_1-q_1,0) \quad \text{  (strings)}\\(q_5,2q_5) \quad\text{  (fivebranes)}
				\end{cases},
			\end{align}
            where $(p,q)$ is the bound state of the associated monopole wrapping the $S^1$.

			As a result, in the 9d theory with the Wilson line turned on, a $(p_1,q_1)$ particle is attached to a $(2p_1-q_1,0)$ string flux tube with tension
			\begin{align}
                \Tfluxstr(p_1,q_1)=\frac{\tildetstr}{\sqrt{\tau_2}}|2p_1-q_1|=\frac{\tildetstr}{\sqrt{\tau_2}}|2\mathcal{Q}_1|,
			\end{align}
			and a $(p_5,q_5)$ four-brane is attached to a $(q_5,2q_5)$ fivebrane with tension
			\begin{align}
				\Tfluxfive(p_5,q_5)=2\tildetfive\sqrt{\tau_2}|q_5|=\tildetfive\sqrt{\tau_2}|2\mathcal{Q}_5|.
			\end{align}

            Some general properties of the flux tubes are as follows. The flux tubes have vanishing charge $Q$ under the unbroken gauge field $A_2$ (as they must; otherwise, charge conservation would prevent them from ending). Additionally, flux tubes are present if and only if the monopole is such that $\mathcal{Q}\neq 0$. That is to say, the flux tubes confine the fractional charge of the monopoles into physical Dirac strings. Specifically, flux tubes carry twice the $\mathcal{Q}$ of the associated monopole, as is expected since one can pair create two monopoles connected by a flux tube. (In fact, we do not expect the flux tube to carry the minimal $\mathcal{Q}$, since those states possess nonzero $Q$, which is forbidden for a flux tube.) This nonzero $\mathcal{Q}$ is in line with the fact that the flux tubes are \textit{not} invariant under the Wilson line.\footnote{Note that it is consistent to have an object that is not invariant under the Wilson line in the spectrum, so long as it is unwrapped. Such behavior merely indicates the object is charged under a broken U(1) symmetry.}

            \subsubsection{Tension and stability of minimal flux tubes} \label{subsec:9dfluxtubestability}
            We now turn to the important question of the stability of the flux tubes, using the findings of \S $\ref{ss.fluxtubestabilitygeneral}$. We will consider only the minimal $\mathcal{Q}$ confined monopole at each site, since any larger monopole of the same $Q$ would decay to an unconfined monopole of charge $Q$ plus one of these minimal monopoles. These minimal monopoles are connected to minimal flux tubes, so we will only consider the stability of minimal flux tubes decaying to the lightest confined monopoles of a given charge.

            We now quantitatively describe the tensions of these minimal flux tubes. For the lightest particle with a given half-integer invariant charge and the lightest fourbrane with a given odd charge, the flux tube which confines them will have tension
			\begin{align}
				\Tfluxstr=
				\tstr \exp\left(-\frac{1}{\sqrt{14}}\rho+\frac 1{\sqrt 2}\phi \right),\qquad 	\Tfluxfive=
				2\tfive \exp\left(-\frac{3}{\sqrt{14}}\rho -\frac 1{\sqrt 2}\phi \right),\label{e.Tflux}
			\end{align}
            respectively, which we note as precisely the same moduli dependence as for the sublattice splittings in \eqref{e.Tsplit}. When the factors in the exponents are much less than 0, the flux tubes will become light.
            
\begin{figure}[h]
				\begin{center}
					\includegraphics[width = 120mm]{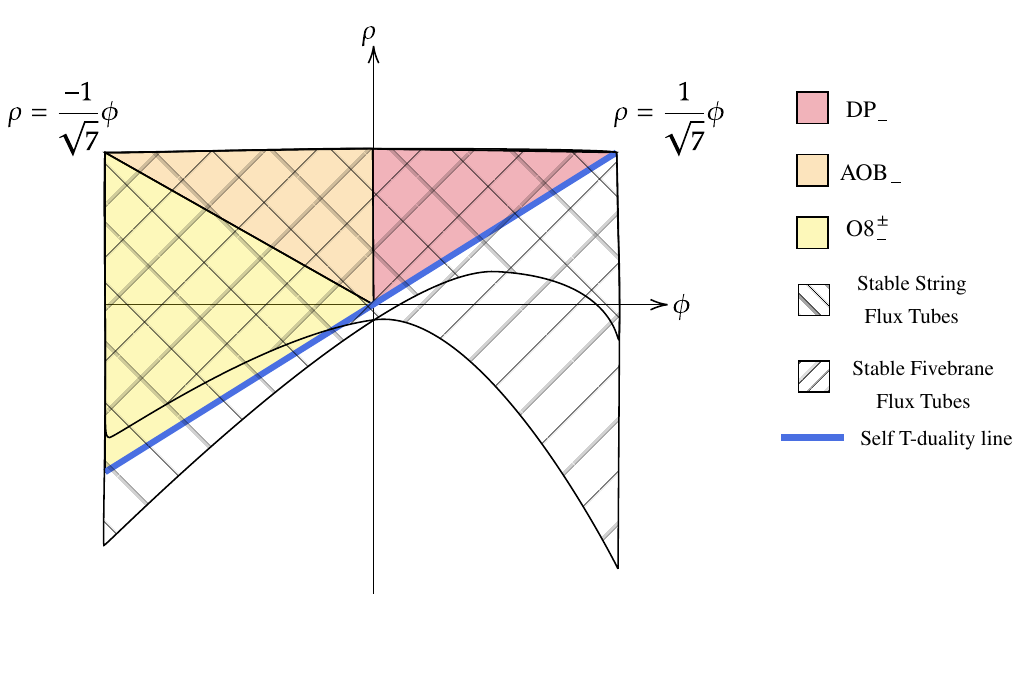}
				\end{center}
				\caption{Regions of moduli space where the string and fivebrane flux tubes are stable. Only the string flux tubes are light in the entirety of the O8$^\pm_-$ region. Below the bottom border of the AOB$_-$ phase, a self T-dual description of the AOB$_-$ phase emerges, which can then be $S$-dualized to another DP$_-$ phase, and then T-dualized again to another O8$^\pm_-$ phase. Therefore, light flux tubes persist across the entire moduli space. We include the stability regions below the self T-duality line merely to show that there are stable flux tubes up to and past the line where self-duality occurs.} 
				\label{f.stabilityoffluxtubesindifferentphases}
			\end{figure}

            The lightness of flux tubes alone does not guarantee their stability, as one must consider the mass of the monopole to which they are attached to determine the rate of pair creation. The stability is determined via the methods of section \S $\ref{ss.fluxtubestabilitygeneral}$, which involves computing the ratio of the mass scales of the monopoles and flux tubes. In this 9d example at hand, the lightest confined 0-brane and 4-brane monopoles are of charge $Q_1=1/2,\,Q_5=1$ respectively, and carry tension and associated mass scales of
            \begin{align}
             m_{1,\,\text{mon}}^2\sim\left(\frac{e^\phi}{4}+e^{-\phi}\right)e^{\frac{6}{\sqrt{14}}\rho}M_9^2, &\quad T_{4,\,\text{mon}}^2\sim\left(\frac{e^\phi}{4}+e^{-\phi}\right)e^{\frac{2}{\sqrt{14}}\rho}M_9^{10} \\
             m_{1,\,\text{mon}}^2\sim\left(\frac{e^\phi}{4}+e^{-\phi}\right)e^{\frac{6}{\sqrt{14}}\rho}M_9^2, &\quad m_{4,\,\text{mon}}^2\sim\left(\frac{e^\phi}{4}+e^{-\phi}\right)^{1/5}e^{\frac{2}{5\sqrt{14}}\rho}M_9^2
             \end{align}
             Meanwhile, the flux tube tensions given in \eqref{e.Tflux} are associated with mass scales
             \begin{align}
				m_{1,\,\text{flux}}\sim e^{-\frac{1}{2\sqrt{14}}\rho+\frac 1{2\sqrt 2}\phi}M_9^2,\qquad 	m_{5,\,\text{flux}} \sim
				e^{-\frac{3}{6\sqrt{14}}\rho -\frac 1{6\sqrt 2}\phi}M_9^2,
			\end{align}
            This gives values for the stability parameters of:
            \begin{align}
        \frac{m_{1,\,\text{mon}}}{m_{1,\,\text{flux}}} \sim \dfrac{\left(\frac{e^\phi}{4}+e^{-\phi}\right)e^{\frac{6}{\sqrt{14}}\rho}}{e^{-\frac{1}{2\sqrt{14}}\rho+\frac 1{2\sqrt 2}\phi}}, \quad \frac{m_{5,\,\text{mon}}}{m_{5,\,\text{flux}}} \sim \dfrac{\left(\frac{e^\phi}{4}+e^{-\phi}\right)^{1/5}e^{\frac{2}{5\sqrt{14}}\rho}}{e^{-\frac{3}{6\sqrt{14}}\rho -\frac 1{6\sqrt 2}\phi}}
            \end{align}
            The flux tubes are stable when these ratios are much greater than $1$. The regions where this occurs are plotted in Figure \ref{f.stabilityoffluxtubesindifferentphases}. We see that both string and fivebrane flux tubes are stable everywhere in the DP$_-$ and AOB$_-$ phases, whereas only the string flux tubes are light across the entirety of the O8$^\pm_-$ region. 

            Finally, the AOB$_-$ phase in the region region $\rho < \phi/\sqrt 7$ is related to the phase with $\rho > \phi/\sqrt 7$
            by the self T-duality of AOB$_-$ \cite{Montero:2022vva}. The former AOB$_-$ phase can then be S-dualized to another $S$-dualized to another DP$_-$ phase, and then T-dualized again to another O8$^\pm_-$ phase. Thus, the region in $\phi$-$\rho$ moduli space with $\rho < \phi/\sqrt 7$ features an identical dual description to the region with $\rho > \phi/\sqrt 7$, so stable string flux tubes persist across the entire moduli space.

            \subsection{Large splitting $\Rightarrow$ low tension flux tubes}

From \eqref{e.Tsplit}, \eqref{e.Tflux}, \eqref{e.Tsplit} and \eqref{e.Tflux}, the sublattice splitting $\Tsplit$ and the flux tube tensions $T_\text{flux}$ are proportional to the tensions of DP-phase fundamental strings $T_\text{DP F1}$ and AOB-phase D5-branes $T_\text{AOB D5}$:\footnote{This is in line with our intuition, since the sublattice splitting $T_\text{split}$ equals the extra contributions of the broken charge to the tension of the subextremal object---and the flux tube is a brane with precisely twice the minimal broken charge. Moreover, the DP F1 string is the minimal flux tube in the DP phase.}
			\begin{align}
				\Tsplitstr=\frac 12 \Tfluxstr=\frac 12 T_\text{DP F1},\qquad 
				\Tsplitfive=\frac 12 \Tfluxfive=\frac 12 T_\text{AOB D5}. \label{e.splitfluxcoincidence}
			\end{align}
           In other words, the DP fundamental string tension is related to both the degree of string LWGC failure $\Tsplitstr$ and the tension of the string flux tubes $\Tfluxstr$ associated with particle confinement. Similarly, the  D5-brane tension in the AOB-phase is related to both the degree of fivebrane LWGC failure $\Tsplitfive$ and the tension of the fivebrane flux tubes $\Tfluxfive$.

           From \eqref{mgamma9d}, \eqref{tildedef}, \eqref{DeltaT9d}, and \eqref{e.Tflux}, we observe that the sublattice splitting and the flux tube tensions satisfy
                \begin{align}
				\Tsplitstr \cdot \Tfluxfive = \Tsplitfive \cdot \Tfluxstr = m_\gamma M_9^{7}\,.
                \label{e.productoftensionsrelatedtophoton}
			\end{align}
                 where $m_\gamma$ is the mass of the lightest KK mode \eqref{mgamma9d} of the non-invariant 2-form gauge field $B_2$. This agrees with our general expectation \eqref{geninv}, with $c=1$.

                 Furthermore, large splitting of the string spectrum occurs when $\Tsplitstr \gg M_9^2$. By \eqref{e.Tflux} and \eqref{e.splitfluxcoincidence}, this occurs when $
                \rho \ll \sqrt{7} \phi$.
                In the duality frames of interest, we necessarily have $\rho \geq \phi/\sqrt{7}$. Together, these inequalities imply $\rho, \phi \gg 0$, which by \eqref{e.Tflux} also implies that $\Tfluxfive \ll M_9^6$: large splitting of the string spectrum implies light flux fivebrane tubes.

                Conversely, if the fivebrane splitting is large, $\Tsplitfive \gg M_9^6$, then by \eqref{e.Tflux} and \eqref{e.splitfluxcoincidence} we must have $3\rho \ll - \sqrt{7} \phi$. In the duality frames of interest, we necessarily have $\rho \geq \phi/\sqrt{7}$. Together, these imply $\phi \ll 0$, and furthermore
                \be
                -\frac{1}{\sqrt{14}}\rho+\frac 1{\sqrt 2}\phi =  -\frac{1}{\sqrt{14}}\rho - \frac 1{\sqrt 2}|\phi| \leq \frac{1}{7 \sqrt{2}}|\phi| - \frac 1{\sqrt 2}|\phi| = - \frac{6}{7 \sqrt{2}}|\phi| \ll 0 \,.
                \ee
By \eqref{e.Tflux}, this tells us that $\Tfluxstr \ll M_9^2$: once again, large splitting implies light flux tubes.
                
\begin{figure}[h]
		\begin{center}
		\includegraphics[width = 80mm]{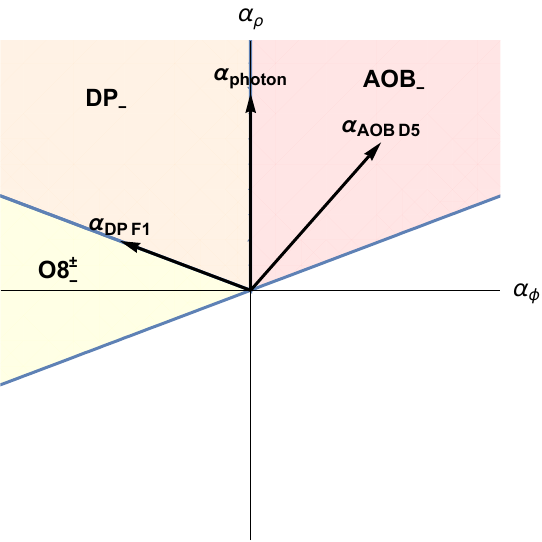}
		\end{center}
		\caption{$\alpha$-vectors for $m_\gamma$, $\Tflux$, and $\Tsplit$. From \eqref{e.splitfluxcoincidence},  $\vec \alpha_{\text{DP F1}} = \vecalphasplitstr = \vecalphafluxstr$ and $\vec \alpha_{\text{AOB D5}} = \vecalphasplitfive = \vecalphafluxfive $. In regions where the DP F1-string becomes high tension (implying large string LWGC failure), the AOB D5-brane becomes low-tension (implying light fivebrane flux tubes), and vice versa.}
				\label{f.9dphases}
			\end{figure}

These calculations can be graphically represented in terms of the $\alpha$ vectors of the branes involved, see Figure \ref{f.9dphases}. The $\alpha$ vector, defined in terms of the tension of a brane $T$, points in the direction of moduli space where that brane becomes lightest \cite{Etheredge:2022opl, Etheredge:2023odp, Etheredge:2023usk, Etheredge:2023zjk, Etheredge:2024amg},
\be
\alpha = - \nabla \log T.
\ee
The $\alpha$ vectors for various objects in the 9d AOB$_-$ and DP$_-$ phases are plotted below. Note that due to \eqref{e.splitfluxcoincidence}, these $\alpha$ vectors contain information about many quantities simultaneously.

Any direction in moduli space which makes an acute angle with a given $\alpha$ vector has the associated object becoming light in those asymptotic directions. Conversely, any direction which makes an obtuse angle sees the object become heavy in that limit of moduli space. In agreement with the above discussion, in Figure \ref{f.9dphases} we see that in any regions where the DP F1 strings become heavy and severely violate the LWGC (the bottom portion of the red AOB$_-$ phase which makes greater than a 90 degree angle with the $\alpha$ vector), the AOB D5 branes comprising the flux tubes become light.\footnote{Note that because the $\alpha$ vectors form an angle strictly less than 180 degrees, there is no direction in moduli space where neither tension flows. In particular, we note that there is a wedge around $\alpha_{\text{photon}}$ where both the splitting becomes small and the flux tube tensions become light, which is only consistent with Dirac quantization because of the enhancement of the gauge group from U$(1)$ to U$(1)^2$.}

The fact that the $\alpha$ vectors satisfy $\alpha_{\mathrm{photon}} = \alpha_{\text{DP F1}} + \alpha_\text{AOB D5}$ is the graphical manifestation of the general expectation of \eqref{geninv}, which is also stated in \eqref{e.productoftensionsrelatedtophoton}.

			\section{8d string theory example}\label{s.8d}
            By reducing the theory from the previous section on a circle, we find an 8d theory LWGC violation for particles, strings, 4-branes, and 5-branes that descend from wrapped/unwrapped strings/fivebranes in 9d. Under T-duality, this theory maps to an 8d compactification of Type IIA string theory, which provides a UV-complete realization of the LWGC violation mechanism described in \S\ref{sec:Zp}. This theory can then be lifted to M-theory, which provides a geometric description of LWGC violation and monopole confinement. In this section, we explore the various descriptions of this theory and its LWGC violation, and we demonstrate that large LWGC violation $\Rightarrow$ light flux tubes.

\subsection{Type IIA description}
            
			We begin by explaining the Type IIA construction of the string theory in question, and we expand on its connection to the discrete gauge theory studied in \S\ref{sec:Zp}.
            
            In ten dimensions, Type IIA string theory possesses a U(1) 1-form gauge symmetry with gauge field $C_1$, under which D0-branes are charged. This symmetry is extended by the action of $(-1)^{F_L}$ to the full 1-form symmetry group $O(2)=U(1) \rtimes \ZZ_2$. After compactifying on a circle $S^1_{F_L}$ with a Wilson line for $(-1)^{F_L}$, $O(2)$ is broken down to a (gapped) $\mathbb{Z}_2$ that counts D0-brane charge mod 2,           leaving us with the 9d AOA background \cite{Aharony:2007du}.\footnote{Note that this background does not violate the LWGC. We will soon modify it to produce an 8d theory that does violate the LWGC.}

            As discussed at length in \S\ref{sec:Zp}, further compactifying this background on a circle $S^1_{\mathbb{Z}_2}$ with a Wilson line $\int C_1 = - \pi$ for the $\ZZ_2$ symmetry yields an 8d theory with LWGC violation and confined monopoles. This 8d theory is related by T-duality to the circle compactification of the AOB$_-$ background studied in the previous section, as the $C_1$ Wilson line dualizes to the nonzero $C_0$ background of AOB$_-$.

        This 8d theory features LWGC violation in both the string and particle spectrum. From the AOB$_-$ perspective, the former descend simply from unwrapped strings in 9d, while the latter descend from wrapped strings. In Type IIA language, LWGC violation for particles results from the twisted boundary conditions for D0-branes along $S^1_{\mathbb{Z}_2}$ imposed by the nontrivial $C_1$ Wilson line. As a result, D0-branes may carry half-integer charges $q \in \mathbb{Z}/2$ under the surviving $U(1)_{\mathrm{KK}}$, whose gauge boson is the graviphoton along $S^1_{\ZZ_2}$.
            Particles with $q \in \frac{1}{2} + \mathbb{Z}$, which come from particles of odd D0-brane charge,
             are subextremal.

            On the magnetic side, the fractionally charged monopoles are KK monopoles, with the 4d Taub-NUT geometry transverse to $S^1_{F_L}$ and with the fiber direction along $S^1_{\mathbb{Z}_2}$. These are confined by flux tubes, which are given by D6-branes wrapping $S^1_{\mathbb{Z}_2}$.

            Meanwhile, subextremal strings come from D2-branes wrapping $S^1_{\mathbb{Z}_2}$. The associated fractionally charged monopoles are given by NS5-branes wrapping both circles, confined by D4-brane flux tubes transverse to both circles. We note here that, while we can deduce this via dualities, the cause of the D2-brane splitting and NS5-brane confinement remains somewhat unclear from the perspective of the Type IIA EFT. Roughly speaking, we expect that the WZ terms induce charge under $C_1$ on the wrapped D2-brane, rendering it subextremal. For the wrapped NS5-branes we observe that there should be two vacua on the brane which are exchanged by the $\mathbb{Z}_2$ gauge symmetry. Additionally, there should be a domain wall separating these two vacua given by the D4-brane ending on the wrapped NS5-brane, hence wrapped NS5-branes are confined via D4-branes. It would be interesting to study these phenomena in more detail.

			\subsection{M-theory lift} \label{subsec:Mlift}
           The M-theory lift of this 8d theory is somewhat subtle due to the presence of the discrete Wilson line and the nonzero $C_1$ background. We will find that the correct description is given by M-theory on a tilted 3-torus, which is then quotiented by a combination of a reflection and a translation.

           The 8d description above can be obtained by beginning with the 9d AOA theory, which is given by M-theory on a Klein bottle in the $(x_9,x_{10})$ directions, obtained by quotienting $\RR^2$ by
           \begin{subequations} \begin{align}
                   \omega_{10}:  x_{10} &\to x_{10}+2\pi R_{10};\\
                   \omega_{9}: x_{9} &\to x_{9}+\pi R_{9}, \quad x_{10} \to -x_{10}.
           \end{align} \end{subequations}
           We then reach the LWGC-violating 8d theory by compactifying another circle given by coordinate $x_8$ and turning on a Wilson line for the $\ZZ_2$ isometry of the Klein bottle:
           \begin{align}
               \omega_8: x_{8} \to x_8 + 2\pi R_{8}, \quad x_{10} \to x_{10}+ \pi R_{10}.
           \end{align}

Consider the double cover of this space, obtained by quotienting $\RR^3$ by $\omega_{10}$, $\omega_9^2$ and $\omega_8$:
\begin{subequations}
    \begin{align}
        \omega_{10}:  x_{10} &\to x_{10}+2\pi R_{10};\\
        \omega_{9}^2: x_{9} &\to x_{9}+2\pi R_{9};\\
        \omega_8: x_{8} &+ 2\pi R_{8}, \quad x_{10} + \pi R_{10}.
    \end{align}
\end{subequations}
which is a tilted torus with flat metric $ds^2 = \sum dx_i ^2$. 

We can equivalently describe the space in terms of three angles $\thetaztwo,\,\thetaFL,\,\theta_M$ which are $2\pi$ periodic $\theta_i \cong \theta_i + 2 \pi$ and are related to the original orthogonal coordinates by\footnote{We will afterwards denote $R_8$ and $R_9$ as $R_\ztwo$ and $R_{F_L}$, respectively.}
\begin{equation}
    x_{10} = R_{10}  \qty( \theta_M - \frac{1}{2} \thetaztwo), \qquad
x_9 = R_9 \thetaFL, \qquad x_8 = R_8 \thetaztwo.
\end{equation}
$\thetaztwo,\,\thetaFL$, and $\theta_M$ will be interpreted respectively as the cycles associated with the nontrivial $\ZZ_2^{(C_1)}$ Wilson line in the 8d theory, the nontrivial $(-1)^{F_L}$ Wilson line in the 8d theory, and the additional emergent direction for which the D0 branes are KK momentum (which we will call the M-theory circle). In terms of these angles, the action of the generators is\footnote{Note that although $\thetaztwo$ is $2\pi$ periodic, $x_8$ has period $4\pi R_8$. This is because shifting $\thetaztwo$ by $2\pi$ also shifts $x_{10}$. We must perform this shift twice to reach the original value of $x_{10}$.}
\begin{subequations}\label{eq.ws}
    \begin{align}
        \omega_{10}&: \theta_M\to \theta_M+2\pi;\label{eq.w10}\\
        \omega_9&: \thetaFL \to \thetaFL+\pi, \quad \theta_M\to-\theta_M+\thetaztwo;\label{eq.w9}\\
        \omega_8&: \thetaztwo \to \thetaztwo + 2\pi. \label{eq.w8}
    \end{align}
\end{subequations}
and the metric is\footnote{This makes manifest that the tilt of the torus is due to the $C_1$ Wilson line in the 8d description, which is in line with intuition. $C_1$ is the graviphoton for the emergent M-theory direction: $d s^2 = R_{\ZZ_2}^2 d \thetaztwo^2 + R_{F_L}^2 d \thetaFL^2 + R_{10}^2 \qty(d\theta_M+ C_1)^2$ and $C_1=-\frac{1}{2} d\theta_{\ZZ_2}$, so the $C_1$ value causes $S^1_M$ to acquire a component along the $S^1_\ztwo$ direction.}
\begin{equation}
    d s ^2 = R_{\ZZ_2}^2 d \thetaztwo^2 + R_{F_L}^2 d \thetaFL^2 + R_{10}^2 \qty(d\theta_M- \frac{1}{2} d \thetaztwo)^2.
\end{equation}

\begin{figure}[H]
                \centering
                \includegraphics[width=\textwidth]{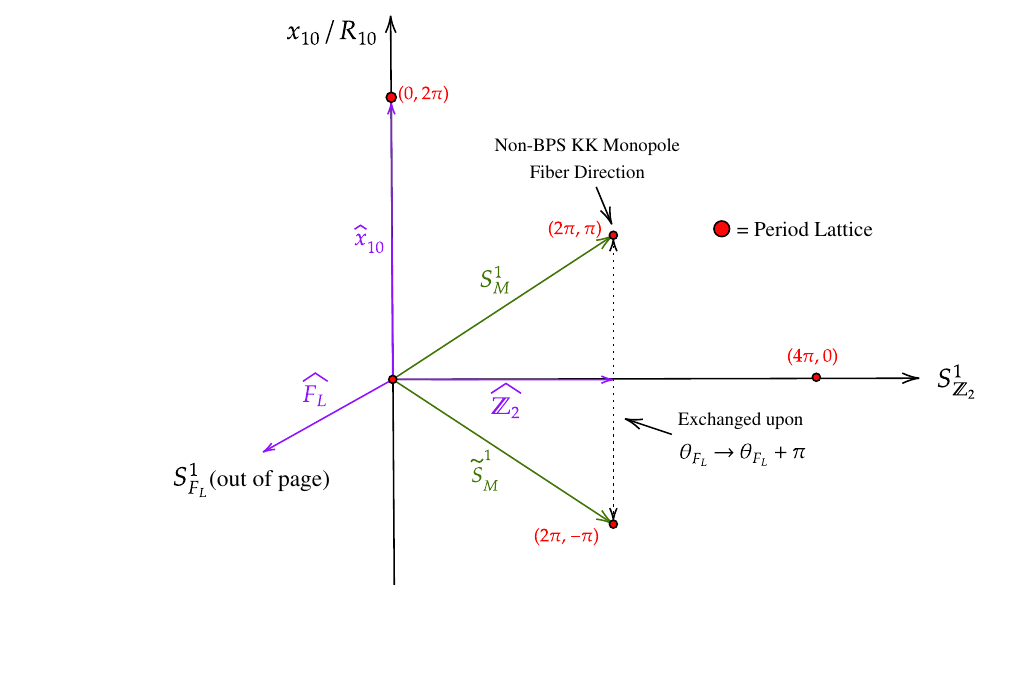}
                \caption{Period lattice for the M-theory torus. $\int C_1=-\pi$ causes non-orthogonality between $S^1_M$ and $S^1_{\mathbb{Z}_2}$. The other effect of the Wilson lines is that $S^1_M$ and $\Tilde{S}^1_M$, pictured in green, are exchanged as $\theta_{F_L} \to \theta_{F_L} + \pi$. The orthogonal basis is shown in purple.}
                \label{periodlattice}
            \end{figure}

\begin{figure}[h]
                \centering
                \includegraphics[width=\textwidth]{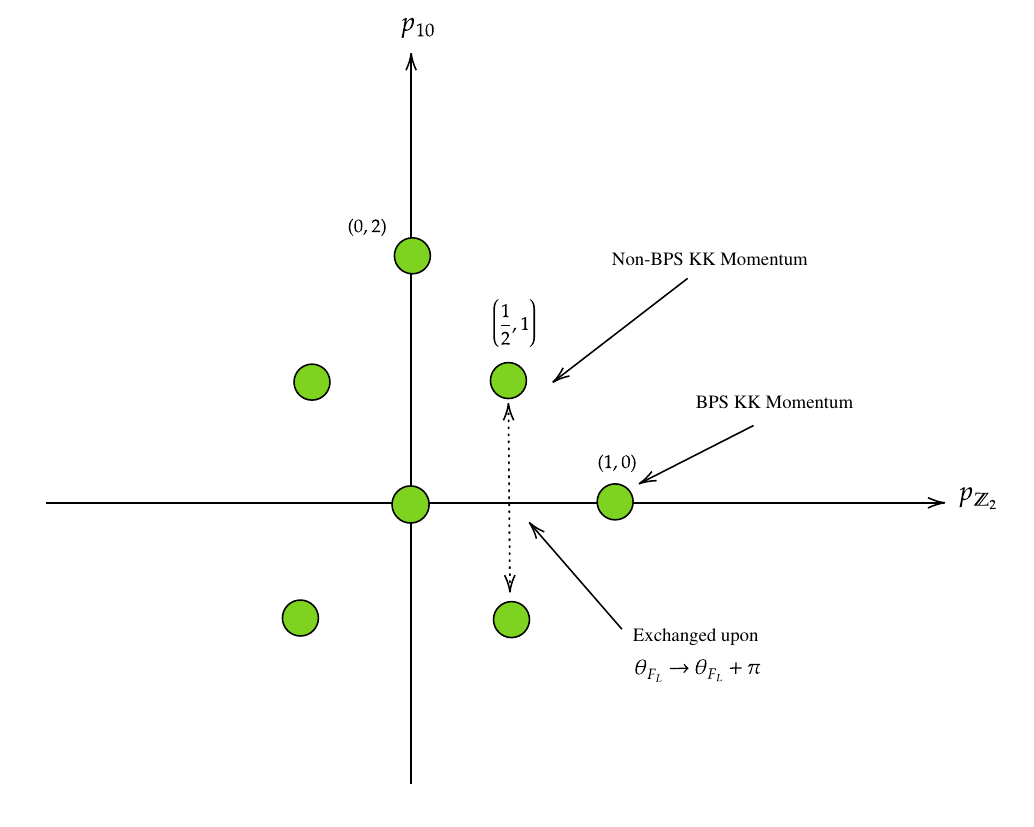}
                \caption{Dual lattice to the period lattice. The surviving graviphotons are along the horizontal axis and out of the plane. As an abuse of language we refer to the $(1/2,1)$-site as KK-momentum along $S^1_{M}$.}
                \label{momentumlattice}
            \end{figure}

           We introduce an orthogonal basis for the period lattice $\{\hat{\mathbb{Z}}_2,\,\hat{F}_L,\,\hat{x}_{10}\}$ and a dual orthogonal basis for the momentum lattice $\{\hat{p}_{\mathbb{Z}_2},\,\hat{p}_{F_L},\,\hat{p}_{10}\}$.  In this basis, the period associated to a single wrapping of $S_M^1$ is $\hat{\mathbb{Z}}_2+\frac{1}{2}\hat{x}_{10}$, the period associated to wrapping $S_{\mathbb{Z}_2}^1$ is $2\hat{\mathbb{Z}}_2$, and the period associated to wrapping S$_{F_L}^1$ is $\hat{F}_L$. We summarize the period lattice and the momentum lattice in Figures \ref{periodlattice} and \ref{momentumlattice}, respectively. 
            
            The two identifications \ref{eq.w10} \& \ref{eq.w9} are, in fact, precisely the ones shown in \cite{Heidenreich:2016aqi} to produce LWGC violation. This can be seen explicitly by setting $w = x_{10}/R_{10}$, $y =  \theta_{\mathbb{Z}_2}$, $z = \theta_{F_L}$.\footnote{It is somewhat unsurprising to find this theory here, as \cite{Montero:2022vva} observed that it can be reached by compactifying the 9d theories discussed in \S\ref{s.9dexample}.} 
        Following \eqref{eq.singleZ2involution}, then, the quotiented 3-torus is equivalently described as a 3-torus quotiented by a different $\ZZ_2$ rototranslation symmetry which exchanges two cycles of the torus upon half-translation along the third cycle. In our case, the cycles corresponding to the two coordinates $a=w-\frac{1}{2}y\equiv \theta_M$ and $b = w+\frac{1}{2}y\equiv\tilde{\theta}_M$ are exchanged upon half-translating along $S^1_{F_L}$. We call the reflected cycle $\tilde{S}^1_M$, as shown in Figure \ref{periodlattice}. The action in this basis is sumarized as
       \begin{equation}
           \theta_M \leftrightarrow \tilde{\theta}_M, \quad \theta_{F_L} \to \theta_{F_L} + \pi.
       \end{equation}

			The 8d theory possesses invariant 1-forms coming from the graviphotons $A_{\mathbb{Z}_2}$, $A_{F_L}$ along $S^1_{\mathbb{Z}_2}$ and $S^1_{F_L}$, respectively, and $B_{\mathbb{Z}_2}$, $B_{F_L}$ from the reductions of the M-theory 3-form $C_3$ along $S^1_M\times S^1_{\mathbb{Z}_2}$ and $S^1_M\times S^1_{F_L}$, respectively. It also features a 2-form $B_2$ coming from $C_3$ reduced along $S^1_M$. The graviphoton along $S_M^1$ is not invariant under the Wilson line and thus is projected out. Consequently, any state with non-zero $p_{10}$ will necessarily be subextremal.
            
           There are subextremal electric states of invariant charge $q_{\mathbb{Z}_2}=1/2$ coming from the KK momenta $(p_{\ZZ_2},p_{10})=(1/2,1)$; these are the D0-branes of the IIA description, lifted into M-theory in the presence of the Wilson lines.\footnote{Recall that without any Wilson lines, a D0-brane lifts to a KK mode around the M-theory circle.} The lowest charge extremal state has KK momentum $(p_{\ZZ_2},p_{10})=(1,0)$.
            
\begin{figure}[h]
                \centering
                \includegraphics[width=.8\textwidth]{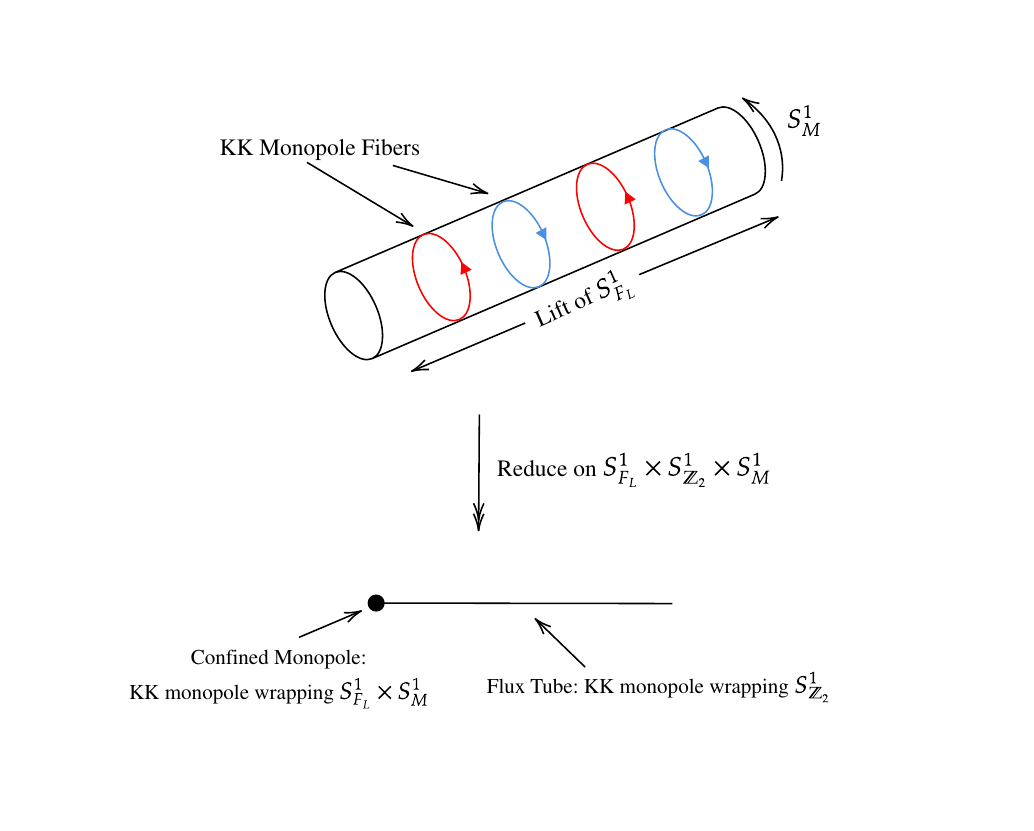}
                \caption{We may lift the configuration of a KK-monopole with fibration associated to $S^1_M$ and NUT transverse to $S^1_{F_L}$ to the universal cover $\mathbb{R}_{F_L}$ to investigate its behavior. Due to the roto-translation, the lift of the KK-monopole is a lattice of alternating orientation monopoles (as depicted by the alternating colors). Thus, the monopole configuration requires a flux tube with fibration associated to $S^1_{\mathbb{Z}_2}$.}
                \label{f.lift}
            \end{figure}

           On the magnetic side, the charge of a KK-monopole is associated with a site in the period lattice, which corresponds to the period of the nontrivial circle fiber in the KK monopole geometry. Since the period of $S_M^1$ is not invariant under translation around $S^1_{F_L}$, a KK-monopole associated with a period of $S_M^1$ cannot be wrapped on  $S^1_{F_L}$ and instead requires a flux tube.\footnote{It is perhaps instructive to think in terms of the universal cover, where we lift $S^1_{F_L}$ to $\mathbb{R}$. Now a KK-monopole associated with the period of $S^1_{M}$ (that is a monopole where $S^1_{M}$ is a non-trivial fibration) lifts to a lattice of KK-monopoles, where every other site is reflected, see Figure \ref{f.lift}. Since the period of $S^1_{M}$ is not invariant, this is not a consistent state of the background.} The charge of this flux tube is determined by the difference in the periods of the KK-monopole and its reflection:
           \begin{equation}
                \left(\hat{\mathbb{Z}}_2+\frac{1}{2}\hat{x}_{10}\right)-\left(\hat{\mathbb{Z}}_2-\frac{1}{2}\hat{x}_{10}\right)=\hat{x}_{10}\,.
           \end{equation}
           Upon reduction to IIA this monopole is precisely the $\text{U}(1)_{\mathbb{Z}_2}$ KK-monopole, and its flux tube is given by a D6-brane wrapping $S^1_{\mathbb{Z}_2}$, as discussed above.

			We can also describe string LWGC violation and the associated confined 3-brane monopoles from the M-theory perspective. Strings descend from M2-branes wrapping 1-cycles of the geometry, which are again associated with the period lattice. An M2-brane wrapping $S^1_{M}$ is subextremal (its tension contains contributions both from the $B_2$ charge associated to wrapping the $
            \hat{x}_{10}$ cycle and the broken charge associated to the reduction of $C_3$ along the $\hat{\mathbb{Z}}_2$ cycle) and carries charge $\frac{1}{2}$ under $B_2$. An M2-brane wrapping the $\hat{x}_{10}$ cycle  has charge 1 under $B_2$ and is extremal. Indeed, this latter string is identified as the BPS fundamental string of Type IIA string theory, while the former string corresponds to a non-BPS D2-brane wrapping $S^1_{\mathbb{Z}_2}$.
            
            The associated monopoles come from M5-branes wrapping 2-cycles. In particular, the confined monopoles come from M5-branes wrapped along the periods associated to $S^1_{M}$ and $S^1_{F_L}$,             and the corresponding flux tubes come from M5-branes wrapping $\hat{x}_{10}$. In type IIA language, these monopoles correspond to NS5-branes wrapping $S^1_{\mathbb{Z}_2}\times S^1_{F_L}$, confined via D4-brane flux tubes.
			
			We may similarly consider the fate of the M2-brane wrapping both $S^1_{M}$ and $S^1_{F_L}$, which yields a confined particle. Its associated monopole is a non-BPS, subextremal M5-brane wrapping $S^1_{M}$, with large LWGC violation when the M-theory circle $S^1_M$ is large.

			Finally, for the sake of completeness, we consider the 1-forms associated to the graviphoton along $S^1_{F_L}$ and to $C_3$ reduced along $S^1_M\times S^1_{F_L}$. The former possesses neither subextremal sites nor confined monopoles. The latter possesses charged states coming from wrapped M2-branes and monopoles coming from M5-branes wrapping $S^1_{\mathbb{Z}_2}$, all of which are extremal.

            \subsection{LWGC violation, flux tube tensions, and photon mass}

            In the AOB$_-$ description of the 8d theory in question, the LWGC violation of particles, strings, fourbranes, and fivebranes descends from LWGC violation of strings and fivebranes in 9d. In \S\ref{s.9dexample}, we saw that the 9d theory satisfies the relation  
        \begin{equation}\label{e.condition}
				\Tsplit T_{\text{flux}} = m_{\gamma}M_d^{d-2} \,.
			\end{equation}
            We will show in this section that pairs of electric and magnetic branes satisfying this condition automatically satisfy it after dimensional reduction, both with and without wrapping. This will imply that the LWGC-violating objects in the 8d phase satisfy the bound as well, since they descend from, or are dual to, branes which satisfy the equality in 9d.\footnote{The following arguments are quite generic and apply equally well to situations when $\Tsplit T_{\text{flux}}\lesssim  c m_\gamma M_d^{d-2}$.}

            In a general circle compactification from $D$ to $d=D-1$ dimensions on a circle of radius $R$, a $p$-brane will yield a $p$-brane and a $(p-1)$-brane in $d$ dimensions with respective tensions
           \begin{subequations}
            \begin{align}
            T_p^{(d)} &= e^{-\frac{p+1}{\sqrt{(d-1)(d-2)}}\rho} T_p^{(D)} \label{preservep}
            \\
            T_{p-1}^{(d)} &= (2 \pi R)e^{\frac{d-p-2}{\sqrt{(d-1)(d-2)}}\rho} T_p^{(D)}\,,
            \label{reducep}
            \end{align}
    \label{e.TensionReduction}
             \end{subequations}
            where $\rho$ is the canonically normalized radion in $d$ dimensions.

            Meanwhile, a massive field of mass $m_\gamma^{(D)}$ in $D$ dimensions will descend to a field of mass
            \begin{align}
                m_{\gamma}^{(d)} = e^{-\frac{1}{\sqrt{(d-1)(d-2)}} \rho } m_{\gamma}^{(D)}\label{gammred}
            \end{align}
            in $d$ dimensions, as can be seen by setting $p=0$ in \eqref{preservep}.

            With this, we may consider the dimensional reduction of LWGC violation for strings and fivebranes. In $D=9$ dimensions, the degree of LWGC violation is measured by the sublattice splitting $\Tsplitp{p}^{(D)}$, where $p=1$ for strings and $p=5$ for fivebranes. The corresponding flux tube tensions---which are responsible for magnetic monopole confinement---are given by $T^{(D)}_{\text{flux}, D-p-3}$. Upon reducing to $d=8$ dimensions, we find LWGC violation in the unwrapped string and fivebrane spectrum, with $\Tsplitp{p}^{(d)}$ related to $\Tsplitp{p}^{(D)}$ via \eqref{preservep}. The associated flux tubes are given by wrapping the flux tubes in 9d around the circle; the associated tensions $T^{(d)}_{\text{flux}, d-p-3}$ are related to $T^{(D)}_{\text{flux}, D-p-3}$ by \eqref{reducep}. Thus we have
            \begin{align}
                \Tsplitp{p}^{(d)} \cdot \Tfluxp{d-p-3}^{(d)} &=  e^{-\frac{p+1}{\sqrt{(d-1)(d-2)}}\rho} T_p^{(D)}  \cdot (2 \pi R) e^{\frac{p}{\sqrt{(d-1)(d-2)}}\rho} T_{D-p-3}^{(D)} \nonumber \\
                &= (2 \pi R) e^{-\frac{1}{\sqrt{(d-1)(d-2)}}\rho}T_p^{(D)} \cdot T_{D-p-3}^{(D)}
            \end{align}
            Combining this result with \eqref{gammred}, setting
            \begin{align}
                T_p^{(D)} \cdot T_{D-p-3}^{(D)} \sim m_\gamma^{(D)} M_{D}^{D-2}
                \label{e.assump}
            \end{align}
            and using $ M_d^{d-2} = (2 \pi R) M_D^{D-2}$, we find
            \begin{align}
                \Tsplitp{p}^{(d)} \cdot \Tfluxp{d-p-3}^{(d)} \sim m_{\gamma}^{(d)} M_d^{d-2}\,,
            \end{align}
            so \eqref{e.condition} is indeed preserved under dimensional reduction, preserving the dimension of the LWGC-violating branes.\footnote{The above argument does not demonstrate strict equality because of the potential presence of additional moduli, such as shape moduli or axions.}

            Similarly, we may consider the case where the LWGC-violating strings and fivebranes are wrapped on the circle, in which case $\Tsplitp{p-1}^{(d)}$ is related to $\Tsplitp{p}^{(D)}$ via \eqref{reducep}. Now, the flux tubes come from unwrapped flux tubes in 9d, and their tensions $T^{(d)}_{\text{flux}, d-p-2}$ are related to $T^{(D)}_{\text{flux}, D-p-3}$ by \eqref{preservep}. Thus we have
            \begin{align}
                \Tsplitp{p-1}^{(d)} \cdot \Tfluxp{d-p-2}^{(d)} &=  e^{\frac{d-p-2}{\sqrt{(d-1)(d-2)}}\rho} T_p^{(D)}  \cdot (2 \pi R) e^{-\frac{d-p-1}{\sqrt{(d-1)(d-2)}}\rho} T_{D-p-3}^{(D)} \nonumber \\
                &= (2 \pi R) e^{-\frac{1}{\sqrt{(d-1)(d-2)}}\rho}T_p^{(D)} \cdot T_{D-p-3}^{(D)}
            \end{align}
By \eqref{gammred} and \eqref{e.assump}, we again find
\begin{align}
                \Tsplitp{p-1}^{(d)} \cdot \Tfluxp{d-p-2}^{(d)} \sim m_{\gamma}^{(d)} M_d^{d-2}\,,
            \end{align}
            so \eqref{e.condition} is also preserved under dimensional reduction for wrapped LWGC-violating branes.

       To conclude this section, we note that the case where string flux tubes are wrapped on the circle yields 1-dimensional flux tubes in 8d, stretching between what we might call ``confined instantons.'' We may interpret these confined instantons as a decay process for the associated flux tube particle. On the dual side we have unwrapped fivebranes, which are codimension-2 in 8d. It is not completely clear if or how the LWGC applies to such branes.

			\section{7d coarseness-3 string theory example}\label{s.7dexample}
			
		We now consider the 7d examples of \cite{Montero:2022vva} that exhibit LWGC violation with coarseness 3. These examples come in essentially two flavors: ones with LWGC violation only for particles, and one with LWGC violation for both particles and strings. In either case, there are confining monopoles associated with the sub-extremal objects. In these examples, we are pinned to strong coupling, but there exist limits where the flux tubes become light and the monopoles deconfine.
			
			These examples are constructed as quotients of IIB on $T^2\times S^1$ with a nontrivial duality bundle. Equivalently, they are quotients of F-theory on $T^4\times S^1$. The coarseness 3 examples come from a $\mathbb{Z}_3$ or $\mathbb{Z}_6$ quotient, which acts by the generator of the cyclic subgroup of order 3 (respectively 6) of SL($2,\mathbb{Z}$) on the two $T^2$'s and rotation by $\frac{1}{3}$ ($\frac{1}{6}$) around the circle. The quotient can be realized explicitly as the equivalence \cite{Montero:2022vva}
			\begin{equation}
				(\vec{x},\theta)\sim(\rho\cdot\vec{x},\theta+\text{ord}(\rho)^{-1})
			\end{equation}
			where $\vec{x}$ is the coordinate on $T^2$, $\theta$ is the coordinate on $S^1$ of periodicity 1, ord$(\rho) \in \{ 3, 6\}$ is the order of the subgroup, and
			\begin{equation}
				\rho=\left\{\begin{array}{cc}\left(\begin{array}{cc} 0&-1\\1&-1\end{array}\right),&~~~~~\text{ord}(\rho)=3, \\
					\left(\begin{array}{cc} 1&-1\\1&0\end{array}\right),&~~~~~\text{ord}(\rho)=6 \,,\end{array} \right.
			\end{equation}
			with the same action on the SL($2,\mathbb{Z}$) duality bundle as on the spacetime $T^2$.
			
			We describe the physics in these backgrounds via a twisted compactification of IIB on $T^2$. We denote the reduction of a $p$-form $\Xi_p$ over the torus with a hat, i.e.,
            \begin{align}
                \hat{\Xi}_p \equiv \int_{T^2} \Xi_p\,,
            \end{align}
            and we similarly use a hat to denote a $p$-brane that has been wrapped over $T^2$ to yield a $(p-2)$-brane in 8d. For example, a $\widehat{\text{D3}}$ is a string in the 8d theory.
            
            The 8d theory then has 2-forms $B_2$, $C_2$, and $\hat{C}_4$ under which F1 strings, D1 strings, and $\widehat{\text{D3}}$ strings are charged. It has 1-forms $\vec{A}=(\int_A C_2,\int_A B_2,\int_B C_2,\int_B B_2)$ under which wrapped $(p,q)$-strings are charged, and it has axions $\vec{\phi}=(\hat{C}_2,\hat{B}_2) \equiv (\phi_C, \phi_B)$. The pair of 2-forms ($C_2,B_2$) and the pair of axions $\vec{\phi}$ each transform as the 2-dimensional representation of the SL($2,\mathbb{Z}$) duality bundle, while $\vec{A}$ transforms under the tensor product of the 2d representation of the SL($2,\mathbb{Z}$) duality bundle and the inverse transpose of the 2d representation of the SL($2,\mathbb{Z}$) of the spacetime torus.\footnote{It was stated in \cite{Montero:2022vva} that $\vec{A}$ transforms under $\rho\otimes\rho$, but under the coordinate transformation above the correct transformation is under $\left(\rho^{-1}\right)^T\otimes\rho$.}
            
            As noted in \cite{Montero:2022vva}, non-zero values for the axions $\vec{\phi}$ induce $(p,q)$-string charge on $\widehat{\text{D3}}$. Hence, for generic $\vec{\phi}$ a $(p,q,n)$-string has charge \begin{align}
				(p-n\phi_C,q+n\phi_B,n) \label{e.7dchargeinduced}
			\end{align}
            under $(B_2,C_2,\hat{C}_4)$. It is therefore helpful to move to a new basis:
			\begin{align}
				\tilde{B}\equiv B_2 \,,~~~~
				\tilde{C}&\equiv C_2 \,,~~~~
				\tilde{C_4}\equiv \hat{C_4}-\phi_C B_2+\phi_B C_2\,,
			\end{align}
			under which a $(p,q,n)$-string has charge $(p,q,n)$.
            
            On the magnetic side, $\widehat{\text{D5}}$, $\widehat{\text{NS5}}$, and D3-branes carry magnetic charge under $C_2$, $B_2$, and $\hat C_4$, respectively.\footnote{Note the magnetic dual of $\hat{C}_4$ is simply $C_4$ due to its self-duality in 10d.} In the presence of nonzero axion vevs, the 3-branes $\widehat{\text{D5}}$ and $\widehat{\text{NS5}}$ carry charge $-\phi_B$ and $\phi_C$ under $C_4$, respectively.            This can be seen by reducing the $B_2\wedge C_4$ terms in the D5-brane action, or carefully tracking the charges under the $\{\tilde{B},\tilde{C},\tilde{C}_4\}$ basis.
			
			We then compactify this 8d theory to 7d, twisting by the action of $\rho$. Since $(\rho-\mathbb{I})$ has trivial kernel, the gauge fields $C_2$ and $B_2$ are projected out of the spectrum, and the complex structure of the torus and the axio-dilaton are pinned to a $k$th-root of unity (where $k$ is the order of the cyclic subgroup). By explicit computation, one can check that, for both $k$=3 and $k$=6, two polarizations of $\vec{A}$ have zero modes that survive, yielding a two-dimensional charge lattice of particles. The axions are also fixed to values that are invariant under the action of $\rho$: for $k$=3, there are three such values: $\vec{\phi} \in \{(0,0), \pm(-\frac1 3,\frac1 3)\}$ are invariant. For $k=6$, we must have $\vec{\phi} = (0,0) $.

			\subsection{LWGC violation for strings}
			Nonzero vevs for the axions $(\phi_C, \phi_B)$ induce a discrete fractional charge $q \in \frac 13 \mathbb{Z}$ on the $\widehat{\text{D3}}$ according to \eqref{e.7dchargeinduced}. The $\tilde{C_4}$ charge of an extremal string is then three times the minimal charge.
            
            On the magnetic side of the story, $(p,q)$-fivebranes wrapping the full internal manifold carry a charge under $\int_{S^1}C_4$ of
            \begin{equation}
                m=-p \phi_B+q \phi_C= \mp \frac{1}{3} (p+q).
            \end{equation}
           Wrappings with $m\notin\mathbb{Z}$ are inconsistent and must be attached to flux tubes. These flux tubes correspond to $(p,q)$-fivebranes wrapping the $T^2$. This follows from the non-invariance of $(p,q)$-fivebranes under the action of $\rho$. For instance, a D5-brane wrapping the circle must be attached to an unwrapped $(\text{D5}+\rho\cdot\overline{\text{D5}})=\text{D5}+\overline{\text{NS5}}$. Any wrapped $(p,q)$-fivebranes with $m\in\mathbb{Z}$ can relax to $m$ D3-branes wrapping the $S^1$.

            It is instructive to note that, in the basis $\{\tilde{C_4},\,\tilde{C_4}-\tilde{B},\,\tilde{C_4}-\tilde{C}\}$, $\rho$ acts via cyclic permutation. This gives a landscape realization of a $U(1)^3\rtimes\mathbb{Z}_3$ model with Wilson line for the $\mathbb{Z}_3$. Note that this enhancement of $U(1)^3$ is special to the pinned value $\tau=e^{i\frac{\pi}{3}}$. In fact, there is a slightly larger gauge symmetry coming from extending $\rho$ by $\left(\begin{array}{cc}
                0 & -1 \\
                -1 & 0
            \end{array}\right)\in GL(2,\mathbb{Z})$, yielding a $U(1)^3\rtimes S_3$ gauge theory with Wilson line for $\mathbb{Z}_3\subset S_3$.\footnote{For this to be a symmetry of the background it must act on both the IIB fields and the torus, the same way as $\rho$.}

            \subsubsection{Splitting and flux tube tensions}
			We now compute the tensions of the $\widehat{\text{D3}}$-strings and $\widehat{5}$-brane monopoles/fluxtubes via reduction from 8d. In 8d our objects are BPS, and thus their tensions are straightforward to compute. The string states come from the F1 strings, D1 strings, and $\widehat{\text{D3}}$ strings. Under the U-duality group SL$(3,\mathbb{Z})\times$SL$(2,\mathbb{Z})$, these strings transform as a $\tilde{\textbf{3}}$ under SL$(3,\mathbb{Z})$ and hence their tensions must be covariant under it.\footnote{The fields transform as a \textbf{3}, so the charged branes transform under the dual representation.} This results in a tension for $\widehat{\text{D3}}$ strings:
			\begin{equation}
				T_{\widehat{\text{D3}}}^2\sim e^{\frac{4}{\sqrt{6}}\rho_{\text{T}^2}}T_{\text{10d D3}}^2+e^{-\frac{2}{\sqrt{6}}\rho_{\text{T}^2}}T_{\text{10d str}}^2\frac{\vert -\phi_c+\tau\phi_b\vert^2}{\text{Im}(\tau)}\,.
			\end{equation}
			Specializing to the case of the non-BPS strings in the 7d theory, this yields
			\begin{equation}
				T_{\widehat{\text{D3}}\text{ 7d}}^2\sim e^{-\frac{4}{\sqrt{30}}\rho_{S^1}}\left(e^{\frac{4}{\sqrt{6}}\rho_{\text{T}^2}}T_{\text{10d D3}}^2+e^{-\frac{2}{\sqrt{6}}\rho_{\text{T}^2}}T_{\text{10d str}}^2\frac{2}{9\sqrt{3}}\right)\,,
			\end{equation}
			so
			\begin{equation}
				\Tsplit\sim e^{-\frac{2}{\sqrt{30}}\rho_{S^1}}e^{-\frac{1}{\sqrt{6}}\rho_{\text{T}^2}}T_{\text{10d str}} \label{7dDelta}\,.
			\end{equation}
			Here, we have suppressed constants, including those that depend on the complex structure of the torus, and utilized \eqref{e.TensionReduction} to compute the dependence on the volume moduli.

			The monopoles and flux tubes come from type IIB $\widehat{\text{D5}}$-branes, $\widehat{\text{NS5}}$-branes, and D3-branes. These transform as a \textbf{3} under SL($3, \mathbb{Z}$), so their tensions are also determined by covariance:
			\begin{equation}
				T^2\sim e^{\frac{2}{\sqrt{6}}\rho_{\text{T}^2}}T_{\text{10d five}}^2+(p\phi_b-q\phi_c)^2e^{-\frac{4}{\sqrt{6}}\rho_{\text{T}^2}}T_{\text{10d D3}}^2\,,
			\end{equation}
			which specializes to 7d as
			\begin{align}
				T^2_{\text{flux}}&\sim e^{-\frac{8}{\sqrt{30}}\rho_{S^1}}e^{\frac{2}{\sqrt{6}}\rho_{\text{T}^2}}T_{\text{10d five}}^2, \label{7dflux} \\
				T^2_{\text{mon}}&\sim e^{\frac{4}{\sqrt{30}}\rho_{S^1}}\left(e^{\frac{2}{\sqrt{6}}\rho_{\text{T}^2}}T_{\text{10d five}}^2+e^{-\frac{4}{\sqrt{6}}\rho_{\text{T}^2}}T_{\text{10d D3}}^2\right).
			\end{align}
			
			Together, these results say that LWGC violation for strings is large when the volumes of the circle and the torus shrink, while the monopoles are heavy with light flux tubes when the circle is large or when the torus shrinks. This means that, for fixed circle volume, when the LWGC violation becomes large, the flux tubes become light.

			In this example, the broken $U(1)$ corresponds to the $B_2$ and $C_2$ fields whose zero modes are projected out by the twisting along the circle. As in previous examples, nonzero KK modes of these fields survive, so the relevant 2-form mass $m_\gamma$ is determined by the KK scale: 
			\begin{equation}
				m_{\gamma}^2\sim e^{-2\sqrt{\frac{6}{5}}\rho_{S^1}}\,.
			\end{equation}
			Together with \eqref{7dDelta} and \eqref{7dflux}, this gives
			\begin{equation}
				\Tsplit T_{\text{flux}}\sim m_\gamma\,,
			\end{equation}
            in accordance with \eqref{geninv}.

			\subsection{LWGC violation for particles}
			In both the order 3 quotient with $\vec{\phi}=0$ and the order 6 quotient, the invariant polarizations of $\vec{A}$ have a coarseness 3 sublattice of BPS particles coming from suitably wrapped $(p,q)$-strings. By direct computation, one can see that the invariant polarizations of $\vec{A}$ are
			\begin{align}
				X_1&\equiv\int_A C_2+\int_B B_2\,, &
				Y_1&\equiv -\int_A C_2-\int_A B_2+\int_B C_2\,.
			\end{align}
            In what follows, we will work in the for the charge lattice where $(1,0)$ corresponds to minimal charge under $X_1$, and $(0,1)$ corresponds to minimal charge under $Y_1$.
            
			Because the charged particles (wrapped strings) are generically not invariant under the duality transformation, they are not all extremal. By direct computation, we find that the charge sites spanned by $(q_X,q_Y)=(2,-1)$ and $(q_X,q_Y)=(1,1)$ possess extremal states. These correspond to the bound states:
			\begin{align}
				p&=\text{D1}_A+\text{F1}_B \,, &
				\tilde{p}&=\text{D1}_A+\overline{\text{F1}}_A+\text{D1}_B\,, \label{ppteq}
			\end{align}
			where the subscript refers to the cycle wrapped by the string.
			
			The magnetic states correspond to suitably wrapped $(p,q)$-fivebranes, namely ones wrapping one cycle of the torus and the $S^1$. The dual field strengths are
			\begin{align}
            X_4&\equiv\int_{B,S^1}C_6-\int_{A,S^1}B_6\,, &
				Y_4&\equiv\int_{A,S^1}C_6-\int_{B,S^1}C_6+\int_{B,S^1}B_6 \,.
			\end{align}
            We work in the analogous basis for the magnetic charge lattice: $(1,0)$ is minimal charge under $X_4$, $(0,1)$ is mimimal charge under $Y_4$. The wrapped NS5-brane $\text{NS5}_{A,S^1}$ is charged under $X_4$, but it is not invariant under the duality transformation, so it must be attached to a $\overline{\text{D5}}_B+\text{NS5}_A+\text{NS5}_B$ flux tube.\footnote{For simplicity we focus on the order 3 quotient. The constituent branes for the order 6 flux tube are the $\overline{\text{D5}}_A+\overline{\text{D5}}_B+\text{NS5}_A$.} A monopole with appropriate charge, such as $\text{D5}_{B,S^1}+\text{NS5}_{A,S^1}$, is invariant under the duality transformation and thus unconfined. We see the unconfined magnetic charges are generated by $(2,-1)$ and $(1,1)$. Note that the Dirac pairing inherited from the 8d theory is not the standard euclidean inner product in this basis (because of the nontrivial projection from the 4d lattice to the surviving 2d lattice), rather it is given by $Q_M^T\left(\begin{array}{cc}1/3&1/3\\-1/3&2/3\end{array}\right)q_e$.\footnote{One can define a new basis for $X_1$, $Y_1$ such that the electric charge lattice is generated by $(3,0)$ and $(0,1)$ and the magnetic charge lattice is generated by $(1/3,0)$ and $(0,1)$, where the Dirac pairing is the standard euclidean inner product.} In accordance with the general argument in \S\ref{ss.generalization} (summarized in Table \ref{glossarydisc}), the unconfined monopole lattice is dual to the complete electric lattice, and the extremal electric lattice is dual to the lattice of confined and unconfined monopoles.

            A precise calculation of the relevant particle masses and brane tensions requires a more careful analysis of the 1/4-BPS spectrum of the theory. For our purpose, however, it suffices to focus on the leading-order moduli dependence of these quantities. We have
            \begin{align}
                \msplit&\sim e^{-\frac{1}{\sqrt{30}}\rho_{S^1}}e^{\frac{1}{\sqrt{6}}\rho_{\text{T}^2}} \,, &
                T_\text{flux}&\sim e^{-\sqrt{\frac{5}{6}}\rho_{S^1}}e^{-\frac{1}{\sqrt{6}}\rho_{\text{T}^2}}\,.
            \end{align}
            Meanwhile, the lightest surviving KK mode of the 1-form gauge field has a mass of
            	\begin{equation}
				m_{\gamma}\sim m_{\rm KK
                } \sim  e^{-\sqrt{\frac{6}{5}}\rho_{S^1}}\,.
\label{mgammas6}
			\end{equation}
            This yields the expected relation
            \begin{equation}
                \msplit T_\text{flux}\sim m_\gamma\,.
            \end{equation}

            Finally, in the order 3 quotient with discrete theta angle, LWGC-violating particles also arise from D3-branes wrapping $T^2 \times S^1$.
            These are confined by F1 string flux tubes due to the ($p,q$)-string charge on the $\widehat{\text{D3}}$ worldsheet, which is not invariant under the $SL(2,\mathbb{Z})$ action along the $S^1$. The corresponding magnetic monopoles are threebranes, namely unwrapped D3-branes as well as ($p,q$)-fivebranes wrapping $T^2$. The D3-branes are BPS, but the ($p,q$)-fivebranes are not BPS. For instance, a wrapped D5-brane has tension
			\begin{equation}
				m^2\sim e^{-\frac{8}{\sqrt{30}}\rho_{S^1}}\left(e^{\frac{2}{\sqrt{6}}\rho_{\text{T}^2}}T_{\text{10d five}}^2+\frac{1}{9}e^{-\frac{4}{\sqrt{6}}\rho_{\text{T}^2}}T_{\text{10d D3}}^2\right)\,,
			\end{equation}
            which gives
            \be
            m_{\rm split}^2 \sim  e^{-\frac{8}{\sqrt{30}}\rho_{S^1}} e^{\frac{2}{\sqrt{6}}\rho_{\text{T}^2}}T_{\text{10d five}}^2\,.
            \ee
			Meanwhile, the flux tube tension is given by 
			\begin{equation}
				T_\text{flux}^2\sim e^{-\frac{4}{\sqrt{30}}\rho_{S^1}}e^{-\frac{2}{\sqrt{6}}\rho_{\text{T}^2}}T_{\text{10d str}}^2\,,
			\end{equation}
			and the photon mass is given by the KK scale as in \eqref{mgammas6}.
            Together, these again give
            \begin{equation}
                \msplit T_\text{flux}\sim m_\gamma.
            \end{equation}

            We now consider the effect of non-zero axion vevs on the particles $p$ and $\tilde p$, as defined above in 
            \eqref{ppteq}. Setting $\vec{\phi}=(-1/3,1/3)$, the particles $p$ and $\tilde p$ naively pick up respective charges $(q_A^{(p)}, q_B^{(p)}) = (\phi_B,-\phi_C)=(1/3,1/3)$ and $(q_A^{(\tilde p)}, q_B^{(\tilde p)}) = (\phi_C,\phi_B-\phi_C)=(-1/3,2/3)$ under the graviphotons $(A_A,A_B)$ associated with the $A$ and $B$ cycles of the torus. However, these graviphotons are projected out of the spectrum, rendering these particles subextremal.

            Thus, only particles with vanishing charge $(q_A, q_B)$ under the broken gauge fields can remain extremal. The extremal states of smallest charge under the surviving gauge fields $X_1$, $Y_1$ are the bound states $z=p+\tilde{p}+\overline{\mathrm{KK}}_B$ and $\tilde{z}=2p-\tilde{p}+\overline{\mathrm{KK}}_A$; these states have respective charges of $(q_X^{(z)},q_{Y}^{(z)} ) =(3,0)$ and $(q_X^{(\tilde z)},q_{Y}^{(\tilde z
            )} ) =(3,3)$.           Thus, extremal particles only exist for charges $(q_X, q_{Y})=(3j,3k)$, $j,k\in\mathbb{Z}$, and the LWGC is violated with coarseness 3. 
            
            Nonzero axion vevs also induce charge under the dual gauge fields $X_4$ and $Y_4$ for the KK-monopoles associated to the graviphotons of the torus. The minimal charge KK-monopoles carry fractional charge and are confined by the same mechanism as the $(p,q)$-fivebranes. The confined monopoles of magnetic charge 1 under $A_A$ and $A_B$, respectively carry 
            charges of
        \begin{align}
            (Q_{X}^{(A)}, Q^{(A)}_{Y} )&= \phi_B \vec Q_{\text{NS5}_{B,S^1}} + \phi_C \vec Q_{\overline{\text{D5}}_{B,S^1}} = \Bigl(\frac13,-\frac23\Bigr) \\
            (Q_{X}^{(B)}, Q^{(B)}_{Y} )&= \phi_B \vec Q_{\overline{\text{NS5}}_{A,S^1}} + \phi_C \vec Q_{{\text{D5}}_{A,S^1}}= \Bigl(-\frac13,-\frac13\Bigr)\,,
            \end{align}
            under the surviving gauge fields.            This means that there exist fractionally charged confined monopoles of charge $Q_X \in \frac13\mathbb{Z}$, $Q_Y \in Q_X + \mathbb{Z}$, and once again the magnetic superlattice generated by the confined monopoles is dual to the superextremal sublattice.

			\section{Higgsing} \label{s.Higgsing}            
			In this section, we explore another method for LWGC violation: Higgsing.
			In \cite{Saraswat:2016eaz}, Saraswat showed that the process of Higgsing can, under certain conditions, lead to a violation of the LWGC (or indeed, of the mild WGC). It is also well known that Higgsing, in general, leads to monopole confinement. In this section, we make the relationship between these more precise, explaining the conditions under which LWGC violation implies fractionally charged confined monopoles (and vice versa) within the context of spontaneous gauge symmetry breaking. We also argue that the relation between LWGC violation, flux tube tension, and photon masses from \eqref{geninv}.
			
			This section is structured as follows: in \S\ref{ss.general}, we analyze a general Higgsing from $U(1)^n$ to $U(1)^{n-1}$. We show that under certain assumptions, LWGC violation is equivalent to the presence of fractionally charged confined monopoles. In \S\ref{ss.pnt}, we focus on the EFT example from \cite{Saraswat:2016eaz} in greater detail, including a quantitative comparison of flux tube tension and LWGC violation. In \S\ref{ss.kkex}, we explore an example of Higgsing and LWGC violation that arises in a simple Kaluza-Klein reduction. Finally, in \S\ref{ss.GMSV}, we look at a UV complete Higgsing example that arises from M-theory compactified on a Calabi-Yau threefold.

			\subsection{General Higgsing}\label{ss.general}
			
			In this subsection, we consider a general model of Higgsing from $U(1)^n$ to $U(1)^{n-1}$. We will see that under certain assumptions, the sLWGC will be satisfied at a sublattice of points which are dual to the lattice associated with confined monopoles. The analysis here is very similar to the analysis of the general dimensional reduction of a gauge theory with a discrete Wilson line, studied above in \S\ref{ss.generalization}.
			
			We begin with an action of the form
			\begin{equation}
				S = \frac{1}{2 \kappa_d^2} \int d^4x \sqrt{-g} R  -  \int \frac{1}{2} \tau_{ij} F^i \wedge \star F_j +   ...\,,
			\end{equation}
			where $\tau_{ij}$, $i,j=1,...,n$ is the gauge kinetic matrix, and $...$ includes the action of charged matter fields, which will play a role in satisfying the WGC and in the Higgsing process. We assume that the gauge fields are normalized to have integral periods, so that the electric charges are specified by a vector of integers $q_i \in \Gamma \simeq \mathbb{Z}^n$. Similarly, the magnetic charges of a monopole are specified by a vector of integers $\tildeq^i \in \tilde \Gamma  \simeq \mathbb{Z}^n$. The inner product between them is thus normalized so that $q_i \tildeq^i \in \mathbb{Z}$ for all $q_i \in \Gamma$, $\tildeq^i \in \tilde \Gamma$.
			
			In what follows, it will usually be helpful to work in an orthonormal basis. We do this by introducing a vielbien
			\begin{equation}
				\tau_{ij} = e_i^a e_j^b \delta_{ab}\,,
			\end{equation}
			which allows us to write
			\begin{equation}
				q_a = e^i_a q_i~~~~,~~~~\tildeq^a = e_i^a \tildeq^i\,.
				\label{ortho}
			\end{equation}
			The total electric charge of a particle is then given by
			\begin{equation}
				|| \bfq ||^2 = q_i \tau^{ij} q_j = q_a \delta^{ab} q_b \equiv  \bfq \cdot \bfq \,,
			\end{equation}
			and similarly for a magnetic monopole we have
			\begin{equation}
				|| \bfQ ||^2 = \tildeq^i \tau_{ij} \tildeq^j = \tildeq^a \delta_{ab} \tildeq^b \equiv \bfQ \cdot \bfQ \,,
			\end{equation}
			
			Next, we make three key assumptions:
			\begin{enumerate}
				\item There exists a scalar field of integral charge $\bfq_H$ which acquires a vev, Higgsing the gauge group from $U(1)^n$ down to $U(1)^{n-1}$.
				\label{A1}
				\item The LWGC is exactly saturated before Higgsing. That is, for all charges $\bfq$, there exists a particle that saturates the WGC bound:
				\begin{equation}
					\frac{|| \bfq ||^2}{m^2}  = \gamma^{-1} \kappa_d^2\,,
					\label{LWGCgn}
				\end{equation}
				where $\gamma$ is taken to be an order-one constant.
				\label{A2}
				\item The process of Higgsing does not affect the masses of the charged particles, nor does it affect the order-one constant $\gamma$.
				\label{A3}
			\end{enumerate}
			
			After Higgsing, the resulting theory will have a charge lattice given by $\Pi(\Gamma)$, where $\Pi(\bfq)$ is the projection of $\bfq$ onto the orthogonal subspace of $\bfq_H$, $\Sigma_\perp \subset \mathbb{R}^n$:
			\begin{equation}
				\Pi(\bfq) \equiv \bfq - \frac{\bfq \cdot \bfq_H}{||\bfq_H||^2}  \bfq_H  \in \Sigma_\perp \,.
				\label{projel}
			\end{equation}
			Given assumptions \ref{A2} and \ref{A3}, the particles that satisfy the sLWGC after Higgsing are precisely those whose charges $\bfq$ lie in the subspace $\Sigma_\perp$. This follows from the fact that the WGC bound after Higgsing takes the form
			\begin{equation}
				\frac{|| \Pi(\bfq) ||^2}{m^2}  \geq \gamma^{-1} \kappa_d^2 = \frac{|| \bfq ||^2}{m^2}  \,,
			\end{equation}
			where on the right-hand side we have used the assumed value of the mass in \eqref{LWGCgn}. This inequality is violated unless $\Pi(\bfq) = \bfq$, which is precisely the condition that $\bfq \in \Sigma_\perp$. Thus we see that the sublattice of particles which satisfy the sLWGC after Higgsing $\Gamma_{\text{ext}}$ is given by
			\begin{equation}
				\Gamma_{\text{ext}} \supseteq \Gamma \cap \Sigma_\perp\,.
				\label{sLcontain}
			\end{equation}
            
			Meanwhile, the magnetic charge of a monopole of charge $\bfQ$ is given, after Higgsing, by $\Pi(\bfQ)$, where $\Pi(\bfQ)$ is again the projection of $\bfQ$ onto the orthogonal subspace of $\bfq_H$. More precisely, we have
			\begin{equation}
				\Pi(\bfQ)^a = \tildeq^a - \frac{\tildeq^b (q_H)_b}{||\bfq_H||^2} q_H^a\,,
				\label{projmag}
			\end{equation}
			where $q_H^a = \delta^{ab} (q_H)_b$. This can be justified using the argument of \S\ref{ss.generalization} above.
            
			Monopoles which map to themselves under the projection, $\bfQ = \Pi(\bfQ)$, remain unconfined, as evidenced by the fact that their Dirac pairing with $\bfq_H$ vanishes, $\tildeq^i (q_H)_i = 0$. Conversely, monopoles with $\bfQ \neq \Pi(\bfQ)$ are confined.
			
			We want to prove the following:
			\begin{equation}
				\Gamma \cap \Sigma_\perp =  \Pi(\tilde \Gamma)^\vee \,,
			\end{equation}
			where as in \S\ref{ss.generalization}, $[\cdots]^\vee$ represents duality of these two $(n-1)$-lattices as subsets of $\Sigma_\perp \simeq \mathbb{R}^{n-1}$ and $[\cdots]^*$ represents duality within the ambient space of the un-Higgsed theory (in particular, $\Gamma = \tilde \Gamma^*$).
			By \eqref{sLcontain}, this establishes that the lattice dual to the lattice of confined monopoles is entirely populated by superextremal particles. Indeed, if the containment in \eqref{sLcontain} is an equality, then these two lattices will be equal to one another, $\Gamma_{\text{ext}} = \Pi(\tilde \Gamma)^\vee$.
			
			We begin by showing that $\Gamma \cap \Sigma_\perp \subseteq  \Pi(\tilde \Gamma)^\vee$. Given $\bfq \in \Gamma \cap \Sigma_\perp$ and $\bfQ \in  \Pi(\tilde \Gamma)$, we know by definition of the projection $\Pi$ that there exists $\bfQ' \in \tilde \Gamma$ and $\alpha \in \mathbb{R}$ such that $\tildeq^a = (\tildeq')^a + \alpha q_H^a$. By Dirac quantization, we have $q_a (\tildeq')^a \in \mathbb{Z}$, and since $\bfq_H \in \Sigma_\perp$, we have $ q_a q_H^a = 0$. Therefore, we have $q_a \tildeq^a = q^a (\tildeq')^a \in \mathbb{Z}$, and therefore $\bfQ \in (\Gamma \cap \Sigma_\perp)^\vee$. This establishes the first direction of the claim.
			
			\begin{table}
				\begin{center}
					\renewcommand{\arraystretch}{1.5}
					\begin{tabular}{|c|c|c|}\hline
						{\bf Symbol} & { \bf Description} & {\bf Relationships} \\\hline
						$\Gamma$ & Un-Higgsed electric charge lattice & $= \tilde \Gamma^*$ \\\hline
						$\tilde \Gamma$ & Un-Higgsed magnetic charge lattice & $= \Gamma^*$ \\\hline
						$\Pi(\Gamma)$ & Electric charge lattice & $= (\tilde \Gamma \cap \Sigma_\perp)^\vee$ \\\hline
						$\tilde \Gamma \cap \Sigma_\perp$ & Magnetic charge lattice & $= \Pi(\Gamma)^\vee$ \\\hline
						$\Gamma_{\text{ext}}$ & Superextremal sublattice & $\supseteq \Gamma \cap \Sigma_\perp$, $\subseteq{\Pi(\Gamma)}$ \\\hline
						$\Gamma \cap \Sigma_\perp$ & Sublattice with charge unaffected by Higgsing & $\subseteq \Gamma_{\text{ext}}$, $= \Pi(\tilde\Gamma)^\vee$ \\\hline
						$\Pi(\tilde \Gamma)$ & Lattice of (possibly) confined monopoles& $\subseteq \tilde\Gamma \cap \Sigma_\perp$, $=^\dagger (\Gamma \cap \Sigma_\perp)^\vee$ \\\hline
					\end{tabular}
					\caption{Glossary of lattices. Any lattices not labeled ``un-Higgsed'' should be understood as lattices in the theory after Higgsing. $^\dagger$Note that if $\Gamma \cap \Sigma_\perp$ has dimension less than $\Sigma_\perp$, then $(\Gamma \cap \Sigma_\perp)^\vee$ is equal to the \emph{closure} of $\Pi(\tilde \Gamma)$. This possibility would only arise if the sLWGC is violated in the Higgsed theory, so we do not expect it to occur in the landscape.}
				\label{glossary}
			\end{center}
			\end{table}
			
			Next, we show the converse, $\Gamma \cap \Sigma_\perp \supseteq  \Pi(\tilde \Gamma)^\vee$. Given $\bfq \in \Pi(\tilde \Gamma)^\vee$, we have by definition that $\bfq$ is an element of $\Sigma_\perp$ such that $q_a \tildeq^a \in \mathbb{Z}$ for all $\bfQ \in \Pi(\tilde \Gamma)$. Then, since $\bfq \cdot \bfq_H = 0$ and every element $\tildeq'$ of $\tilde \Gamma$ can be written as $ (\tildeq')^a = \tildeq^a + \alpha q_H^a$ for some $\bfQ \in \Pi(\tilde \Gamma)$ and some real $\alpha$, we have that $q_a (\tildeq')^a \in \mathbb{Z}$ for all $(\tildeq')^a \in \tilde\Gamma$. This implies that $\bfq \in {\tilde \Gamma}^* = \Gamma$, so $\bfq \in \Gamma \cap \Sigma_\perp$, establishing the result.
			
			Thus, we have shown that $\Gamma \cap \Sigma_\perp =  \Pi(\tilde \Gamma)^\vee$. The last step is to take the dual of this equation, which naively gives $(\Gamma \cap \Sigma_\perp)^\vee =  \Pi(\tilde \Gamma)$. Indeed, this equation is true if $\Gamma \cap \Sigma_\perp$ has dimension $n-1$, which is precisely the statement that the sLWGC is satisfied along this sublattice. If the sLWGC is not satisfied, however, then the sublattice $\Gamma \cap \Sigma_\perp$ has dimension less than $n-1$. In this case, the dual of this sublattice in $\Sigma_\perp$ consists not only of isolated points, but rather of manifolds of dimension $n-1- \text{dim}(\Gamma \cap \Sigma_\perp)$. Dualizing $\Pi(\tilde \Gamma)^\vee$ yields not $\Pi(\tilde \Gamma)$, but rather the closure $\overline{\Pi(\tilde \Gamma)}$, so the correct statement is $(\Gamma \cap \Sigma_\perp)^\vee =  \overline{\Pi(\tilde \Gamma)}$. The condition that the closure of $\overline{\Pi(\tilde \Gamma)}$ has nonzero dimension is equivalent to the statement that $\Pi(\tilde \Gamma)$ forms a dense subset of a submanifold of $\Sigma_\perp$. Thus we learn that sLWGC violation implies a dense set of confined monopole charges.

            The list of all of these lattices---and the relationships we have derived between them---can be found in Table \ref{glossary}, which may be compared with the analogous results of Table \ref{glossarydisc} above.

			\subsubsection{sLWGC constraints on the charge lattice}\label{sss.constraint}
			
			In the landscape, we expect that sLWGC violation will not occur. This imposes constraints on the geometry of the lattice, as it requires that $\Gamma \cap \Sigma_\perp$ is nonempty. This translates into a simple constraint on the geometry of the charge lattice, as we now explain.
			
			For the purpose of pedagogy, let us first assume that the parent theory has $n=2$ $U(1)$s. Let $\bfq_H$ denote the charge of the particle that acquires a VEV, and let us assume that $\bfq_H$ is primitive. (If the charge of the Higgsed particle is not primitive, then we may simply define $\bfq_H$ to be the charge of the primitive vector in the same direction as this charge.) Let $\bfq \in \Gamma$ be such that $\{ \bfq, \bfq_H \}$ generate the lattice $\Gamma$.
			
			Then, suppose that the ratio of $\bfq \cdot \bfq_H  / ||\bfq_H||^2$ is rational, so that
			\begin{equation}
			\frac{ \bfq \cdot \bfq_H }{||\bfq_H||^2} = \frac{p}{k}\,,
			\label{pq}
			\end{equation}
			with $p$, $k$ relatively prime integers, and $k >0$. We claim that $k$ is precisely the coarseness of the sublattice $\Gamma \cap \Sigma_\perp$ inside the electric charge lattice of the Higgsed theory, $\Pi(\Gamma)$.
			
			To this, consider the vector $\bfq' = p \bfq_H - k \bfq$. This satisfies
			\begin{equation}
			\bfq' \cdot \bfq_H = (p \bfq_H - k \bfq) \cdot \bfq_H = 0\,,
			\end{equation}
			by \eqref{pq}. This means that $\bfq'$ is an element of $\Gamma \cap \Sigma_\perp$.
			
			Conversely, suppose $\bfq'' = a \bfq_H - b \bfq$ is a nonzero element of $\Gamma \cap \Sigma_\perp$. This implies $\bfq'' \cdot \bfq_H = 0$, which means that $a/b = p / k$. Since $p$, $k$ have been assumed to be relatively prime, this implies $|b| > k$, and in fact $b$ is a multiple of $k$. This implies that $\bfq''$ is an integer multiple of $\bfq'$, so $\bfq'$ is a generator of $\Gamma \cap \Sigma_\perp$.
			
			Meanwhile, $\Pi(\bfq)$ is a generator of $\Pi(\Gamma)$, since $\bfq$ is a generator of $\Gamma$. Since $\bfq' = k \Pi(\bfq)$ we find that indeed, $k$ is precisely the coarseness of the sublattice $\Gamma \cap \Sigma_\perp$. Note that this result is independent of the choice of generator $\bfq$.
			
			More generally, for an $n$-dimensional lattice with $\bfq_H$ the primitive charge of the Higgsed field, we may define a basis $\{ \bfq_\alpha, \bfq_H \}$ of $\Gamma$, with $\alpha = 1, ..., n-1$. Defining integers $p_\alpha$, $k_\alpha$ by
			\begin{equation}
			\frac{\bfq_H \cdot \bfq_\alpha }{||\bfq_H||^2} = \frac{p_\alpha}{k_\alpha} \,,
			\end{equation}
			we can show by a similar argument that the coarseness of $\Gamma \cap \Sigma_\perp$ is given by $k = \text{lcm}(k_\alpha)$.
			
			If $\bfq_H \cdot \bfq_\alpha / ||\bfq_H||^2$ is irrational for any $\alpha$, then $\Gamma \cap \Sigma_\perp$ is not a finite index sublattice of $\Pi(\Gamma)$, and the sLWGC is likely violated. The rationality of $\bfq_H \cdot \bfq_\alpha / ||\bfq_H||^2$ puts a strong geometric constraint on the lattice $\Gamma$ along the Higgs branch. In the examples of \S\ref{ss.kkex} and \S\ref{ss.GMSV}, we will see that this constraint is obeyed.

			\subsection{EFT example}\label{ss.pnt}
			
			In this section, we review and expand upon the analysis of \cite{Saraswat:2016eaz}, showing how it fits into the general analysis of the preceding subsection. To this end, consider a 4d theory with two gauge fields $A$ and $B$, with action
			\begin{equation}
			S = \frac{1}{2 \kappa_4^2} \int d^4x \sqrt{-g} R  - \int \frac{1}{2 g_A^2} F \wedge \star F  - \int \frac{1}{2 g_B^2} H \wedge \star H\,,
			\end{equation}
			where $F = \rmd A$, $H= \rmd B$. For simplicity, let us assume to begin that $g_A = g_B \equiv g$: we will later relax this assumption.
			
			The electric charges of the theory are labeled by a two-vector $q_i = (q_A, q_B) \in \mathbb{Z}^2$. Given the gauge kinetic matrix $\tau_{ij} = \text{diag}(g^{-2}, g^{-2})$, the electric charge vector $q_a$ defined in \eqref{ortho} is given simply by $q_a  = g \delta^i_a q_i  = g(q_A, q_B)$. Similarly, on the magnetic side we have $\tildeq^i  = (\tildeq^A, \tildeq^B)\in \mathbb{Z}^2$, with $\tildeq^a = \frac1g  \delta_i^a \tildeq^i = \frac{1}{g} (\tildeq^A, \tildeq^B) $.

			In this context, the assumptions \ref{A1}-\ref{A3} translate to the following:
			\begin{enumerate}
			\item There exists a scalar field of integral charge $(q_H)_i = (q_A, q_B) = (Z, 1)$ which acquires a vev, Higgsing the gauge group from $U(1) \times U(1)$ down to $U(1)$.
			\item The LWGC is exactly saturated before Higgsing. That is, for all charges $(q_A, q_B)$, there exists a particle that saturates the WGC bound:
			\begin{equation}
				\frac{g^2 (q_A^2+q_B^2)}{m^2}  = \gamma^{-1} \kappa_4^2\,,
				\label{LWGCpnt}
			\end{equation}
			where $\gamma$ is an order-one constant.
			\item The process of Higgsing does not affect the masses of the charged particles, nor does it affect the order-one constant $\gamma$.
			\end{enumerate}

			As shown in \cite{Saraswat:2016eaz}, these assumptions imply a violation of the LWGC after Higgsing, for $Z \neq 0$. One way to see this is by changing the basis of gauge fields to a Higgsed combination and a surviving combination:
			\begin{align}
			A^H_\mu =  \frac{1}{Z^2+1} ( Z A_\mu + B_\mu ) \,,~~~ C_\mu =  \frac{1}{Z^2+1} (  A_\mu - Z B_\mu )\,.
			\end{align}
			Here, the surviving gauge field has gauge coupling
			\begin{align}
			g_C^2 = g^2 / (Z^2+1) \,.
			\end{align}
			A particle of charge $q_i  = (q_A, q_B)$ before Higgsing acquires a charge $q_C = q_A - Z q_B$ after Higgsing. This charge can be understood as the projection \eqref{projel} of $\bfq$ onto the subspace $\Sigma_\perp \simeq \mathbb{R}$, which is generated by the vector $q_a =  (1, -Z)$.
			
			By assumption, such a particle saturates the WGC bound \eqref{LWGCpnt} before Higgsing. After Higgsing, the WGC bound is given by
			\begin{equation}
			\frac{g_C^2 q_C^2}{m^2} \geq \gamma^{-1} \kappa_4^2 \,,
			\label{LWGCC}
			\end{equation}
			sing \eqref{LWGCpnt} and plugging in $q_C = q_A - Z q_B$, the WGC bound becomes
			\begin{equation}
			\frac{1}{Z^2+1} (q_A - Z q_B)^2 \geq q_A^2 + q_B^2\,, \qquad \text{or equivalently} \qquad 0 \geq (Z q_A + q_B)^2 \,.
			\end{equation}
			This is satisfied only when $q_B = - Z q_A$, i.e., if the particle carries vanishing charge under the Higgsed gauge field, so $q_a \in \Sigma_\perp$. Such a particle has $q_C = q_A (1+Z^2)$, and since $q_A \in \mathbb{Z}$, we find that $q_C$ is an integer multiple of $Z^2+1$. Thus, the LWGC is violated with coarseness $1+Z^2$.
			
			Meanwhile, on the magnetic side of the story, we have that a monopole of magnetic charge $\tildeq^i = (\tildeq^A, \tildeq^B)$ carries magnetic charge $\tildeq^C = \frac{1}{1+Z^2} (\tildeq^A - Z \tildeq^B)$ under the surviving gauge field $C_\mu$. This charge can be understood as the projection \eqref{projmag} onto the subspace orthogonal to $(q_H)_a$. For $\tildeq^B = - Z \tildeq^A$, we have that $\tildeq^C$ is an integer: these are precisely the unconfined monopoles, which carry zero magnetic charge under the Higgsed gauge field $A^H_\mu$. For other values of $\tildeq^B \neq - Z \tildeq^A$, the monopole in question is confined, and its charge $\tildeq^C$ is an integer multiple of $1/(1+Z^2)$. This precisely matches the coarseness of the sLWGC-satisfying sublattice $\Gamma \cap \Sigma_\perp$, and indeed this sublattice is dual to the superlattice of confined monopoles, $\Pi(\tilde\Gamma)^\vee$, as expected from the general analysis.
			
			Let us next consider the relationship between flux tube tension, LWGC violation, and the Higgsed photon mass in this example. From \eqref{LWGCpnt} and \eqref{LWGCC}, we have that the difference between the mass-squared of a charged particle and the mass-squared of an extremal particle after Higgsing is given by
			\begin{equation}
			\msplit^2 = \gamma M_{\rm Pl}^2 \left(g^2   (q_A^2+q_B^2) - g_C^2 q_C^2 \right) = \frac{g^2}{1+Z^2} \gamma M_{\rm Pl}^2 (Z q_A + q_B)^2\,,
			\end{equation}
			where we have used $g_C^2 = g^2/(1+Z^2)$ and $q_C = q_A - Z q_B$. Meanwhile, for a general abelian Higgs model, we expect the flux tube tension will scale as
			\begin{equation}
			T \sim v^2\,,
			\end{equation}
			where $v$ is the vev of the Higgs scalar field, and the photon mass is given by
			\begin{equation}
			m_{\gamma}^2 = || \bfq_H ||^2 v^2 = (1+Z^2) g^2 v^2\,.
			\end{equation}
			Thus we have
			\begin{equation}
			\frac{\msplit T}{m_\gamma M_{\rm Pl}^2} \sim \frac{\sqrt{\gamma}}{1+Z^2} \frac{ v}{M_{\rm Pl}}\,.
			\end{equation}
			In accordance with \eqref{geninv}, this is much less than 1 provided $v \ll M_{\rm Pl}$, as required for control of the EFT. We see that the inverse relationship between flux tube tension and LWGC violation occurs automatically in this example, without the need to impose it as an additional UV constraint.
			
			Before concluding our analysis of this EFT example, let us consider what happens when we drop the assumption that $g_A = g_B$. In this case, the surviving gauge field is given by
			\begin{equation}
			C_\mu = \frac{1}{g_B^2 + Z^2 g_A^2} (g_B^2 A_\mu - Z g_B^2 B_\mu)\,.
			\end{equation}
			Its gauge coupling is given by
			\begin{align}
			\frac{1}{g_C^2} = \frac{1}{g_A^2} + \frac{Z^2}{g_B^2}\,.
			\end{align}
			A particle of charge $(q_A, q_B)$ before Higgsing still acquires a charge $q_C = q_A - Z q_B$ after Higgsing. The WGC bound after Higgsing then takes the form
			\begin{equation}
			0 \geq \msplit^2 = M_{\rm Pl}^2 \gamma \left (g_A^2 q_A^2 + g_B^2 q_B^2 - g_C^2 q_C^2 \right)= \frac{M_{\rm Pl}^2 \gamma}{g_B^2 + Z^2 g_A^2} (Z g_A^2 q_A + g_B^2 q_B)^2\,.
			\end{equation}
			This is satisfied only if $Z g_A^2 q_A = - g_B^2 q_B$, which is impossible if $g_A^2/g_B^2$ is irrational! Thus, as discussed in \cite{Saraswat:2016eaz}, an EFT that satisfies the LWGC can, under Higgsing, lead to a theory that violates not only the LWGC but even the sLWGC and the mild version of the WGC. However, there is no evidence that this possibility ever occurs in the landscape of quantum gravity. Indeed, in what follows we will see examples of Higgsing in which ultraviolet effects conspire to forbid the possibility of sLWGC violation.
			
			On the magnetic side of things, we find (possibly confined) monopoles with charges
			\begin{align}
			\tildeq^C = \frac{1}{Z^2 g_A^2 + g_B^2} (g_B^2 \tildeq^A - Z g_A^2 \tildeq^B)\,.
			\end{align}
			If $\tildeq^B = - Z \tildeq^A$, its Dirac pairing with the Higgs charge $q^i_H = (Z, 1)$ vanishes, and $\tildeq^C$ is an integer: these are the unconfined monopoles. If $\tildeq^B \neq - Z \tildeq^A$, however, the monopole is confined. Furthermore, if $g_B^2 / g_A^2$ is irrational, then the set of possible confined charges $\tildeq^C$ is dense in the real line. As anticipated from the general Higgsing analysis above, sLWGC violation implies a dense collection of confined monopole charges.

			\subsection{KK example}\label{ss.kkex}
			
			So far in this section, we have focused our attention on EFT examples, which are not necessarily realizable in the landscape of quantum gravity.			In this subsection, we consider an example of Higgsing that arises naturally in Kaluza-Klein theory, finding that LWGC violation implies the existence of fractionally charged confined monopoles, but not vice versa. This example is very similar to the BF reduction example given above in \S\ref{sec:Zp}.
			
			We begin in $D$ dimensions with a U(1) gauge field coupled to a massless scalar field of charge $p$. Upon reduction to $d=D-1$ dimensions on a circle of radius $R$, this scalar field gives rise to a KK tower of particles of mass
			\begin{equation}
			m_{H,d}^2 = \frac{1}{R^2 } \left( n - \frac{p \theta}{2 \pi} \right)^2\,,
			\end{equation}
			where $n$ is the KK mode number and $\theta \in [ 0 , 2 \pi) $ is the axion that comes from integrating the gauge field around the circle. If $\theta = 2 \pi n / p$, the $n$th Kaluza-Klein mode is massless, and a Higgs branch opens up. In particular, Higgsing is only possible when $\theta$ is an integer multiple of $2 \pi /p$.
			
			This in turn places restrictions on the shape of the charge lattice when Higgsing occurs. The lower dimensional theory has two U(1) gauge fields, one which descends from the higher-dimensional U(1) gauge field, the other of which is the KK photon. The gauge kinetic matrix in $d$ dimensions takes the form (see e.g.~\cite{Heidenreich:2019zkl}):
			\begin{align}
			\tau_{ij} = \left( 
			\begin{array}{cc}
				\frac{1}{e_d^2}  & \frac{1}{e_d^2}  \left( \frac{\theta}{2\pi} \right) \\
				\frac{1}{e_d^2}  \left( \frac{\theta}{2\pi} \right) & \frac{1}{e_{\rm KK}^2}  + \frac{1}{e_d^2}  \left( \frac{\theta}{2\pi} \right)^2
			\end{array}
			\right)\,,
			\label{taured}
			\end{align}
			where $e_d^2 = (2 \pi R) e_D^2$ is the $d$-dimensional gauge coupling for the descendent U(1) and $e_{\rm KK}^2 = 2 \kappa_d^2  / R^2$ is the Kaluza-Klein gauge coupling. 
			The inverse of the gauge kinetic matrix is given by
			\begin{align}
			\tau^{ij} = \left( 
			\begin{array}{cc}
				e_d^2  + e_{\rm KK}^2  \left( \frac{\theta}{2\pi} \right)^2 &  - e_{\rm KK}^2  \left( \frac{\theta}{2\pi} \right) \\
				- e_{\rm KK}^2  \left( \frac{\theta}{2\pi} \right) &  e_{\rm KK}^2 
			\end{array}
			\right)\,.
			\label{invtau}
			\end{align}
			The Higgs field has charge $q_i^{H} = (p, n)$. After Higgsing, the surviving charge of a particle is given by its projection along the orthogonal subspace $\Sigma_\perp$ to $q^H$.  This subspace is generated by the vector $q^\perp_i = (0, 1)$, since
			\begin{equation}
			q^\perp_i \tau^{ij} q_i^{H} = 0\,,
			\end{equation}
			for $\theta = 2 \pi n/p$.
			Thus, the surviving charge $q$ of a particle of initial charge $q_i$ is given by
			\begin{equation}
			q  = \frac{p}{e_{\rm KK}^2} q_i \tau^{ij} q_j^\perp =   p q_2 -  n q_1 \in \mathbb{Z}  \,.
			\label{chargeafterH}
			\end{equation}
			
			KK modes of the graviton have $q_1 = 0$ and therefore $q \in p \mathbb{Z}$. These KK modes are extremal, with masses given by $m_k^2 = (k/R)^2$, where $k = q_2$. However, a massive particle in 5d of mass $m_D$ and charge $q_1$ gives rise to a tower of KK modes of mass
			\begin{equation}
			{m_d}^2 = {m_D}^2 + \frac{1}{R^2 } \left( n' - \frac{q_1 \theta}{2 \pi} \right)^2 = {m_D}^2 + \frac{1}{R^2 } \left( n' - \frac{q_1 n }{ p } \right)^2  \,, ~~~n' \in \mathbb{Z}\,,
			\end{equation}
			and charge given by \eqref{chargeafterH}:
			\begin{align}
			q  =   p n' - q_1 n \,.
			\end{align}
			For $m_D > 0$, these KK modes are subextremal after Higgsing. In the absence of a massless particle of charge $p'=1$, therefore, we expect to find LWGC violation with coarseness $p$.
			
			It is interesting to compare this with the general analysis above, which argued that the coarseness of the sLWGC lattice is determined by the denominator of the fraction $\bfq \cdot \bfq_H/||\bfq_H||^2$, when written in lowest form (see \eqref{pq} and the surrounding discussion). Here, we have $q_{H,i} = (p,n)$, and setting $q_i=(q_1, q_2)$, we find
			\begin{align}
			\frac{\bfq \cdot \bfq_H}{||\bfq_H||^2} = \frac{q_1}{p}\,,
			\label{KKrat}
			\end{align}
			whose denominator $p$ indeed matches the coarseness of the sublattice. For this to work, it is crucial that the value of $\theta$ is fixed uniquely on the Higgs branch in terms of the integers $p$, $n$: for general $\theta \in [0, 2 \pi)$, there is no such quantization condition on the left-hand side of \eqref{KKrat}, and the sLWGC may be violated. We see, therefore, that Kaluza-Klein theory places nontrivial constraints on the geometry of the charge lattice, ensuring that the sLWGC is satisfied.
			
			On the magnetic side of the story, the unconfined monopoles are those whose charge $\tildeq_0^i$ has vanishing Dirac pairing with the Higgsed field:
			\begin{equation}
			\tildeq_0^i q^H_i = \tildeq_0^1 p + \tildeq_0^2 n = 0  ~~ \Rightarrow ~~ \frac{\tildeq_0^1}{\tildeq_0^2} = -\frac{n}{p} \,.
			\end{equation}
			The surviving magnetic charge $\tildeq$ of a monopole of charge $\tilde {q}^i$ is determined by projecting onto the subspace spanned by the charge of an unconfined monopole $\tildeq_0^i = (-n, p)$:
			\begin{equation}
			\tildeq =  e_{\rm KK}^2 \tilde {q}^i \tau_{ij} \tildeq_0^j = \frac{1}{p}  \tildeq^2 \,,
			\end{equation}
			where we have used the formula for $\tau_{ij}$ in \eqref{taured} and again set $\theta = 2 \pi n/p$. Here, we see that the unconfined monopole of charge $\tildeq^i = ( -n ,p)$ has charge $1$, whereas the confined monopoles have fractional charge $\tildeq \in \mathbb{Z}/p$. Thus, in agreement with the general analysis of \S\ref{ss.general}, the confined monopole lattice has coarseness $1/p$ and is dual to the sLWGC-satisfying sublattice provided there are no massless, charge 1 particles.
			
			Note that if massless particles of charge $1, ..., p-1$ do exist, however, then it is possible for the LWGC to be satisfied even though fractionally charged confined monopoles exist. Thus, in general, the dual of the confined monopole lattice may be a proper subset of the sLWGC-satisfying sublattice.

			\subsubsection{F-theory construction}
			
			This Kaluza-Klein analysis suggests a simple way to generate LWGC violation, by giving a vev to a massless charge
			$p$ scalar field in a $U(1)$ gauge theory without massless matter for some charge $q < p$.
			
			A number of explicit examples that fit this description were discovered in \cite{Raghuram:2018hjn}, which used F-theory to construct examples of 6d supergravity theories with massless matter in exotic representations of nonabelian gauge groups, which can be Higgsed to construct U(1) gauge theories with massless particles of large charge. The largest charge constructed in this way was $q=26$, which arises in a 6d theory with gauge group $U(1)$ and massless hypermultiplets of charge $q \in \{1, 4, 5, 6, 9, 10, 11, 12, 15, 16, 20, 21 \} \equiv \mathcal{Q}_0$.
			
			Following the steps outlined in the KK analysis above, we can compactify this theory on a circle to five dimensions, turning on a Wilson line $\theta = 2 \pi / 21$ and giving a vev to the massless KK mode of a particle with charge $p=21$. This vev breaks the gauge group $U(1) \times U(1)_{\rm KK
			}$ down to a $U(1)$ subgroup, and it results in a collection of extremal KK particles of charge $q_{\rm ext} = \pm q' + 21 n' $ for all $q' \in \mathcal{Q}_0$, $n' \in \mathbb{Z}$.
			This spectrum of superextremal particles violates the LWGC, and it satisfies the sLWGC only with a sublattice of coarseness $21$ or larger, which is by far the largest coarseness observed in any LWGC-violating example studied so far.

			It should be noted that this 6d theory is quite mysterious from a geometric perspective,\footnote{We thank Lara Anderson, James Gray, Andrew P. Turner, and Xingyang Yu for discussions on this point.} since direct analyses of Calabi-Yau compactifications in F/M-theory have yielded massless matter only up to charge 6 in both 6d \cite{Cianci:2018vwv} and 5d \cite{Collinucci:2019fnh}.\footnote{On the other hand, it has also been shown that there exist infinite families of 6d supergravity theories that are consistent with anomaly cancellation and feature massless fields with arbitrarily large $U(1)$ charges \cite{Taylor:2018khc}. Compactifying these theories with nonzero Wilson line $\theta$ could in principle lead to arbitrarily large violations of the sLWGC, though it is clear that all but a finite number of these theories reside in the swampland.} Relatedly, $\mathbb{Z}_6$ is the largest discrete symmetry group constructed in F-theory \cite{Aspinwall:1998xj, Anderson:2019kmx}. Furthermore, it has been proven that the Mordell-Weil torsion of an elliptically fibered Calabi-Yau threefold with non-constant $j$-invariant cannot be larger than $\mathbb{Z}_6$, suggesting that this is the largest discrete group achievable in F-theory \cite{Hajouji:2019vxs}. However, giving a vev to a charge $p=21$ field generates a $\mathbb{Z}_{21}$ subgroup, which seems to contradict this bound.
			
			It should be emphasized that the construction of the massless particles of large charge in \cite{Raghuram:2018hjn} involves a combination of F-theoretic geometric engineering and field-theoretic Higgsing processes, so an explicit F-theory realization of these models is still lacking, though it is difficult to see why the field theory analysis should fail to give an accurate description of the effective field theory. Clearly, more analysis is needed.

			For our purposes, let us ignore these puzzles and simply remark that although the LWGC is evidently violated with a large coarseness in this example, the WGC is satisfied at a large fraction of charge sites---namely, all except those of charge $q \equiv 3$, $7$, $8$, $13$, $14$, or $18$ mod $21$. Thus, there are many superextremal particles whose charges do not lie along the distinguished sLWGC-satisfying sublattice.

			\subsection{M-theory example}\label{ss.GMSV}
			
			In this subsection, we consider a UV-complete model of Higgsing in 5 dimensions, which arises from M-theory compactified on the Greene-Morrison-Vafa-Strominger (GMSV) geometry of \cite{Greene:1995hu, Greene:1996dh}. More details on this theory in the context of the WGC can be found in \cite{Alim:2021vhs}.
			
			The theory in question has a prepotential given by
			\begin{equation}
			\calF = \frac{5}{6} X^3 + 2 X^2 Y \,,
			\end{equation}
			with $X, Y >0$ in the phase of interest. The theory has two U(1) gauge fields.
			The vector multiplet moduli space is one-dimensional, given by the slice of the prepotential with $\calF=1$. 
			This prepotential determines the gauge kinetic matrix of the theory as
			\begin{equation}
			a_{ij} = \calF_i \calF_j - \calF_{ij}
			\label{aIJ}
			\end{equation}
			where $\calF_i = \partial_i \calF$, $\calF_{ij} = \partial_i \partial_j \calF$, and $I, J=1,2$.
			
			The mass of a BPS particle of charge $q_i=(q_1,q_2)$ is given by
			\begin{equation}
			m = \frac{g_5}{\sqrt{2}\kappa_5} |q_i Y^i|=\frac{g_5}{\sqrt{2}\kappa_5} |q_1 X + q_2 Y|\,,
			\label{5dBPS}
			\end{equation}
			where $g_5^2 = (2 \pi)^{4/3} (2 \kappa_5^2)^{1/3}$.
			The spectrum of BPS states can be computed using Gopakumar-Vafa invariants \cite{Gopakumar:1998jq, Gopakumar:1998ii}. As shown in \cite{Alim:2021vhs}, the BPS states saturate the LWGC in the directions of the charge lattice where BPS black holes exist;\footnote{More precisely, the BPS states saturate the LWGC for the finite set of charges computed. It is very likely but unproven that the LWGC is satisfied for all charges within the movable cone of the geometry, i.e., in the directions of the charge lattice for which BPS black holes exist.} in other directions, satisfying the LWGC requires non-BPS particles, the spectrum of which is highly nontrivial to compute.
			
			When $Y=0$, 16 curves in the geometry shrink and a conifold singularity develops. Physically, 16 BPS particles of charge $q_i=(0,1)$ become massless, and a Higgs branch opens up. Giving a vev to a particle of charge $q_{H,i}=(0,1)$ breaks the gauge group from $U(1)^2$ down to $U(1)$.
			
			From \S\ref{sss.constraint}, we naively expect that after Higgsing, the theory will satisfy the sLWGC with coarseness determined by the denominator of 
			\begin{align}
			\frac{q_i \tau^{ij} q_{H,j}}{||\bfq_H||^2} = \frac{q_i a^{ij} q_{H,j}}{q_{H,i} a^{ij} q_{H,j}} \,.
			\end{align}
			At the conifold locus $Y=0$, the inverse of gauge kinetic matrix \eqref{aIJ} is given by
			\begin{equation}
			a^{ij} = \left(\frac{5}{6}\right)^{1/3} \left( \begin{array}{cc}
				\frac35 & -\frac14 \\
				-\frac14 & \frac{5}{16}
			\end{array} \right) \,.
			\label{aGMSV}
			\end{equation}
			Setting $q_{H,i}=(0,1)$ and taking $q_i=(1,0)$ without loss of generality, we find
			\begin{align}
			\frac{q_i a^{ij} q_{H,j}}{q_{H,i} a^{ij} q_{H,j}} = -\frac{4}{5} \,.
			\end{align}
			Thus, by \eqref{pq}, we expect that the sLWGC will be satisfied with coarseness 5 after Higgsing.
			
			However, this naive conclusion is wrong, due to additional constraints imposed by supersymmetry. In fact, the LWGC is satisfied in this theory after Higgsing, and the resulting BPS spectrum matches that of M-theory compactified on the quintic \cite{Candelas:1990rm}.
			
			To see this, let us compare the WGC bound and the mass of the BPS particles after Higgsing. To begin, we have that the manifold $
			\Sigma_\perp$ is generated by $q_{\perp,i} = (5, 4)$, since this satisfies
			\begin{align}
			\bfq_\perp \cdot \bfq_H = 0\,,
			\end{align}
			by \eqref{aGMSV}. The BPS particles of charge $(5,4)$ are exactly extremal after Higgsing.
			
			Meanwhile, the surviving charge $q$ of a particle of original charge $q_i$ is then given by its projection along this direction,
			\begin{align}
			\Pi(q_i) = q_i - \frac{\bfq \cdot \bfq_H}{||\bfq_H||^2 }\bfq_H = \biggl(q_1, \frac{4}{5} q_1\biggr)\,.
			\end{align}
			So, the surviving charge is determined entirely by the charge $q_1$. After Higgsing, the mass of particle of charge $q_i$ is given by \eqref{5dBPS}:
			\begin{align}
			m = \frac{g_5}{\sqrt{2}\kappa_5}|q_1 X| = 
			m_{\rm ext} \,,
			\end{align}
			where crucially, we have set $Y=0$ at the conifold locus. Such a particle is exactly extremal, as evidenced by the fact that both its mass and its charge depend only on $q_1$ and not on $q_2$. This is a consequence of supersymmetry and ensures that the LWGC is satisfied even after Higgsing.
			
			The magnetic side of the story is a bit more complicated. Naively, by our general analysis above, we would expect to have confined monopoles of fractional charge $\tildeq \in \frac{1}5 \mathbb{Z}$. However, our general analysis above assumed that the gauge kinetic matrix was constant, whereas in this case the gauge kinetic matrix depends on massless moduli, which could ostensibly play a role in the solitonic flux tube solutions. The existence or nonexistence of fractionally charged confined monopoles in this theory remains an open question, which we leave for further analysis.

\section{Confined monopoles and non-invertible symmetries} \label{s.noninvertibles}

So far, we have argued that significant LWGC violation is associated with
fractional charged monopoles confined by low tension flux tubes. We now
highlight some intriguing parallels between these phenomena and non-invertible
symmetries. These parallels might just be coincidences, but equally they might
indicate a deeper connection that is still to be understood.\footnote{We thank Jacob McNamara for invaluable discussions and feedback on the ideas expressed in this section.}

\begin{figure}[h]
	\begin{center}
		\includegraphics[width = 100mm]{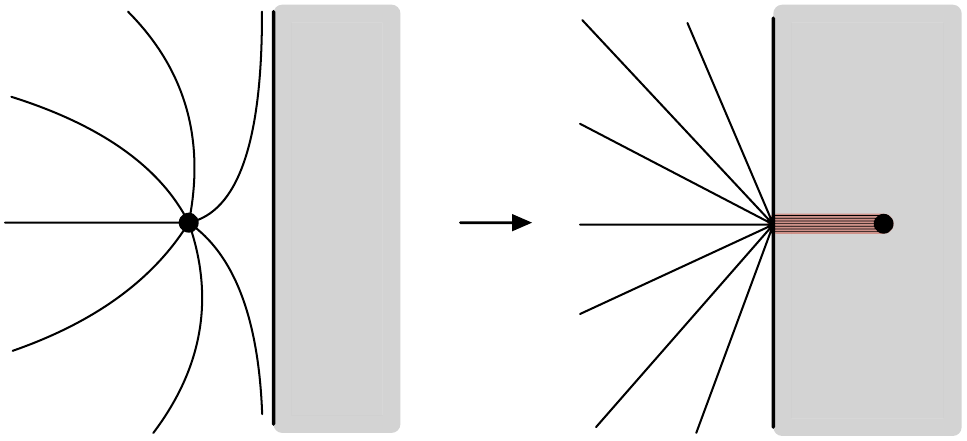}
	\end{center}
	\caption{The end result of pushing a monopole into a superconductor. The Meissner effect confines the monopole's field lines to one or more vortices connecting it to the surface of the superconductor.}
	\label{f.noninvertibles-Meissner}
\end{figure}

First, consider the behavior of a monopole near a domain wall connecting
confined and unconfined phases. As discussed in \S\ref{sec:Zp}, the
monopole cannot freely pass through the domain wall from the unconfined phase
into the confined phase. One can force it through by dragging a tube of the
domain wall into the confining phase, which collapses into the flux tube that
is attached to the monopole in the confined phase, see Figure~\ref{f.flux tubes.push}. This is very like what
would happen if a real monopole were pushed into a superconductor (without
mechanically damaging it): the magnetic field generated by the monopole would
initially disrupt the superconducting state, but eventually an equilibrium
would be reached with the monopole flux carried out of the superconductor
along one or more vortices, see Figure~\ref{f.noninvertibles-Meissner}.

Now we observe that this behavior is analogous to what happens when an
operator charged under a non-invertible symmetry is moved across the the
symmetry defect. As before, one can understand such a process by first pushing
the defect out of the way by forming a tube and then collapsing the tube to a
line. Because the symmetry defect is non-invertible, the result of this fusion
is not the identity but rather a sum of line defects tethering the charged
operator to the original symmetry operator.\footnote{This can be understood from the perspective that symmetry defects are maps between twisted Hilbert spaces, and in the case of a non-invertible defect this can map local operators into disorder operators.}

\begin{figure}[h]
	\begin{center}
		\includegraphics[width = 120mm]{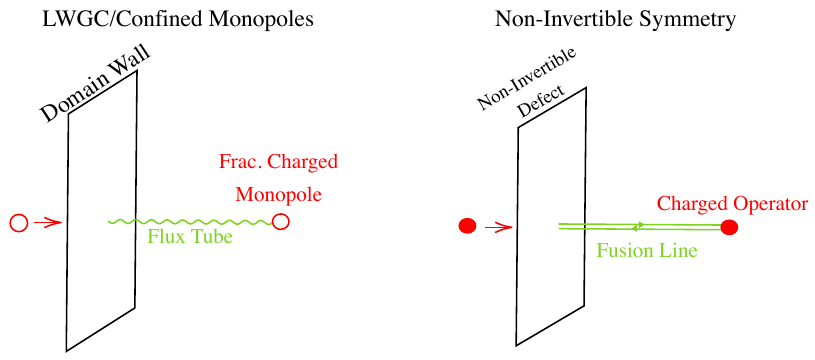}
	\end{center}
	\caption{Here we see the similar behavior between passing a fractionally charged monopole through the domain wall and passing a charged operator through a non-invertible defect.}
	\label{f.noninvertibles-monopoles-Schematic}
\end{figure}

This analogy is summarized in Figure \ref{f.noninvertibles-monopoles-Schematic}.\footnote{Of course, the two sides of a non-invertible symmetry
	operator are (non-trivially) isomorphic to each other, unlike the domain wall
	we are discussing. This is one of several points where our analogy fails to be
	precise.}
 Note that, since we are
comparing dynamical objects with operators, it is difficult to be precise
about what this analogy means. Moreover, the same analogy could be applied to
\emph{any} domain wall linking a confined phase with an unconfined phase,
so this observation is not yet specific to the theories with LWGC violation
that we are most interested in.

\begin{figure}[h]
	\begin{center}
		\includegraphics[width = 100mm]{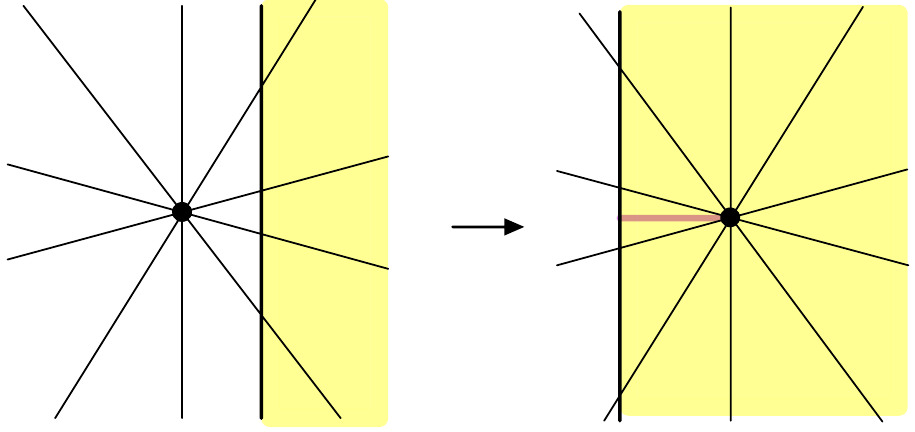}
	\end{center}
	\caption{In the case of fractional confinement, there is no Meissner effect. The magnetic field lines can spread out in both phases, yet the monopole becomes attached to a tube in one phase due to the existence of fractional electric charges (in units of the Dirac quantum associated to the monopole) in that phase.}
	\label{f.noninvertibles-FractionalConfinement}
\end{figure}

However, if we limit our attention to certain kinds of domain walls that
switch on/off LWGC violation while making minimal other changes, then we can
push this analogy further. Consider, for instance, the scenario described in
\S\ref{sec:Zp}, where the domain wall (composed of a $\mathbb{Z}_k$
twist vortex not wrapping the compact $S^1$) switches on/off the
$\mathbb{Z}_k$ discrete Wilson line. As described previously, a KK monopole
that begins in the phase without the $\mathbb{Z}_k$ Wilson line becomes
attached to a tube (comprised of a $\mathbb{Z}_k$ twist vortex wrapping the
compact $S^1$) upon being pushed across the domain wall into the phase with
the $\mathbb{Z}_k$ Wilson line. However, in some respects this is quite
\emph{different} from the generic scenario described above: the photon
under which the KK monopole is magnetically charged \emph{remains
	massless} in the partially confined phase with the $\mathbb{Z}_k$ Wilson line.
Thus, unlike a real monopole near a superconductor, the magnetic field lines
are not expelled from the partially confined phase, i.e., there is no Meissner
effect, see Figure~\ref{f.noninvertibles-FractionalConfinement}. This also implies that the tube to which
the KK monopole is attached \emph{carries no magnetic flux}, i.e., it is
not a ``flux tube'' at all.

The reason why the monopole must become attached to a tube after crossing the
domain wall is that in the phase with the $\mathbb{Z}_k$ Wilson line there are
additional, fractionally-charged electric particles that were absent in the
original phase. Since these new electric charges do not satisfy the Dirac
quantization condition with the KK monopole, they reveal a physical ``Dirac
string'' around which they have a non-trivial Aharonov-Bohm phase, i.e., the
tube connecting the monopole to the domain wall. (For this reason, one might
more correctly call this object an ``Aharonov-Bohm tube'' rather than a ``flux
tube''.)

A different phenomenon occurs when we try to push a fractional electric charge across
the domain wall into the phase without the $\mathbb{Z}_k$ Wilson line. The domain wall carries a worldvolume scalar (controlling the position of the twist vortex along the compact $S^1$) with a brane-localized St\"uckelberg coupling to the graviphoton. As the fractional electric charge passes through the domain wall, it excites this field, leaving behind a diffuse, fractional electric charge on the domain wall and emerging into the other phase with integral electric charge. While at first glance this seems to contradict our analogy with non-invertible symmetries, not all fusions in a symmetry category will be non-invertible, and we can interpret the electrically charged states as operators whose fusion with the non-invertible defect maps itself into the same Hilbert space.

\begin{figure}[h]
	\begin{center}
		\includegraphics[width = 50mm]{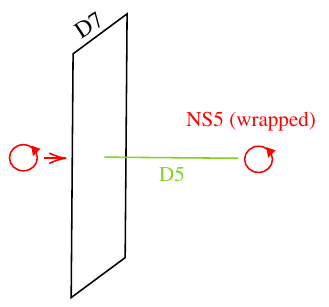}
	\end{center}
	\caption{In landscape examples such as the one discussed in~\S\ref{s.9dexample}, fractional monopole confinement is realized by the Hanany-Witten effect, pictured here for the AOB$_+$/AOB$_-$ D7-brane domain wall.}
	\label{f.noninvertibles-HW-Schematic}
\end{figure}

Domain walls of this type are not limited to the scenario discussed in
\S\ref{sec:Zp}. For instance, in the 9d theories discussed in
\S\ref{s.9dexample}, an unwrapped D7 brane creates a domain wall
connecting AOB$_+$ with AOB$_-$ or DP$_+$ with DP$_-$, and these domain walls
exhibit similar phenomena to those described above. Note that in this case,
the fact that a monopole pushed across the domain wall becomes attached to a
tube is simply the Hanany-Witten effect in disguise, see Figure \ref{f.noninvertibles-HW-Schematic}, where several past works \cite{Apruzzi:2022rei, Heckman:2022muc, r7branesaschargeconjugation} have pointed out connections
between the Hanany-Witten effect and non-invertible symmetries. In particular, in a holographic setting where symmetry operators of the boundary theory are dual to branes at infinity, non-invertible fusions such as in Figure \ref{f.noninvertibles-monopoles-Schematic} will be dual to Hanany-Witten brane configurations such as in Figure \ref{f.noninvertibles-HW-Schematic}.\footnote{If we take the strong view that \textit{every} topological operator in a holographic CFT is associated to a brane ``at infinity''~\cite{Sfolds, NonInvertFromHolographySakura, inakinoninvertible} then every non-invertible symmetry in such a theory must originate from bulk branes exhibiting Hanany-Witten-like behaviors.}

We do not presently know how to make the parallels highlighted above more
precise. We can speculate that there is some actual connection between the
physics we have described and non-invertible symmetries in some form---either
gauged or broken, since we are discussing theories with gravity---but there is
no certainty that such a connection exists. However, we find these parallels
to be intriguing. For instance, despite the fact that all the objects we
consider are dynamical, with finite masses or tensions, there is something
``almost topological'' about many of these phenomena, such as the
Aharonov-Bohm tubes which form, not due to the Meissner effect, but because of
the presence of fractional charges. This raises interesting dynamical
questions about the nature of the interaction between the domain wall and a
nearby monopole that we will not attempt to answer here.\footnote{This might
	be connected to another puzzle: in theories with a large sublattice splitting,
	even \emph{black holes} with charges off the extremal sublattice must be
	heavier than classically expected. What mechanism affects their mass? One
	might imagine that the existence of confined, fractional monopoles is
	important here, but why this ends up mattering is an enduring mystery.}	
			
We conclude this speculation by noting that in all string examples, confinement is imposed by the inconsistency of brane wrappings. It is therefore tempting to connect all of the above to a sort of generalized Freed-Witten anomaly, where the flux tubes allow for anomaly inflow to ameliorate the wrapping. A careful analysis of which generalized Freed-Witten anomalous configurations are amenable to anomaly inflow may yield an understanding of when fractionally confining monopoles source LWGC violation.	

\section{Conclusions\label{s.conclusions}}

In this paper, we have proposed that LWGC violation implies the existence of fractionally charged monopoles that are confined by flux tubes. We have seen that this proposal is satisfied in a number of examples in effective field theory, Kaluza-Klein theory, and string/M-theory.

We have further observed that in limits where sublattice splitting $\Tsplit$ becomes large, the tension of the flux tubes $T_\text{flux}$ becomes small. One can view this limit as a sort of dynamical change of the global structure of the gauge group, as certain electric charges become heavy and exit the spectrum while magnetic monopoles become unconfined and enter the spectrum.

The cases of LWGC violation we have seen are also characterized by the presence of massive gauge bosons. In such cases, the inverse relationship between $\Tsplit$ and $T_\text{flux}$ also manifests in the relation:
\begin{align}
	 \Tsplit T_\text{flux} \leq 
 c m_\gamma \,,
 \label{conceq}
	\end{align}
for some constant $c$, which is order-one in Planck units.

It is possible that the above statements could be strengthened and sharpened. For example, in all of the string theory examples explored here, we found that \eqref{conceq} was approximately saturated. The example of U(1) Higgsing in \S\ref{ss.pnt} did not saturate this bound, but we do not yet have an explicit realization of this scenario in string theory. It would be worthwhile to explore the conditions under which \eqref{conceq} could be promoted to an approximate equality and to pin down the value of the order-one constant $c$. More generally, it would be helpful to study monopole confinement and flux tube tensions in additional examples of LWGC violation, such as those that appear in the forthcoming work \cite{Baykara:Islands}.

In the 9d example of LWGC failure we have examined, $\Tsplit$ was either the tension of an unwrapped fundamental string, or an unwrapped D5-brane in the AOB phase. Upon dimensional reduction of this theory, $\Tsplit$ is still the tension of a fundamental brane. It would be interesting to see whether $\Tsplit$ more generally represents the tension of some brane of significance. If so, this may point towards a proof of our proposals above or enable stronger versions of these proposals.

In our analysis, we have focused on examples of LWGC failure where the WGC fails for infinitely many charges. In cases where LWGC failure occurs at only a finite number of lattice sites, such as the heterotic orbifold examples considered in \cite{Heidenreich:2016aqi}, \eqref{conceq} is satisfied trivially because $\msplit = 0$, according to our definition in \eqref{Deltadef}. However, there may be another definition of $\msplit$ in these cases that satisfies \eqref{conceq} in a less trivial manner. Further study of these examples is needed.

It would also be interesting to understand how the above story generalizes to the WGC at dimension 0 and codimension 2. At present, there is no precise, universally accepted statement of the WGC for these dimensions \cite{Arkanihamed:2006dz, Heidenreich:2015nta,Hebecker:2017wsu, Andriolo:2020lul, Harlow:2022ich}, so it is not entirely clear when LWGC violation has occurred. However, our work above suggests that such violations may be characterized by fractionally charged magnetic branes, which may therefore serve as a proxy for LWGC violation. This is especially important because violations of the LWGC at dimension 0 (also known as the axion LWGC) may be necessary to realize large-field natural inflation in string theory \cite{rudelius:2015xta, Montero:2015ofa, Brown:2015lia, Heidenreich:2019bjd}.

It would be worthwhile to further develop the connection between our work and non-invertible worldsheet symmetries \cite{Heckman:2024obe}. It is possible that there exist dual descriptions of the phenomena described in this paper in which non-invertible worldsheet symmetries are relevant. This might provide a new perspective on why the $\Tsplitstr$ and $\Tfluxstr$ in the 9d example examined in Section \ref{s.9dexample} are proportional to the DP-phase fundamental string tension.

Finally, it is tantalizing to speculate that a suitably defined version of the LWGC, as opposed to the sLWGC, may be true after all. All of the examples of LWGC violation we have seen involve gauge fields that have been given masses through Higgs/Stueckelberg mechanisms or else projected out of the spectrum through an orbifold action. In fact, in the landscape examples we studied the contributions to $\msplit$ come entirely from discrete gauge charge. With a suitable extension of the LWGC to massive and/or discrete gauge fields, it is possible that the full LWGC could be resuscitated.

			\section*{Acknowledgements}
			
			We are grateful for conversations with Lara Anderson, James Gray, Jacob McNamara, Miguel Montero, Hector Parra de Freitas, Andrew P. Turner and Xingyang Yu. ME was supported in part by the Heising-Simons Foundation, the Simons Foundation, and grant no.\ PHY-2309135 to the Kavli Institute for Theoretical Physics (KITP). ME, BH, NP, and SR received support from NSF grants PHY-2112800 and PHY-2412570. The work of TR was supported in part by STFC through grant ST/T000708/1. MR is supported in part by the DOE Grant DE-SC0013607. Part of this work was carried out at the 2024 Simons Summer Physics Workshop at the Simons Center for Geometry and Physics, Stony Brook University, and at the ``Landscape vs.~the Swampland'' workshop at the Erwin Schr\"odinger Institute in Vienna.
			
			\appendix
				
			\section{Conventions and useful formulas \label{s.conventions}}
			
			\subsection{$\text{GL}(2,\mathbb Z)$ on $\tau$, 2-form potentials, and 1-branes}
			
			Following the conventions of \cite{Polchinski:1998rr}, the $SL(2,\mathbb{Z})$ action on the two-form field doublet is 
			\begin{equation}
			\begin{pmatrix}
				C_2\\
				B_2
			\end{pmatrix}  \xrightarrow{\Lambda }  
			\begin{pmatrix}
				C_2'\\
				B_2'
			\end{pmatrix} = \underbrace{\begin{pmatrix}
					a & b\\
					c & d
			\end{pmatrix}}_{\equiv \Lambda} \begin{pmatrix}
				C_2\\
				B_2
			\end{pmatrix},
			\label{e.CBtrans}
			\end{equation}
			where $\text{det}(\Lambda)=1$.
			
			The corresponding transformation on the axiodilaton $\tau$ which leaves the IIB effective action invariant is given by\footnote{Often the transformation on $\tau$ is specified first, but we should actually define the transformation by the action on the two-form fields, since $\tau$ transforms in a projective representation where different actions on the two-form fields can yield the same effect on $\tau$.}
			\begin{equation}
			\Lambda:\tau \rightarrow \frac{a \tau + b}{c \tau + d} = \begin{pmatrix}
				a & b\\
				c & d
			\end{pmatrix} \tau.
			\end{equation}

			To incorporate worldsheet parity $\Omega$, we extend SL$(2,\mathbb Z)$ to include actions within $\mathrm{GL} (2,\mathbb Z)$. Elements $\Lambda \in \mathrm{GL} (2,\mathbb Z)$
			with $\det (\Lambda) = \pm 1$ act differently on $\tau$---the ones with positive or
			negative determinants are respectively holomorphic or antiholomorphic,
			\begin{equation}
			\tau \rightarrow \frac{{a \tau }^{} + b}{c \tau  + d}^{} \quad \mathrm{for}
			\det (\Lambda) = + 1, \quad \tau \rightarrow \frac{{a \tau^{\ast}}^{} +
				b}{{c \tau^{\ast}}^{} + d}^{} \quad \mathrm{for} \det (\Lambda) = - 1
				\end{equation}
				$\Omega$ lies in the latter component, since $\Omega$ reflects $C_0
				\rightarrow - C_0$ while leaving the dilaton $\Phi$ invariant, which when
				combined into the axiodilaton field $\tau = C_0 + i e^{- \Phi}$, produces an
				antiholomorphic action:
				\begin{equation}
			\Omega : \quad \tau \rightarrow - \tau^{\star}, \quad C_0 + i e^{- \Phi}
			\rightarrow - C_0 + i e^{- \Phi},\quad \begin{pmatrix}
				C_0\\C_2\\B_2\\C_4
			\end{pmatrix}\rightarrow\begin{pmatrix}
				-C_0\\C_2\\-B_2\\-C_4
			\end{pmatrix} .
			\end{equation}

			\subsection{String transformations under SL$(2,\mathbb{Z})$}
			The worldsheet of a bound state of $p$ fundamental strings and $q$ D1 strings, i.e. a $(p,q)$ string, is
			\begin{align}
			\int p_1B_2+q_1C_2.
			\end{align}

			The worldsheet action for $(p_1,q_1)$-strings should remain invariant between different dual descriptions, so
			\begin{align}
			\begin{pmatrix}
				q_1&p_1
			\end{pmatrix}\begin{pmatrix}
				C_2\\B_2
			\end{pmatrix}=\begin{pmatrix}
				q_1'&p_1'
			\end{pmatrix}\begin{pmatrix}
				C_2'\\B_2'
			\end{pmatrix}=\begin{pmatrix}
				q_1'&p_1'
			\end{pmatrix}\Lambda\begin{pmatrix}
				C_2\\B_2
			\end{pmatrix}.
			\end{align}
			Thus,
			\begin{align}
			\begin{pmatrix}
				q_1&p_1
			\end{pmatrix} =\begin{pmatrix}
				q_1'&p_1'
			\end{pmatrix}\Lambda,
			\end{align}
			or,
			\begin{align}
			\begin{pmatrix}
				p_1'\\q_1'
			\end{pmatrix}=\begin{pmatrix}
				\frac{a p_1-b q_1}{ad-bc}\\\frac{d q_1-c p_1}{ad-bc}
			\end{pmatrix}=\frac 1{ad-bc}\begin{pmatrix}
				a&-b\\-c&d
			\end{pmatrix}\begin{pmatrix}
				p_1\\q _1
			\end{pmatrix}=\frac 1{\det \Lambda}\begin{pmatrix}
				1&0\\0&-1
			\end{pmatrix}\Lambda \begin{pmatrix}
				1&0\\0&-1
			\end{pmatrix}\begin{pmatrix}
				p_1\\q_1
			\end{pmatrix}.
			\end{align}

			For the specific $GL(2,\mathbb{Z})$ transformation implemented by the $\Omega T$ Wilson line, we thus have
			
			\begin{align}
				\Lambda=\Omega T=\begin{pmatrix}
					1&0\\0&-1
				\end{pmatrix}\begin{pmatrix}
					1&1\\0&1
				\end{pmatrix},\qquad\Rightarrow\qquad 
				\begin{pmatrix}
					p_1'\\q_1'
				\end{pmatrix}
				=
				\begin{pmatrix}
					q_1-p_1\\q_1
				\end{pmatrix}.
			\end{align}

			\subsection{Fivebrane transformations}
			Let us refer to the bound state of $p_5$ D5-branes with $q_5$ NS5-branes as a $(p_5,q_5)$-fivebrane.

			Unlike defining the electric charges of strings through their couplings, we will define the magnetic charges of fivebranes under the two-form fields via flux integrals:
			\begin{align}
				\begin{pmatrix}
					p_5\\q_5 
				\end{pmatrix}\propto 
				\begin{pmatrix}
					\oint_{S^3}F_3\\-\oint_{S^3} H_3
				\end{pmatrix},
			\end{align}
			where locally $F = d C_2$, $H = d B_2$, and the $S^3$ links the fivebranes.
			
			Some brief comments are in order as to why we chose our conventions of putting the RR field first for the fivebranes, and the appearance of the additional minus sign when defining $q_5$. By choosing to have $p_5$ label the D5 charge, the tension formula $T \sim |p+\tau q|/\sqrt{\tau_2}$ applies to both strings and fivebranes. By including the minus sign, we ensure that both strings and fivebranes transform with the matrix $\begin{pmatrix}
				a & - b\\
				- c & d
			\end{pmatrix}$, up to the lack of additional overall $-1$ for the action of disconnected components on the 1-brane charge vector.\footnote{This extra $-1$ occurs due to the appearance of a $\text{det}(\Lambda^{-1})$ factor, which equals $-1$ in the disconnected component, when reading off string charges from their coupling to $(C_2,B_2)$.}
			
			We will quickly illustrate the utility of the minus sign: under $\mathrm{GL} (2, \mathbb Z)$, the flux integrals and thus magnetic charges transform in the same way that the two-form \textit{gauge fields} do (rather than the same way that the electric charges transform, which involves the inverse matrix),
			\begin{equation}
				\begin{pmatrix}
					\left( \oint F_3 \right)'\\
					\left( \oint H_3 \right)'
				\end{pmatrix} = \begin{pmatrix}
					a & b\\
					c & d
				\end{pmatrix} \begin{pmatrix}
					\oint F_3\\
					\oint H_3
				\end{pmatrix}
			\end{equation}
			If we choose to define $q_5 = - \oint H_3$, this relation is written in terms of charges as
			\begin{equation}
				\begin{pmatrix}
					p_5'\\
					q_5'
				\end{pmatrix} = \begin{pmatrix}
					a & - b\\
					- c & d
				\end{pmatrix} \begin{pmatrix}
					p_5\\
					q_5
				\end{pmatrix}=\begin{pmatrix}
					1&0\\0&-1
				\end{pmatrix}\Lambda \begin{pmatrix}
					1&0\\0&-1
				\end{pmatrix} \begin{pmatrix}
					p_5\\
					q_5
				\end{pmatrix}, \label{e.pqfivebranetrans}
			\end{equation}
			which enjoys the same matrix, up to the aforementioned $-1$, as the 1-brane charge vector.
			
			Thus, under the Wilson line $\Omega T$, in our conventions, the fivebrane charge vector transforms as
			\begin{align}
				\begin{pmatrix}
					p_5'\\q_5'
				\end{pmatrix}=\begin{pmatrix}
					p_5-q_5\\-q_5
				\end{pmatrix}
			\end{align}

\section{Coordinates on moduli space}\label{ss.coordinatesof9dphases}
            \subsection{SL$(2,\ZZ)$ duality between DP$_-$ and AOB$_-$}

            In this appendix, we describe several useful sets of coordinates on the moduli space and their relationships to physical parameters. Above, we have used the  canonically normalized dilaton  $\phi$ and radion $\rho$ of the DP$_-$ phase to describe the entire moduli space. However, other coordinates may be more useful for certain calculations.

            To begin, we identify the appropriate strong-weak $\mathrm{SL}(2,\ZZ)$ transformation between DP$_-$ and AOB$_-$.\footnote{This will differ from the transformation given in \cite{Montero:2022vva} due to our different $C_0$ value.} As usual for dualities, this allows one to realize any phenomena that occur at strong DP string coupling in terms of the weakly-coupled AOB dual.
            
            The transformation exchanges strong string coupling with weak string coupling but maintains the axion background $C_0 = - 1 / 2$ (in particular, this means the strong-weak duality is not merely $S \in \mathrm{SL}(2,\ZZ)$). An appropriate element $S'$ of $\mathrm{SL}(2,\ZZ)$ is given by
            \be
S'=\begin{pmatrix}
                -1 & - 1\\
                2 & 1
              \end{pmatrix}
              \,.
\ee
For reference, we note that this can be written as a sequential action of $\mathrm{SL}(2,\ZZ)$ generators as $S T T S T$, where 
\be
S=\begin{pmatrix}
                0 & - 1\\
                1 & 0
              \end{pmatrix}
            \,,~~~~T=\begin{pmatrix}
                1 & 1\\
                0 & 1
              \end{pmatrix}\,.
\ee
            The corresponding action of $S'$ on the axiodilaton is then given by
        \begin{equation}
              \tau = -\frac{1}{2}+i e^{-\Phi} \rightarrow \tau' = \begin{pmatrix}
                - 1 & - 1\\
                2 & 1
              \end{pmatrix} \tau = - \frac{1}{2} + i\frac{1}{4}e^{\Phi}\,,
            \end{equation}
            from which we can read off the string coupling in the new frame in terms of the old frame
            \begin{equation}
              e^{-\Phi_{\text{AOB}}} = \frac{1}{4}e^{\Phi_{\text{DP}}} \quad \iff \quad g_{S,\text{AOB}}^{-1}=\frac{1}{4}g_{S,\text{DP}}\,.
            \end{equation}     

            \subsection{Normalization conventions}
We may then write our canonically normalized dilaton as
\be
\phi = \frac{1}{\sqrt{2}} (\Phi_{\rm DP} - \log 2)\,,
\ee
where the constant $\log 2$ is chosen so that the tension of the AOB and DP fundamental strings are equal when $\phi=0$:
\be
T_{\rm AOB} = \tilde t_1 e^{\Phi_{\rm AOB}/2} = \tilde t_1 e^{\Phi_{\rm DP}/2} = T_{\rm DP} ~~~\Leftrightarrow ~~~ \phi = 0\,.
\ee
This fixes the dilaton's additive constant in the sense that we declare the dilaton such that the tensions of the F-strings are $e^{\Phi/2}$ in both the DP and AOB frames---then, the constraint that $\Phi_\text{AOB}$ is related in a definite way to $\Phi_\text{DP}$ via SL$(2,\ZZ)$ fixes the additive constant.

Similarly, the canonically normalized radion is given by
\begin{align}
\rho(x) = \dfrac{1}{\kappa_d} \sqrt{\dfrac{d-1}{d-2}} \log \dfrac{R}{R_0}\,,
            \end{align}
where $R$ is the (dimensionful) radius of the compactification circle of the DP$_-$ phase and $R_0$ is chosen so that the self T-dual line is located at $\rho_\text{DP} =\phi_\text{DP}^{(9)}/\sqrt 7$.

			\bibliographystyle{JHEP}
			\bibliography{ref}
		\end{document}